\documentclass[twocolumn,tighten]{aastex63}
\usepackage{graphicx}
\usepackage{gensymb}
\usepackage{amsmath}
\usepackage{amsmath}

\def\O3{[\ion{O}{3}]}

\begin{document}
\title{Deep Optical Emission-Line Images of Nine Known and Three New Galactic Supernova Remnants}

\author[0000-0003-3829-2056]{Robert A.\ Fesen}
\affiliation{6127 Wilder Lab, Department of Physics and Astronomy, Dartmouth College, Hanover, NH, 03755, USA}

\author[0000-0002-7855-3292]{Marcel Drechsler}
\affiliation{\'Equipe StDr, B{\"a}renstein, Feldstraße 17, 09471 B{\"a}renstein, Germany}

\author[0000-0002-3172-965X]{Xavier Strottner}
\affiliation{\'Equipe StDr, Montfraze, 01370 Saint Etienne Du Bois, France}

\author{Bray Falls}
\affiliation{Sierra Remote Observatories, Auberry, CA, 93602, USA}

\author{Yann Sainty}
\affiliation{YSTY Astronomy, 54000 Nancy, Lorraine, France}

\author{Nicolas Martino}
\affiliation{Various Amateur Observatory Sites, Lorraine, France}

\author{Richard Galli}
\affil{Oukaimeden Observatory, Oukaimeden, Morocco}

\author{Mathew Ludgate}
\affil{Ross Creek Observatory, Dunedin 9010, New Zealand}

\author{Markus Blauensteiner}
\affil{Fachgruppe Astrofotografie, PO Box 1169, D-64629, Heppenheim, Germany}

\author[0000-0002-5313-6409]{Wolfgang Reich}
\affil{Max-Planck-Institut  f\"ur Radioastronomie,
Auf dem H\"ugel 69, 53121 Bonn, Germany}

\author[0000-0003-2692-2321]{Sean Walker}
\affil{MDW Sky Survey, New Mexico Skies Observatory, Mayhill, NM 88339, USA}

\author[0000-0003-1235-7173] {Dennis di Cicco}
\affil{MDW Sky Survey, New Mexico Skies Observatory, Mayhill, NM 88339, USA}

\author{David Mittelman}
\affil{MDW Sky Survey, New Mexico Skies Observatory, Mayhill, NM 88339, USA}

\author{Curtis Morgan}
\affiliation{Sierra Remote Observatories, Auberry, CA, 93602, USA}

\author{Aziz Ettahar Kaeouach}
\affil{High Atlas Observatory, Oukaimeden Observatory, Oukaimeden, Morocco}

\author{Justin Rupert}
\affil{MDM Observatory, Kitt Peak National Observatory, 950 N. Cherry Ave., Tucson, AZ 85719, USA}

\author{Zouhair Benkhaldoun}
\affil{Oukaimeden Observatory, High Energy Physics and Astrophysics Lab, Faculty of Sciences Semlalia, \\ Cadi Ayyad University, Marrakech, Morocco}


\begin{abstract}

Deep optical emission-line images are presented for nine known plus three new  Galactic supernova remnants (SNRs), all but one having at least one angular dimension greater than one degree. 
Wide-field images taken in H$\alpha$ and \O3 $\lambda$5007
reveal many new and surprising remnant
structures including large remnant shock  extensions and `breakout' features not seen in published optical or radio data. These images represent over 12,000 individual images totaling more than 1000 hours of exposure time
taken over the last two years mainly using small aperture telescopes which detected fainter nebular 
line emissions than  published emission-line images. During the course of this imaging program, we discovered three new SNRs, namely
G107.5-5.1 (the Nereides Nebula), G209.9-8.2, and G210.5+1.3,
two of which have diameters $>1.5\degr$. 
Besides offering greater structural detail on the nine already known SNRs, 
a key finding of this study is the importance of \O3 emission-line imaging for mapping the complete shock emissions of Galactic SNRs.

\end{abstract}
\bigskip


\section{Introduction}

\subsection{Background and History} 
At present there are just over 300 confirmed Galactic supernova remnants (SNRs) 
cataloged \citep{Safi2012,Green2019}. Most are less than a degree in angular size, 
located within five degrees of the Galactic plane, more than 1 kpc distant, and have typical estimated ages between 10$^{4}$ and 10$^{5}$ yr \citep{Green2004,Shan2018,Leahy2022}. 
The vast majority are discovered through radio observations due to their characteristic synchrotron nonthermal emission associated with shocked gas 
leading to a power law flux density, S, with S $\varpropto$  $\nu^{-\alpha}$ where $\alpha$ is the emission spectral index with typical values between 
0.3 and  0.7
\citep{Reynolds2011,Dubner2015}. 


Only a fraction of Galactic SNRs exhibit optical emissions.
\citet{vdb1973} was the first to publish an  atlas of optical images for 23 of the 24 Galactic SNRs known at that time to exhibit optical emission. Subsequently, focused searchers for optical SNRs used a variety of telescopes and wide-field Schmidt cameras to search for optical SNR emissions (\citealt{vdb1978a,Zealey1979}).  
Coincidence of optical emission with that of a remnant's nonthermal radio emission 
using the first and second Palomar Observatory Sky Survey (POSS~I, POSS~II) images and their digitized versions (DSS1, DSS2), plus Schmidt images of the southern hemisphere led to several new optical SNR detections. 

The first comprehensive optical survey of the Galactic plane
to uncover new optical SNRs 
was the photographic Emission-Line Survey of the Milky Way 
(ELSMW; \citealt{Parker1979}). 
This survey used a small
commercial camera lens mounted ahead of a two-stage image intensifier
to generate moderately deep images in H$\beta$, 
[\ion{O}{3}] $\lambda$5007, H$\alpha$ + [\ion{N}{2}] $\lambda\lambda$6548,6583, and [\ion{S}{2}] $\lambda\lambda$6716,6731 line emissions. 
Along with the discovery of new optical SNRs, this survey led to the 
discovery of several large planetary nebulae (PNe), Wolf-Rayet and OB star ring nebulae.
However, with resolution of only  $\sim30''$ and because it was published in book form 
and not later scanned and digitized, its usefulness was  limited.



Today, the identification of a Galactic SNR's optical emission is most often done
by either focused imaging searches using 
narrow emission-line bandpass filters (e.g., \citealt{Mav2001,Mav2005,Mav2009,Boumis2002,Boumis2009,Sezer2012, How2018, Fesen2019,Fesen2020,Bakis2023}) or using recent H$\alpha$ surveys of the Galactic plane. 

Such surveys include the Marseille H$\alpha$ Survey (MHS; \citealt{LeCoarer1992, Georgelin1994,Georgelin1996, Georgelin2000A, Russeil1998}),
the Virginia Tech Spectral-line Survey (VTSS; \citealt{Dennison1998}), the Southern H$\alpha$ Sky Survey Atlas (SHASSA; \citealt{Gaustad2001}),
the Wisconsin H$\alpha$ Mapper (WHAM; \citealt{Haffner2003}),
the AAO/UKST SuperCOSMOS H$\alpha$ survey \citep{Parker2005}, and the
Issac Newton Telescope Photometric H$\alpha$ Survey of the Northern Galactic Plane 
(IPHAS; \citealt{Drew2005}).
These surveys have served to identify optical emissions from known SNRs (e.g., \citealt{Walker2001, Stupar2008,Stupar2011}) or aid in discovering new ones  \citep{Stupar2007,Stupar2012,Stupar2018,Sabin2013,Fesen2015,Fesen2021}

While positional coincidence of optical emission with a remnant's radio emission is suggestive, 
firm identification of a Galactic remnant's associated optical emission is mainly done through 
H$\alpha$ and [\ion{S}{2}] emission-line images and follow-up optical spectra.
In older SNRs whose optical emissions are not dominated by SN ejecta, identifying optical emission arising from shocked interstellar gas (as opposed to photoionized gas) can be established through spectra showing a line ratio of I([\ion{S}{2}])/I(H$\alpha$) $\geq 0.4$ \citep{Blair1981,Dopita1984,Fesen1985,Leonidaki2013,Long2017}.

Unlike the case of steady state photoionization like that
seen in H~II regions, optical shock emission comes from postshock regions which have
extended cool recombination zones rich in S$^{+}$ ions.
An empirical [\ion{S}{2}] $\lambda\lambda$6716,6731  vs.\ H$\alpha$ intensity ratio has proven quite useful in identifying the shocked emission of SNRs in both the Milky Way and nearby local group galaxies \citep{Leonidaki2013,Kop2020,Long2017}.

Despite numerous searchers, today only about 90 SNRs or roughly 30\% of the currently cataloged Galactic remnants 
(\citealt{Green2019} and on-line updates)
exhibit any appreciable associated optical emission. Many of these are weak or partial detections and/or limited to just H$\alpha$ emission. This is unfortunate since 
a remnant's optical emission can be useful in both identifying regions of higher-velocity shocks via the presence of high ionization emission lines such as [\ion{O}{3}] and aiding in defining a remnant's overall size and morphology. Moreover, a remnant's optical emission can often help define the structure and extent of some SNRs, especially for very faint radio SNRs.

\begin{deluxetable*}{lccccccc}[ht]
\tablecolumns{8}
\tablecaption{Comparison of Wide-Field Galactic Emission-Line Imaging Surveys}
\tablewidth{0pt}
\tablehead{ 
\colhead{Survey}& \colhead{Telescope}& \colhead{FOV}       & \colhead{Resolution} & \colhead{Emission Line} & \colhead{Bandpass} 
& \colhead{Exposure} & \colhead{Approx. Depth}    \\
\colhead{Name}  & \colhead{Aperture} & \colhead{(degrees)} & \colhead{(arcsec)}   & \colhead{Filter(s)} &  \colhead{(FWHM \AA)}  
& \colhead{Time (hr)} & \colhead{(Rayleighs$^{\star}$)}  }
\startdata
{\underline{\bf{Professional}}} &               &          &          &                            &             &       &        \\    
 ELSMW$^{1}$   & Nikkor 10 cm lens  &  7.1     & 30 -- 40 & H$\alpha$, [O III], [S II] & 75, 28, 50  &  0.25 & $\sim$10 -- 30 \\  
 VTSS$^{2}$   &  Nikkor 5 cm lens  & $10 \times  10$   &  96   &  H$\alpha$             & 17  & 2.0     & 2  \\
 SHASSA$^{3}$ &  Canon 3.3 cm lens &   $13 \times 13$  & 48    &  H$\alpha$             &   32  & 1.7        & 0.5 -- 2  \\
 WHAM$^{4}$   & 0.6 m lens         &    1             & 3600  &  H$\alpha$             & 0.25  & 0.003   & 0.15    \\ 
 IPHAS$^{5}$  &   INT 2.5 m         & $0.3 \times 0.3$  &  $\simeq$ 1    &  H$\alpha$    & 95   &  0.07 &  3  \\
 SHS$^{6}$    &  AAO/UKST 1.2 m    & $5.5 \times 5.5$  &  $\simeq$ 1  &  H$\alpha$             & 80    &  3.0      &  $\leq$5  \\
 Condor$^{7}$ & $6 \times 18$ cm lenses & $2.3 \times 1.5$ & $\simeq$ 1  & H$\alpha$, [O III], [S II] &  40       & 1 -- 20 & $\sim$ 3 -- 5   \\
 {\underline{\bf{Amateur}}} &                         &                 &       &                            &              &         &     \\             
 MDW$^{8}$    & Astro-Physics 13 cm& $3.4 \times 3.2$  & $\simeq$ 1 -- 2 & H$\alpha$, [O III]     & 30    & 4.0    &  $\sim$ 3 -- 5 \\
 This work    & 8 cm -- 40 cm           & $\sim2$ --  $10$   & 1 -- 5 & H$\alpha$, [O III]        & 30 -- 50 & 5 -- 110 & 0.5 -- 5 \\
\enddata
\tablenotetext{}{$^{\star}$Rayleigh (R) = $3.715 \times 10^{-14}$ $\lambda^{-1}$ erg cm$^{-2}$ s$^{-1}$ arcsec$^{-2}$; R (H$\alpha$) = $5.67 \times 10^{-18}$ erg cm$^{-2}$ s$^{-1}$ arcsec$^{-2}$.}
\tablenotetext{}{$^{1}$\citet{Parker1979}; $^{2}$\citet{Dennison1998}; $^{3}$\citet{Gaustad2001}; $^{4}$\citet{Haffner2003};
$^{5}$\citet{Drew2005}; $^{6}$\citet{Parker2005}; $^{7}$\citet{Lanzetta2023}, $^{8}$https://www.mdwskysurvey.org/ }
\end{deluxetable*}

\subsection{Wide-Field Imaging with Small Telescopes}

Some 45 remnants or roughly 15\% of the 300+ confirmed Galactic SNRs have angular dimensions greater than one degree, with more than a dozen larger than 2 degrees. Such large remnants are difficult to adequately image using large telescopes with typical fields of view (FOV) less than half a degree.
Although wide-field imaging of the sky with FOVs of several degrees has long been possible
using various Schmidt-type cameras, such optical systems cannot be easily adapted for use with narrow bandpass filters (FWHM $<$ 100 \AA) centered on bright nebular emission lines \citep{Meaburn1978}. 


On the other hand, small aperture telescopes with fast optics have both much larger fields of view than large telescopes and are capable of detecting very faint emission nebulae.
Since surface brightness is independent of telescope aperture, and because detected surface brightness is a function of pixel size on
the sky, the number of pixels in an object, and 
the exposure time, long series of exposures taken with small telescopes with large FOVs and pixel scales $\sim2''$ have been shown to detect faint emission from large, extended objects as good or even better than using large telescopes (e.g., 
\citealt{Java2016,Martinez2020}).  

This realization has lead professional investigators
to use amateur size telescopes or small cameras in arrays to image
faint galactic halos, inter-cluster light and outer regions around bright spiral galaxies \citep{Abraham2014,Martinez2021,Gilhuly2022} and 
an assortment of large-scale tidal structures around nearby massive galaxies \citep{Martinez2015}.
There are now several arrays of small, amateur sized telescopes built
seeking to detect low surface brightness features around nearby galaxies with selected broadband filters. These
include the Dragonfly Array with 48 cameras \citep{Abraham2014,Lanzetta2023}, the KiloDegree Survey (KiDS: \citealt{Kuijken2019}), the Huntsman Telescope \citep{Spitler2019} and the Large Array Survey Telescope (LAST) \citep{Ben-Ami023}.

\subsection{Nebular Emission-Line Imaging}

Following the use of small telescopes and camera arrays by professional 
astronomers, the development of affordable large-format CMOS detectors and 
extremely high transmission (T $\geq95$\%) narrow bandpass 
filters (FWHM $\approx$ 30 \AA) has led to a recent revolution in deep emission-line imaging
of Galactic nebulae by amateurs.
The combination of literally hundreds of exposures taken
using small telescopes equipped
with high-throughput emission-line filters and sensitive
digital detectors have enabled amateur astronomers to detect previously unknown large and
faint Galactic emission line nebulae. And, unlike several professional
sky surveys, such imaging has included a variety of nebular emission lines and covered regions both near and far from the Galactic plane.

Table 1 shows a comparison of
optical emission-line surveys divided into professional
and amateur equipment. In most cases, the values listed come from the published papers on the respective surveys. The approximate imaging sensitivity or depth in Rayleighs for the ELSMW and MDW surveys comes from comparisons of their images with other H$\alpha$ surveys.

As can be seen in this table comparison, amateurs using small telescopes with narrow,
high throughput bandpass filters
plus long series of exposures totaling tens and even hundreds of hours
can reach sensitivity levels equal to or exceeding that of large professional Galactic nebulae surveys. Indeed, images taken by amateurs can be as deep
as that cited for IPHAS, SHASSA or SHS images in H$\alpha$. 
Consequently, it has been said that images taken by amateurs can ``far out-distance
the images produced by most professional astronomers'' \citep{Abraham2017}.
Moreover, small aperture 
[\ion{O}{3}] imaging with total exposure times of $\sim50 - 100$ hr, have shown that brightness levels as low as $\simeq$ 0.2 Rayleigh is possible \citep{Fesen2023} thereby approaching the depth of WHAM in H$\alpha$ but with a resolution of a few arcsec instead of WHAM's $\sim$1 degree resolution.

Since small telescope/camera systems have angular resolutions comparable to the
large scale professional surveys; namely, 1--3 arcsec but FOVs $\sim$2--10 degrees, this makes them ideal for imaging large Galactic nebulae.
Consequently, with sufficiently long exposures times, small telescopes with fast optics and large pixel scales equipped with narrow bandpass
filters (FWHM $\simeq$ 30 -- 50 \AA) plus 
efficient low readout CMOS detectors are powerful new tools for imaging especially large and faint nebulae including Galactic SNRs. 


\bigskip

\bigskip

\subsection{Aim of this Study}

Because many large angular Galactic SNRs are relatively nearby (d $\lesssim$ 3 kpc), they
offer higher spatial resolution of a remnant's interstellar shock structure than more distant objects.
Furthermore,
as a SNR ages into the later radiative phase of evolution, both their radio and optical
emissions often become more filamentary due to unstable post-shock cooling leading to small scale shock features \citep{Duin1975,Gaetz1988,Raymond2020}.

Unfortunately, the existing quality and coverage of optical images for large, 
Galactic SNRs is highly inhomogeneous with few well studied
large remnants, such as the Cygnus Loop and IC443. Published images 
of many large Galactic SNRs are often of relatively low resolution and S/N or offer incomplete image coverage.

The aim of this work is to present high-quality, wide FOV optical emission-line images 
of several of the largest known but poorly studied Galactic SNRs in order to characterize their overall morphology and fine-scale structure. Our survey also
provides a test on the value of deep emission-line imaging of Galactic remnants not known to exhibit any strong optical emission.
During the course of this imaging program, we discovered three new SNR candidates (namely, G107.5-5.1, G209.9-8.2, and G210.5+1.3) and we present our image results on these remnants as well.

The outline of the paper is as follows: $\S2$ describes the imaging data 
along with limited optical spectra taken of the new SNRs,
$\S3$ presents the image results along with brief reviews of earlier 
published images (where present).
We then briefly review some of the most interesting properties of these 12 remnants in $\S$4 along with some key findings and conclusions from this imaging survey.

\begin{deluxetable}{lccc}[ht]
\tablecolumns{4}
\tablecaption{Data on the Galactic SNRs in our Survey}
\tablewidth{0pt}
\tablehead{ 
\colhead{SNR} & \multicolumn{2}{c}{\underline {Approx. Center (J2000) } }& \colhead{Angular Size}  \\
\colhead{ID}  & \colhead{RA} & \colhead{Dec}  & \colhead {(degrees)}  } 
\startdata
{\underline{Known SNRs:}}   &           &          &                   \\
G13.3$-$1.3  &  18:19:20 & $-18$:00 & $0.8\times 1.2$     \\
G70.0$-$21.5   &  21:24:00 & $+19$:23 & $4.0 \times 5.5$   \\
G82.2+5.3    &  20:19:00 & +45:30   & $1.0 \times 1.6$   \\
G89.0+4.7    &  20:45:00 & +50:35   & $1.5 \times 2.0$   \\
G119.5+10.2   &  00:06:40 & +72:45   & $1.5 \times 1.7$   \\
G150.3+4.5    &  04:27:00 & +55:28   & $2.5 \times 3.0$   \\
G181.1+9.5    &  06:26:40 & +32:30   & $1.2 \times 1.2$   \\
G288.8$-$6.3  &  10:30:22 & +65:13   & $1.6 \times 1.8$  \\
G321.3$-$3.9  &  15:32:14 & $-60$:52 & $1.1 \times 1.8$   \\
{\underline{New SNRs:}}    &           &          &                   \\
G107.7$-$5.1   &  23:03:48 & +54:33    & $2.4 \times 2.7$   \\
G209.9$-$8.2   &  06:16:30 & $-01$:03  & $1.7 \times 1.8$   \\
G210.5+1.3     &  06:51:50 & +02:45    & $0.4 \times 0.3$ \\
\enddata
\end{deluxetable}

\section{Observations}

\subsection{Optical images}

The 12 Galactic SNRs for which we obtained optical images are listed in Table 2. They are divided by the nine known and 
confirmed remnants listed in  \citet{Green2019} catalog (with on-line updates; Dec 2022) plus three new Galactic remnants which we have found to exhibit optical emissions indicative
of a SNR nature. The nine confirmed remnants were chosen partially based on their large angular size and thus were viewed as good targets for imaging systems with large FOVs. 

Details concerning instrument, image FOV and resolution, along with filter exposure details are given in 
Table 3.
Our images were taken during the last two years from ten different observing sites in Europe, Africa, New Zealand, and the USA. Over 12,000 individual emission-line filter images were obtained on these 12 objects with a total exposure time exceeding 1000 hours, or roughly 1.5 months of open shutter time (see Table 3).

All wide-field images shown in this paper are the products of significant 
post-processing using various commercial image software
including Photoshop, and PixInsight. However, we also
made use of AI to remove noise from the data
(e.g., Topaz Labs DeNoise AI), to sharpen slightly, and to correct distorted stars. After using these programs, we checked  the results against the raw images for accuracy. Other processing techniques developed by our team, such as a ``dynamic level-weighted layered subtraction processing'' (DLWLSP) isolated \O3 and H$\alpha$ features more effectively.
Software processing of the images also enables clean removal of stars in the images thereby making some remnant emission structures more readily visible. A final step included WCS coordinates applied to these images using on-line astrometry software\footnote{https://nova.astrometry.net}.

In addition to wide FOV images taken with small aperture telescopes, we also
obtained images of selected SNRs using 
the Hiltner 2.4 m telescope at the MDM Observatory at
Kitt Peak, Arizona using the Ohio State Multi-Object
Spectrograph (OSMOS; \citealt{Martini2011}) in direct imaging
mode. With a $4096 \times 4096$ CCD, this telescope/camera system
yielded a clear FOV of $18' \times 18'$. On-chip  $2 \times 2$
pixel binning yielded a spatial resolution of $0.55''$.
One to three exposures of 1200--1500 s using 
H$\alpha$ + [\ion{N}{2}] and/or \O3 filters  
were taken to provide higher resolution images of selected portions
of some remnants. However, because these filters' bandpasses were broader than used in the wide FOV images taken with small telescopes, the images were less sensitive to fainter and more extended emissions.

\begin{deluxetable*}{lllccl}[ht]
\footnotesize
\tablecolumns{6}
\tablecaption{Observers, Observing Sites, Imaging Equipment, and Exposure Details }
\tablehead{  \colhead{SNR} &  \colhead{Observers \& Observing Sites} & \colhead{Telescope} &
\colhead{FOV/Pixel scale} & \colhead{Filters} & \colhead{Exposures (hr)} }
\startdata
 G13.3-1.3    & B.\ Falls                      &Takahashi FSQ-106ED &  3.1$\degr$/3.8$''$ & [O III]           & 10.8 ($65 \times 600$s) \\
              & Sierra Remote Obs. USA         &                    &                     & H$\alpha$         &  4.0 ($24 \times 600$s) \\
              &                                &                    &                     & RGB            &  4.5 ($54 \times 300$s) \\
              & B.\ Falls                      &  RCOS 0.4m         &  $20'$/0.50$''$     & [O III]           & 16.7 ($100 \times 600$s) \\
              & Sierra Remote Obs. USA         &                    &                     & RGB             & 5.0 ($30 \times 600$s) \\ 
G70.0-21.5    & B.\ Falls                      &Takahashi FSQ-106ED &  4.0$\degr$/5.6$''$ & H$\alpha$         & 45.8 ($550 \times 300$s)  \\
              & Sierra Remote Obs. USA         &                    &                     & [O III]           & 63.3 ($760 \times 300$s)  \\
              &                                &                    &                     & RGB             & 11.3 ($135 \times 300$s)  \\    
G82.2+5.3    & Y.\ Sainty                     &Takahashi FSQ-106ED & 3.5\degr$/2.0''$   & [O III]          & 21.8 ($131 \times 600$s) \\
              & Oukaimeden Obs., Morocco       &                    &                     & H$\alpha$         & 26.5 ($159 \times 600$s) \\
              &                                &                    &                     & RGB               & 9.0 ($180 \times 180$s) \\
G89.0+4.7    & B.\ Falls                      &Takahashi FSQ-106ED & 2.8$\degr$/2.1$''$  & [O III]           & 41.6 ($500 \times 300$s) \\
              & Sierra Remote Obs. USA         &                    &                     & H$\alpha$         & 45.8 ($550 \times 300$s) \\
              &                                &                    &                     & RGB               & 10.0 ($120 \times 300$s) \\
G107.5-5.1    & B.\ Falls                      &Takahashi FSQ-106ED &  2.5$\degr$/2.0     & [O III]           & 13.0 ($157 \times 300$s) \\   
              &  Sierra Remote Obs. USA        &                    &                     & H$\alpha$         & 36.4 ($437 \times 300$s) \\
              &                                &                    &                     & RGB             & 6.8 ($135 \times 180$s) \\
              & Y.\ Sainty                     &Takahashi FSQ-106ED & 3.5$\degr$/2.0$''$  & [O III]           & 52.3 ($627 \times 300$s) \\
              & Oukaimeden Obs., Morocco:      &                    &                     & H$\alpha$         & 23.7 ($284 \times 300$s) \\
              &                                &                    &                     & RGB             & 11.1 ($223\times 180$s) \\
              & Y.\ Sainty, N.\ Martino, R.\ Galli&Takahashi FSQ-106ED &2.5$\degr$/2.0$''$ & [O III]         & 22.0 ($132 \times 600$s) \\
              &  Moselle, France               &                    &                     & H$\alpha$         & 49.0 ($294 \times 600$s) \\
G119.5+10.2   & Y.\ Sainty                     &Takahashi FSQ-106ED & 3.5$\degr$/2.0$''$  & [O III]           & 20.0 ($240 \times 300$s) \\   
              & Oukaimeden Obs., Morocco       &                    &                     & H$\alpha$         & 35.8 ($430 \times 300$s) \\
              &                                &                    &                     & RGB             & 27.0 ($540 \times 180$s) \\    
G150.3+4.5    & N.\ Martino, Y.\ Sainty        &Takahashi FSQ-106ED & 3.5$\degr$/2.0$''$  & [O III]           & 10.0 ($60 \times 600$s) \\
              & Haute-Alpes, France            &Takahashi FSQ-85ED  &  2.7$\degr$/4.6$''$ & H$\alpha$+[O III] & 13.8  ($83 \times 600$s) \\
              &                                &                    &                     & RGB             & 22.5 ($270 \times 300$s) \\
G181.1+9.5    & S.\ Walker                     & Astro-Physics 130mm&  3.4$\degr$/3.2$''$ & H$\alpha$         & 3.9 ($7 \times 2000$s) \\
              & New Mexico Skies Obs. USA      &                    &                     & [O III]           & 8.7 ($26 \times 1200$s) \\
G209.9-8.2    &  Y.\ Sainty, R.\ Galli,        &Takahashi FSQ-106ED & 3.5$\degr$/2.0$''$  & H$\alpha$         & 86.1 ($1033 \times 300$s) \\ 
              &  A.E.\ Kaeouach                & Takahashi FSQ-85ED &  3.0$\degr$/1.7$''$ & [O III]           & 98.7 ($1185 \times 300$s) \\
              & Oukaimeden Obs., Morocco       &                    &                     & RGB            & 30.4 ($1824 \times 60$s) \\
              & B.\ Falls,  C.\ Morgan         &Takahashi FSQ-106ED &  2.5$\degr$/2.0$''$ & H$\alpha$         & 35.8 ($429 \times 300$s) \\
              &  Sierra Remote Obs. USA        &                    &                     & RGB             & 16.2 ($324 \times 180$s) \\  
              &                                & PW Delta Rho 350   & 2.0$\degr$/0.7$''$  & [O III]           & 16.6 ($249 \times 420$s) \\ 
G210.5+1.3    & M.\ Dreschsler, M.\ Blauensteiner& Lacerta 25cm     & 1.2$\degr$/1.1$''$ & H$\alpha$         & 18.0 ($108 \times 600$s) \\
              & Ursa Major Obs., ROSA, France  &                    &                     & [O III]           & 20.3 ($122 \times 600$s) \\
              &                                &                    &                     & RGB             & 7.5 ($90 \times 300$s) \\
G288.7$-$6.3  & B.\ Falls.                     & Takahashi FSQ-85ED & 2.7$\degr$/4.6$''$  & H$\alpha$+[O III] & 5.9 ($89 \times 238$s) \\
              & Hakos Astrofarm, Namibia       &                    &                     & [O III]           & 7.1 ($85 \times 300$s) \\ 
              &                                &                    &                     & RGB             & 3.8 ($45 \times 300$s) \\
G321.3$-$3.9  & M.\ Ludgate                    & Nikkor AF-S 400mm  & 5.1$\degr$/2.0$''$  & [O III]           & 26.0 ($156 \times 600$s) \\
              & Ross Creek Obs.,  New Zealand. &                    &                     & H$\alpha$         & 17.8 ($107 \times 600$s) \\
              &                                &                    &                     & [S II]            & 11.2 ($67 \times 600$s) \\ 
              &                                &                    &                     &  RGB            & 14.3 ($171 \times 300$s)  \\
\enddata
\label{Table_3}
\end{deluxetable*}

\subsection{Optical Spectra}

In a few cases, low-dispersion optical spectra of filaments of these remnants were obtained with the MDM 2.4m Hiltner telescope using Ohio State Multi-Object Spectrograph (OSMOS; \citealt{Martini2011}). Employing a blue VPH grism (R = 1600) and a 1.2 arcsec wide slit,  exposures of $2 \times 1200 $ s were taken covering 4000--6900 \AA \ with a spectral resolution of 1.2 \AA \ pixel$^{-1}$ and a FWHM = 3.5 \AA.  Spectra were extracted from regions clean of appreciable emission along each of the $15'$ long slits. 

Spectra were reduced using using Pyraf software and OSMOS reduction pipelines\footnote{https://github.com/jrthorstensen/thorosmos} in Astropy and PYRAF\footnote{PYRAF is a product of the Space Telescope Science Institute, which is operated by AURA for NASA.}. L.A. Cosmic \citep{vanDokkum2001} was used to remove cosmic rays and calibrated using a HgNe or Ar lamp and spectroscopic standard stars \citep{Oke1974,Massey1990}.

\section{Results}

Below we present our imaging results on the nine known SNRs,
giving a brief background review followed by a short description of our results for each remnant, contrasting our images with those published where prior images exist. This is followed by short descriptions of the three emission nebula which we view are new Galactic SNRs.

Because these remnants are large in angular size and show considerable fine-scale structure, we have chosen to show the images in large formats so that a reader can better appreciate each remnant's morphology and fine detail.  In addition, in some cases we present 
versions where the stars are been removed by software routines to enhance the visibility of nebular structures.

\subsection{Known Galactic SNRs}

\begin{figure*}[t]
\includegraphics[angle=0,width=18.5cm]{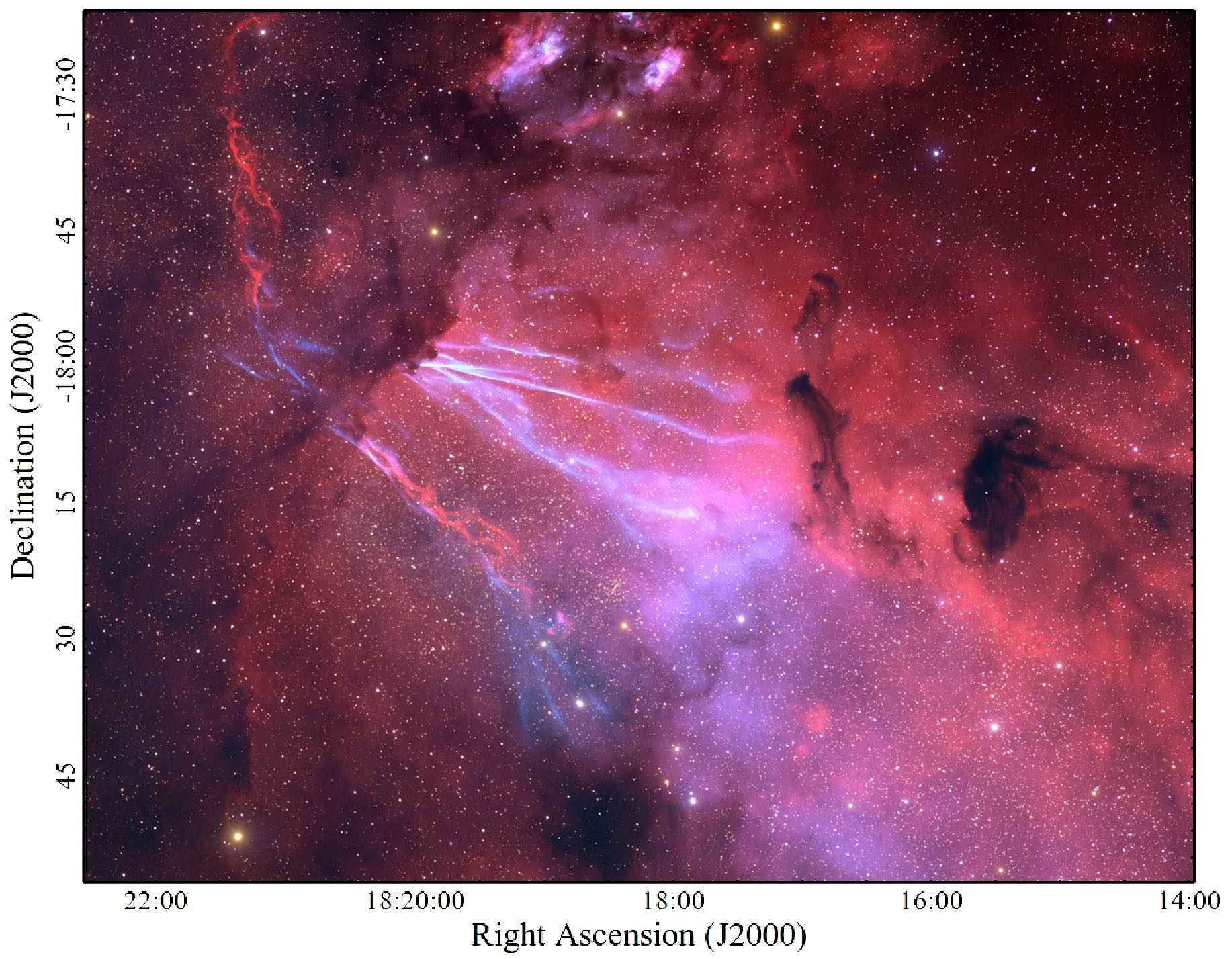} 
\caption{Color composite of H$\alpha$ (red), [\ion{O}{3}] $\lambda$5007 (blue), and broad RGB images
of the G13.3-1.3 remnant. 
Only the short line of red filaments seen here along the remnant's southeastern region were known previously.
\label{G13_color} 
} 
\end{figure*}

\subsubsection{G13.3-1.3}

This remnant was discovered as a diffuse, soft X-ray ROSAT source  by \citet{Seward1995} who
reported coincident optical emission 
visible on the broadband red Palomar Sky Survey images along 
the south central boundary of the X-ray emission. Follow-up MDM 1.3~m 
H$\alpha$ images revealed a complex of thin filaments extending some 20$'$ in length along the remnant's southern boundary.
 
Spectra of one optical filament confirmed its shock nature through
strong [\ion{S}{2}] $\lambda\lambda$6716,6731 
([\ion{S}{2}]/H$\alpha$ = 0.84)
and [\ion{O}{1}] $\lambda\lambda$6300,6364  emissions. The lack
of [\ion{O}{3}] $\lambda\lambda$4959,5007 line emission for this one filament was seen as suggesting a shock velocity $\sim$60 km s$^{-1}$.

Our H$\alpha$ and [\ion{O}{3}] images reveal a far more extensive optical emission structure than previously realized, one
dominated by a tight group of $10' - 30'$ long [\ion{O}{3}] bright filaments which terminate in a broad region of
diffuse [\ion{O}{3}] emission. This is shown in Figure~\ref{G13_color} which is a color composite of
deep H$\alpha$ (red) and [\ion{O}{3}] (blue) images together with short exposure broadband RGB images. These long and bright [\ion{O}{3}] filaments were not been previously reported and their presence was an unexpected discovery. They seem to emanate from behind a dark cloud at their eastern end creating a fan or spray-like morphology.

The location of these [\ion{O}{3}] filaments concentrated near the remnant's center is unusual among the optical emissions of SNRs. There is no strong analogue of these \O3 filaments with the possible exception of the long N-S arrangement of filaments in the  so-called Pickering's Triangle of the Cygnus Loop. However, in that case, the [\ion{O}{3}] filaments appear sharper, are not as  extensive, and do not terminate into broad areas of diffuse [\ion{O}{3}] emission. Neither do they display a clustering from a small region like seen here in G13.3-1.3.

\begin{figure*}[ht]
\begin{center}
\includegraphics[angle=0,width=14.5cm]{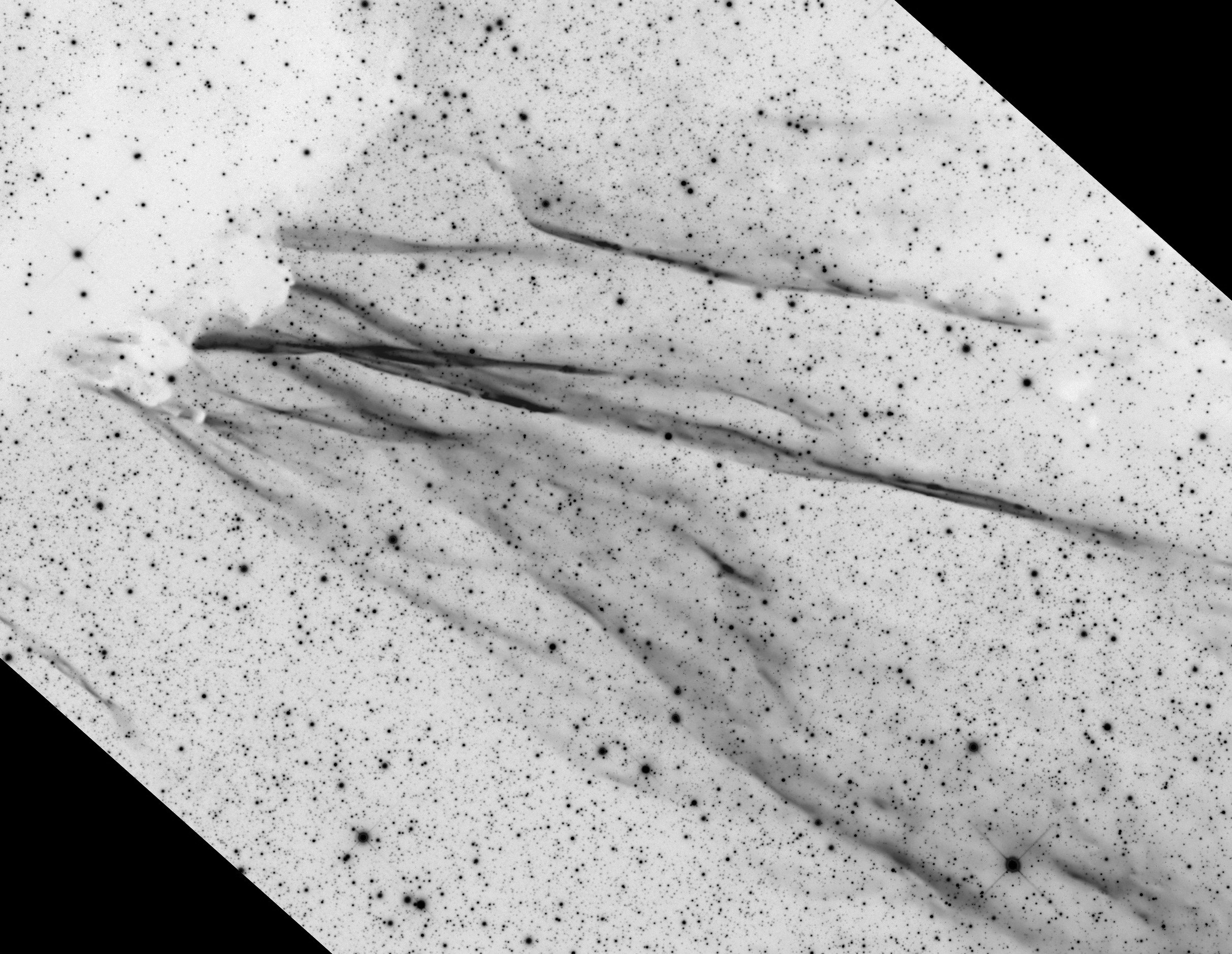}
\includegraphics[angle=0,width=15.5cm]{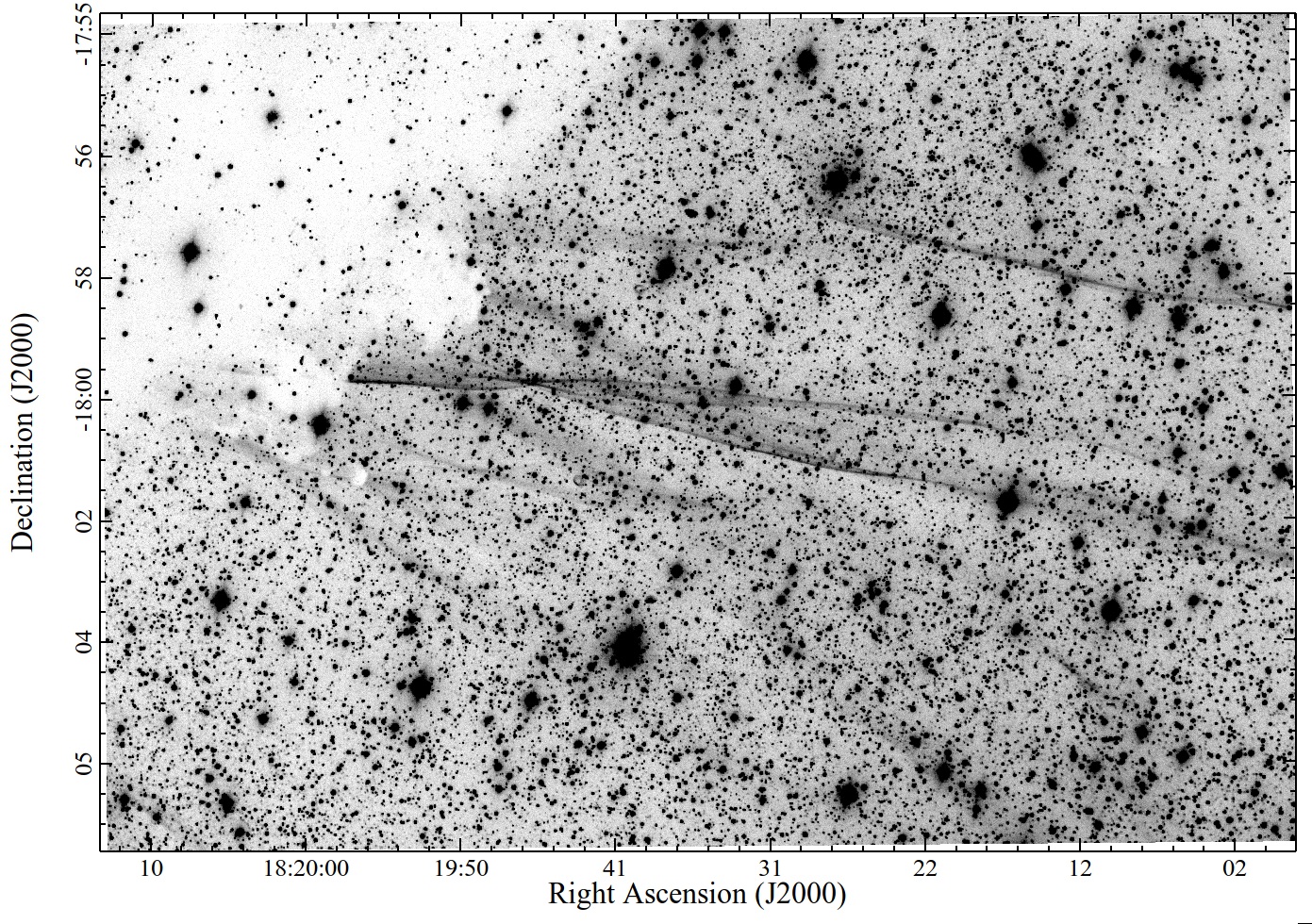} \\
\caption{Close-up views  of G13.3-1.3's unusual [\ion{O}{3}] $\lambda$5007 emission filaments.
Top: [\ion{O}{3}] emission image plus broadband RGB filters. Bottom: Higher resolution MDM [\ion{O}{3}] image showing finer filament details.
\label{G13_streaks} 
} 
\end{center}
\end{figure*}

\begin{figure*}[t]
\begin{center}
\includegraphics[angle=0,width=17.0cm]{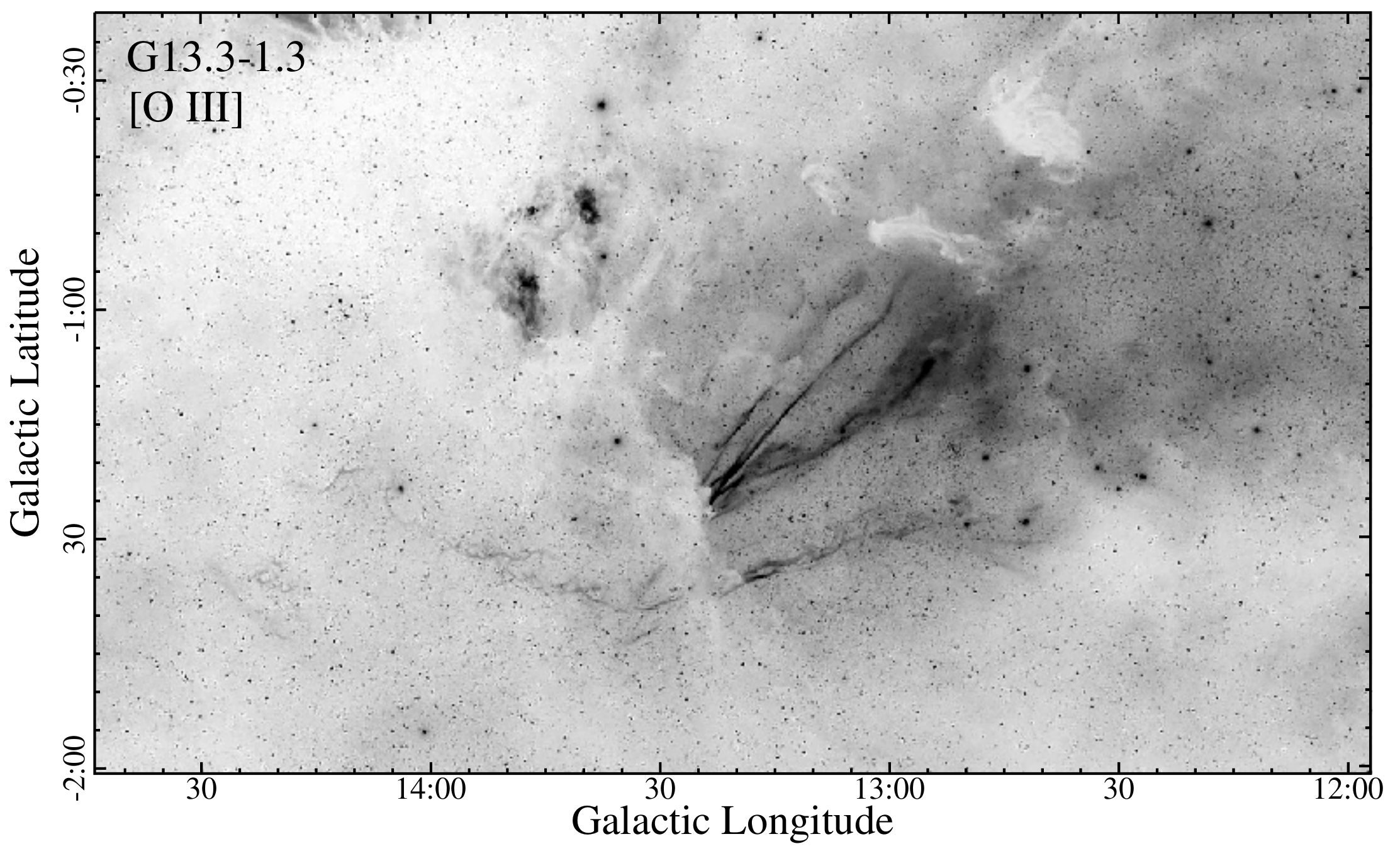} \\
\includegraphics[angle=0,width=8.0cm]{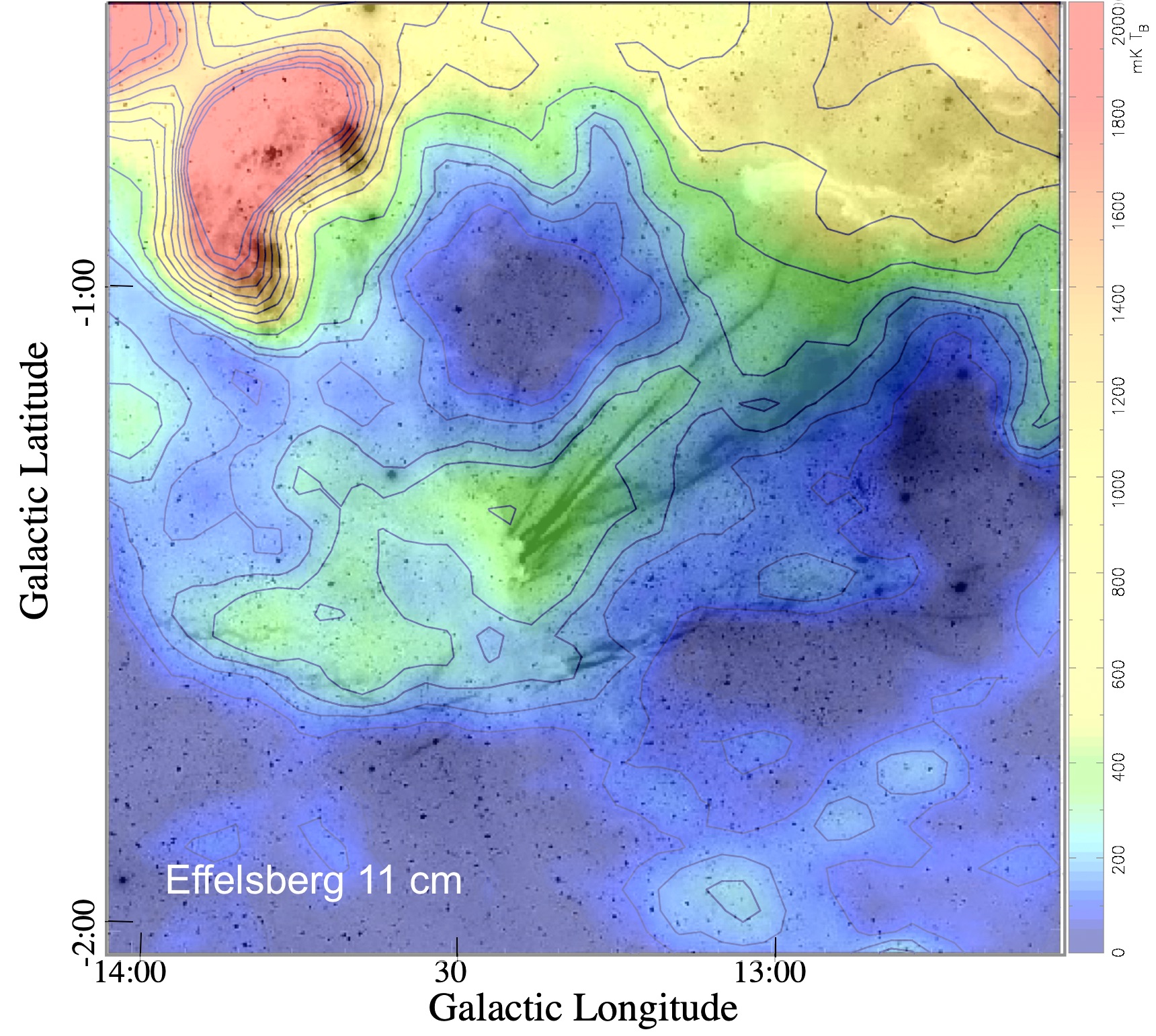}
\includegraphics[angle=0,width=8.0cm]{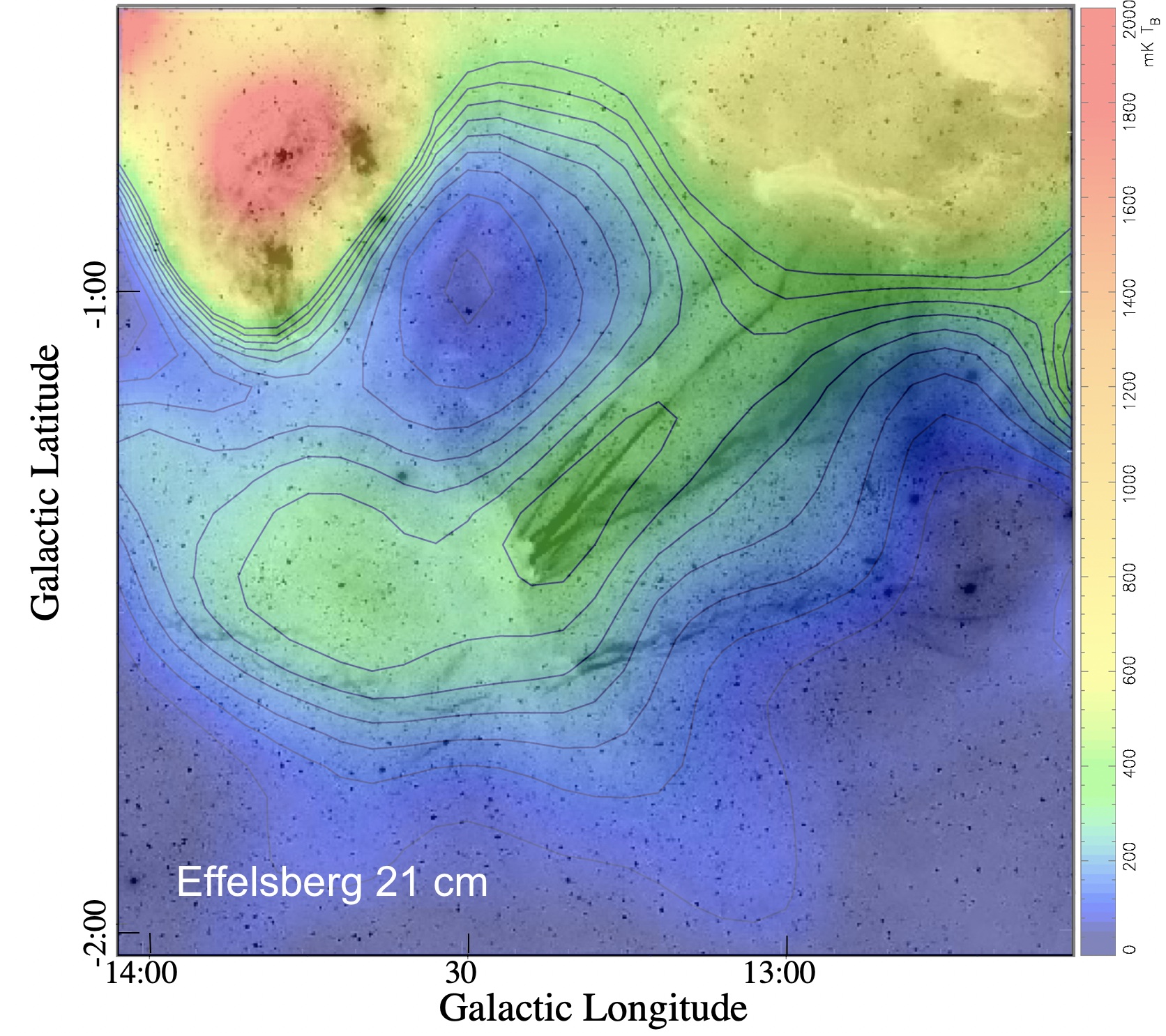}
\caption{Top: Wide FOV [\ion{O}{3}] image of the G13.3-1.3 now shown in Galactic coordinates. 
Bottom: Cropped [\ion{O}{3}] image with colored 11 cm (left) and 21 cm (right) radio overlays showing
the coincidence of the remnant's bright [\ion{O}{3}] filaments with the remnant's radio emission with fainter filaments outlining the radio extension toward $l$ = 14$\degr$.
\label{G13_Galactic} 
} 
\end{center}
\end{figure*}

Better views of the  [\ion{O}{3}] filaments are shown in Figure~\ref{G13_streaks} where we present close-up views of the main group of [\ion{O}{3}] emission filaments. Many individual filaments are visible along with extended diffuse emission\footnote{The orientation of the original top image led to unusual cropping.}.
The bottom panel shows a smaller FOV but higher resolution image, highlighting that the filamentary emissions are sharpest along their southern edges, marking the start of a gradual fainting of diffuse emission.
This emission morphology is similar to that seen in a few other SNRs and is suggestive of an extended sheet of shocked gas viewed edge-on along the line of sight.
Low dispersion spectra indicate a I([\ion{O}{3}] $\lambda$5007)/I(H$\beta$) $\simeq$ 7 suggesting
shocks in excess of 100 km s$^{-1}$ 
\citep{Raymond1983,Raymond2020}.

The nature of the diffuse \O3 emission seen in the bottom-center portion of Figure~\ref{G13_color} is presently unclear. Its connection to the remnant is suggested both by its location near the western end of the bright \O3 filaments and the H$\alpha$ bright filaments at RA = 18:19, Dec = $-$18:30. The diffuse \O3 emission is also coincident with faint ROSAT detected X-ray emission extending to RA = 18:16, Dec = $-$19:00 \citep{Seward1995}.

The red H$\alpha$ filaments along the remnant's southeastern edge shown in Fig.~\ref{G13_color} are some of the same ones that were presented and discussed by \citet{Seward1995} except now the line of filaments are seen to extend farther to the north and east, crossing behind a dark column of obscuring dust.
These filaments also appear to change into [\ion{O}{3}] bright filaments along their southwestern extent.
We note that these filaments, shown here in red, roughly coincide with the 
southeastern edge of the
remnant's X-ray emission as detected in ROSAT data. Similar filaments are seen along the remnant's
eastern border above a thin N-S aligned dark cloud.

There are no published radio emission maps of G13.3-1.3 to inform us concerning possible radio-optical correlations. 
We therefore examined published  100~m
Effelsberg surveys at
11 cm (resolution = 4.3$'$) and 21 cm (resolution = 9.4$'$)
\citep{Reich90a,Reich90b}
and removed the dominating
large-scale Galactic emission by `unsharp masking' \citep{Sofue1979}
from the survey maps to separate the
emission from G13.3-1.3.
In Figure \ref{G13_Galactic}, we show
our \O3 image of G13.3-1.3 (top panel) along with the 11 cm and 21 cm radio maps colored for flux in the bottom two panels.  

Preliminary analysis using the Effelsberg 21 cm data along
with 6 cm data (resolution 9.5') from the Urumqi Sino-German
6 cm survey \citep{Sun2011} indicates
excellent correlation of the remnant's bright \O3 optical emission filaments with a radio `spur' visible in both radio maps.
The Urumqi data were
processed in the same way as the Effelsberg survey data.

The faint bend in the radio emission, best seen in the 21 cm map, roughly matches the fainter optical filaments at $l$ = 13.5$\degr$. On the other hand, there is no
such correlation with the filaments at $b$ = $1.5$ and $l$ = 12.5$\degr$.
Analysis of the Effelsberg
data together with 6 cm Urumqi data indicates the spur is non-thermal with a spectral index
$\alpha$ = $0.65 \pm 0.18$, a bit steeper than the canonical $0.5$ value for shell-type SNRs.

\newpage

\subsubsection{G70.0-21.5}

With angular dimensions of roughly $4.0\degr  \times 5.5\degr$, the G70.0-21.5 remnant is one of the largest Galactic remnants known. Located at $b = -21.5\degr$ it is also among the farthest off the Galactic plane. 
It was first discovered by \citet{Boumis2002}
who obtained H$\alpha$ + [\ion{N}{2}] images across a region some
$7.5\degr \times 8.5\degr$ in Pegasus, along with some [\ion{S}{2}] and \O3 images of small selected regions. They reported finding numerous thin H$\alpha$ filaments, strong in [\ion{S}{2}] emission in some regions, and faint, soft X-ray emission in the ROSAT All-sky Survey. This led them to suggest
that much of the optical emission in this region is part of one or
more previously unrecognized supernova remnants.

Guided by VTSS H$\alpha$ images, an 
independent imaging and spectral
investigation by \citet{Fesen2015} found
that H$\alpha$ filaments centered around RA = 21:24:00, Dec = +19:23 (J2000)
comprised a coherent H$\alpha$ and near-UV  emission structure  
with coincident faint X-ray emission along its northern limb.
Follow-up, low and high resolution spectra suggest a fairly  nearby and old SNR (d = 1 kpc; age $\sim$90,000 yr) with modest shock speeds ($70 - 110$ km s$^{-1}$), the presence of \O3 bright filaments in some regions, and an unusually thin postshock zone where preshock neutral atoms are rapidly excited and ionized \citep{Raymond2020}.

\begin{figure}[ht]
\centerline{\includegraphics[angle=0,width=9.0cm]{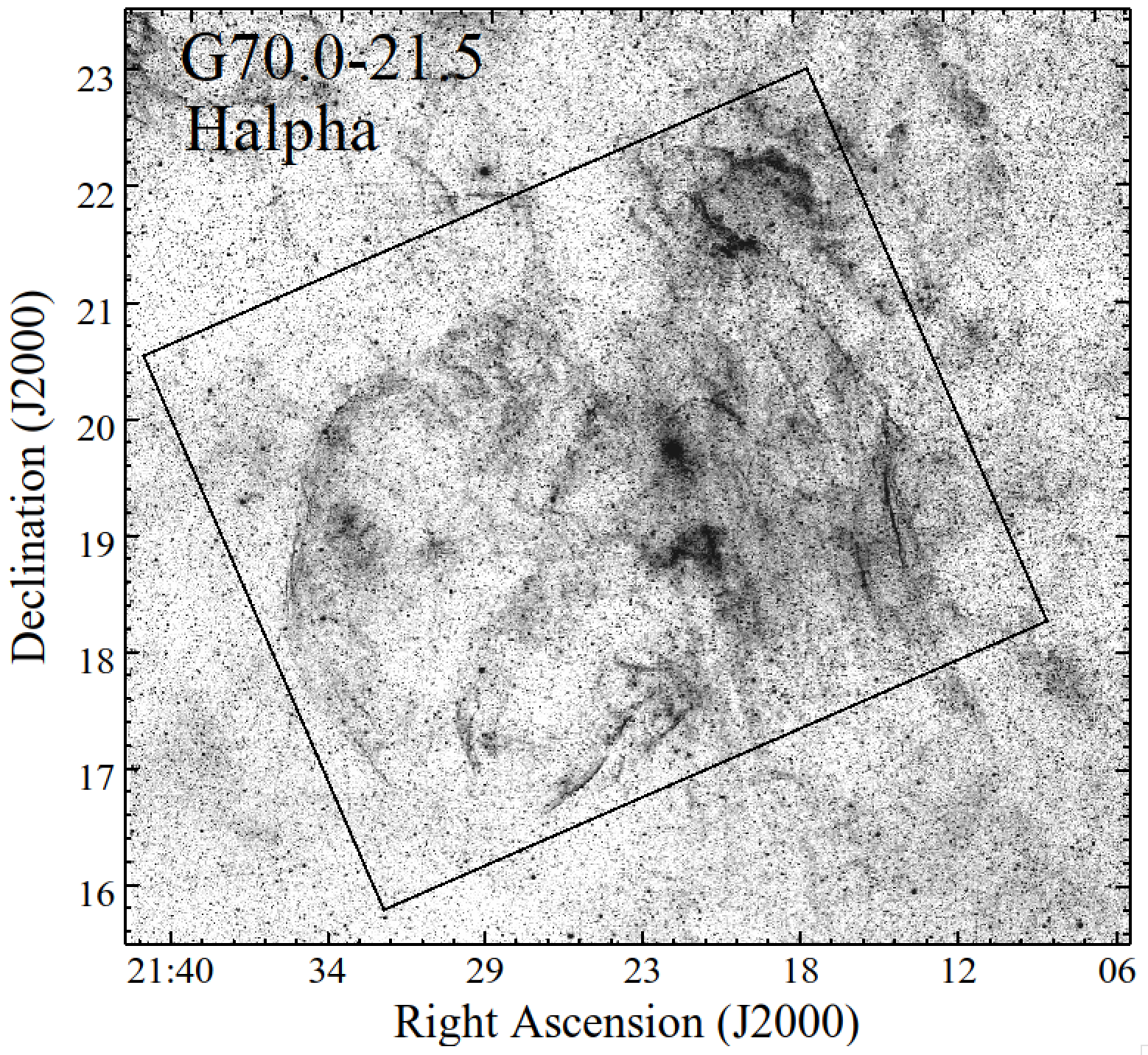}}
\caption{MDW $9.5\degr \times 8.5\degr$ H$\alpha$ image of G70.0-21.5 showing the
region and orientation of our images (black box).  \label{G70_box}  }
\end{figure}

Because of the remnant's large angular size, our H$\alpha$ and \O3 images were taken with an orientation that most efficiently covering most of the remnant with the fewest images. This is shown in Figure \ref{G70_box} where a large MDW H$\alpha$ mosaic image of this remnant is shown  along with our $5.2\degr \times 6.2\degr$ imaging  region. As can be seen, our imaged region encompassed the majority of emission features of the G70.0-21.5 remnant.

\begin{figure*}[ht]
\begin{center}
\includegraphics[angle=0,width=18.0cm]{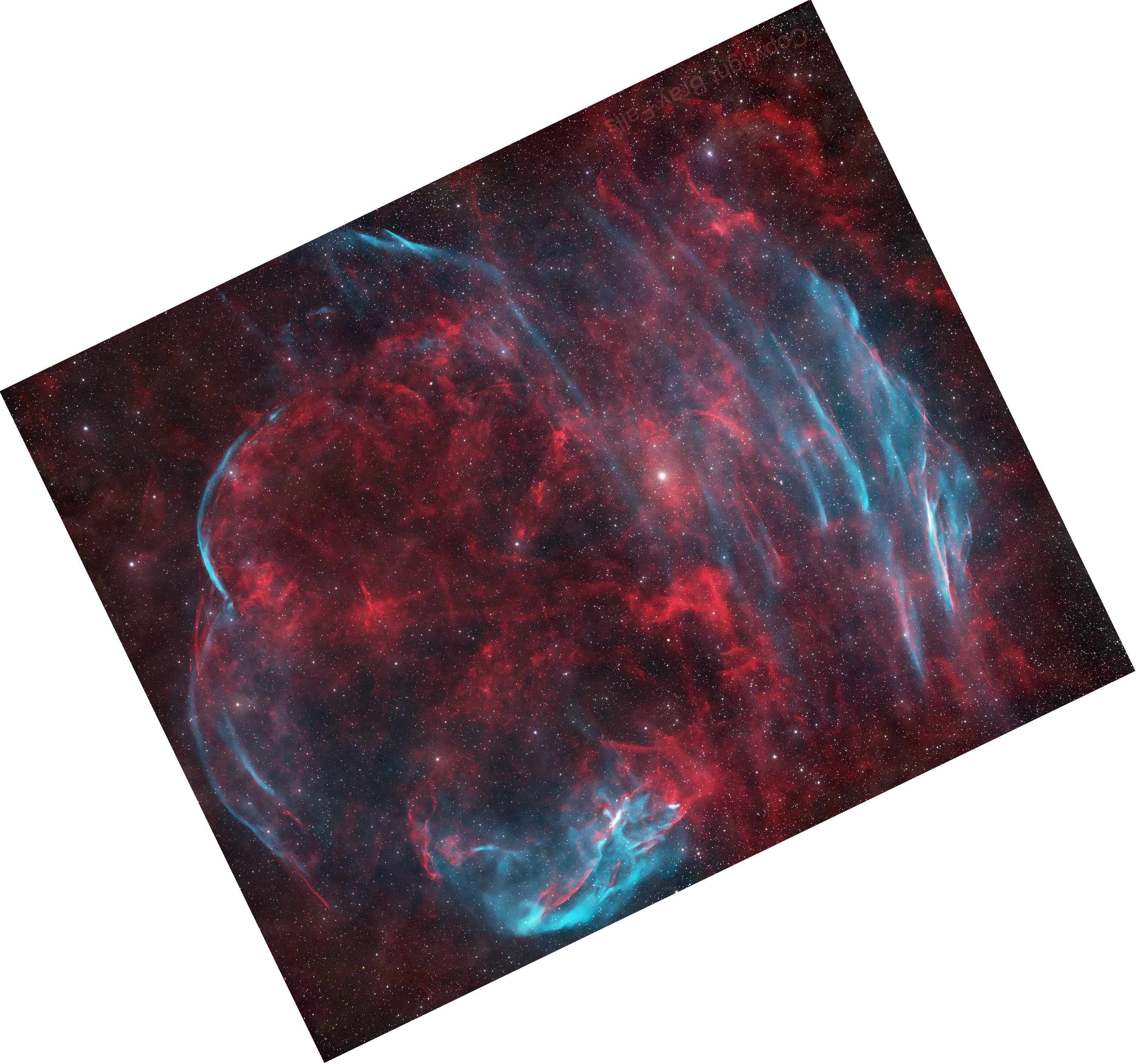}
\caption{ Color composite of H$\alpha$ (red), [\ion{O}{3}] $\lambda$5007 (blue), and broad RGB images
of the G70.0-21.5 remnant. Image FOV is $5.2\degr \times 6.2\degr$. North is up, east to the left.
\label{G70_COLOR}
} 
\end{center}
\end{figure*}

The result of 110 hours of H$\alpha$ and \O3 exposures is shown in the color composite of Figure~\ref{G70_COLOR}.
Because this image covers an area of more than 30 square degrees, much of the 
remnant's fine detail emission structure is lost in this image. Nonetheless, the remnant's gross emission morphology is clearly visible,  namely extensive H$\alpha$ emission throughout the remnant's structure, with strong \O3 emission predominantly found along its outer edges. 

Whereas its H$\alpha$ emission is largely diffuse in appearance as seen in this low resolution image with only a few thin filaments visible, just the opposite is the case in regard to the remnant's \O3 emission  which is highly
filamentary, arranged in curved filaments in the east, with 
long and gentle arcs in the west. However, we note that higher resolution images show the remnant displays a numerous fine H$\alpha$ filaments in certain regions \citep{Boumis2002,Fesen2015,Raymond2020}, especially along its eastern and southern borders as well as in certain areas in its western half.

\begin{figure*}[ht]
\begin{center}
\includegraphics[angle=0,width=17.5cm]{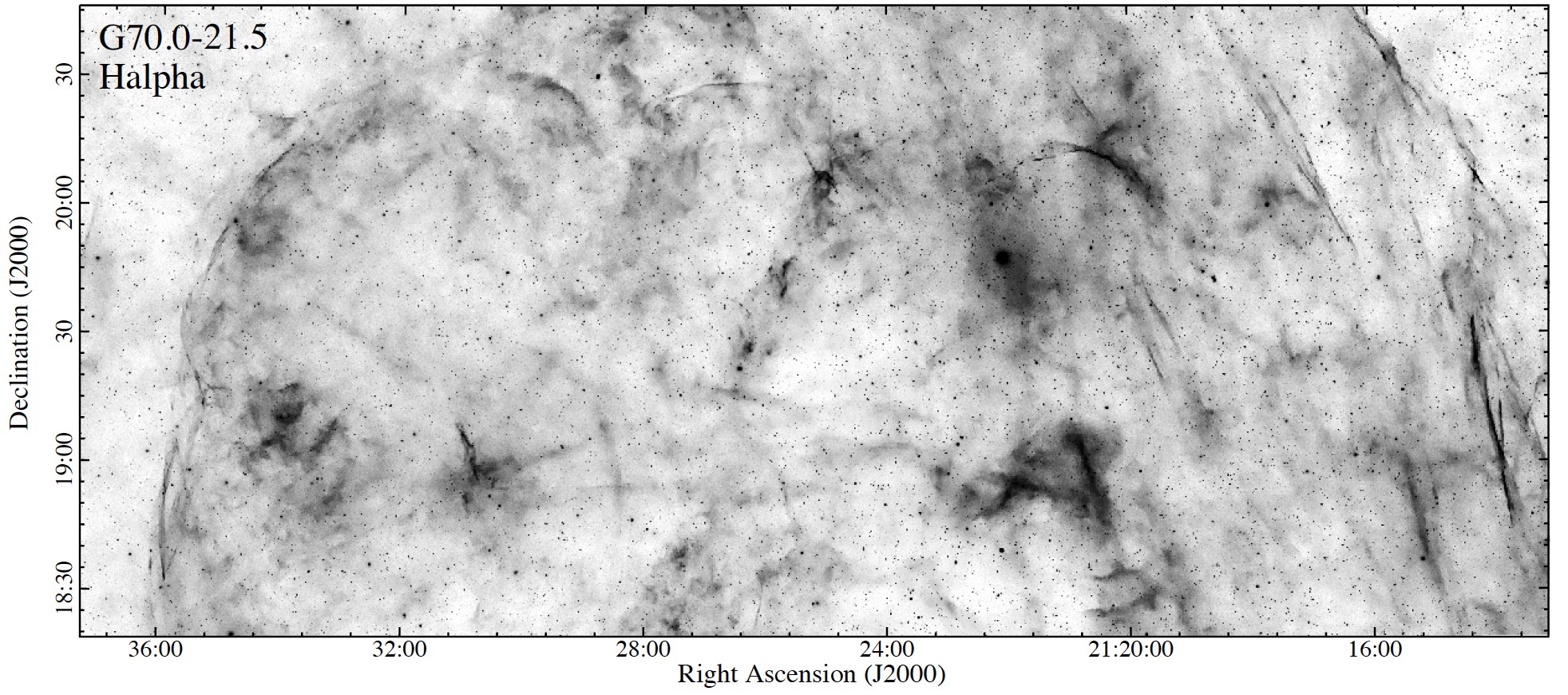}
\includegraphics[angle=0,width=17.5cm]{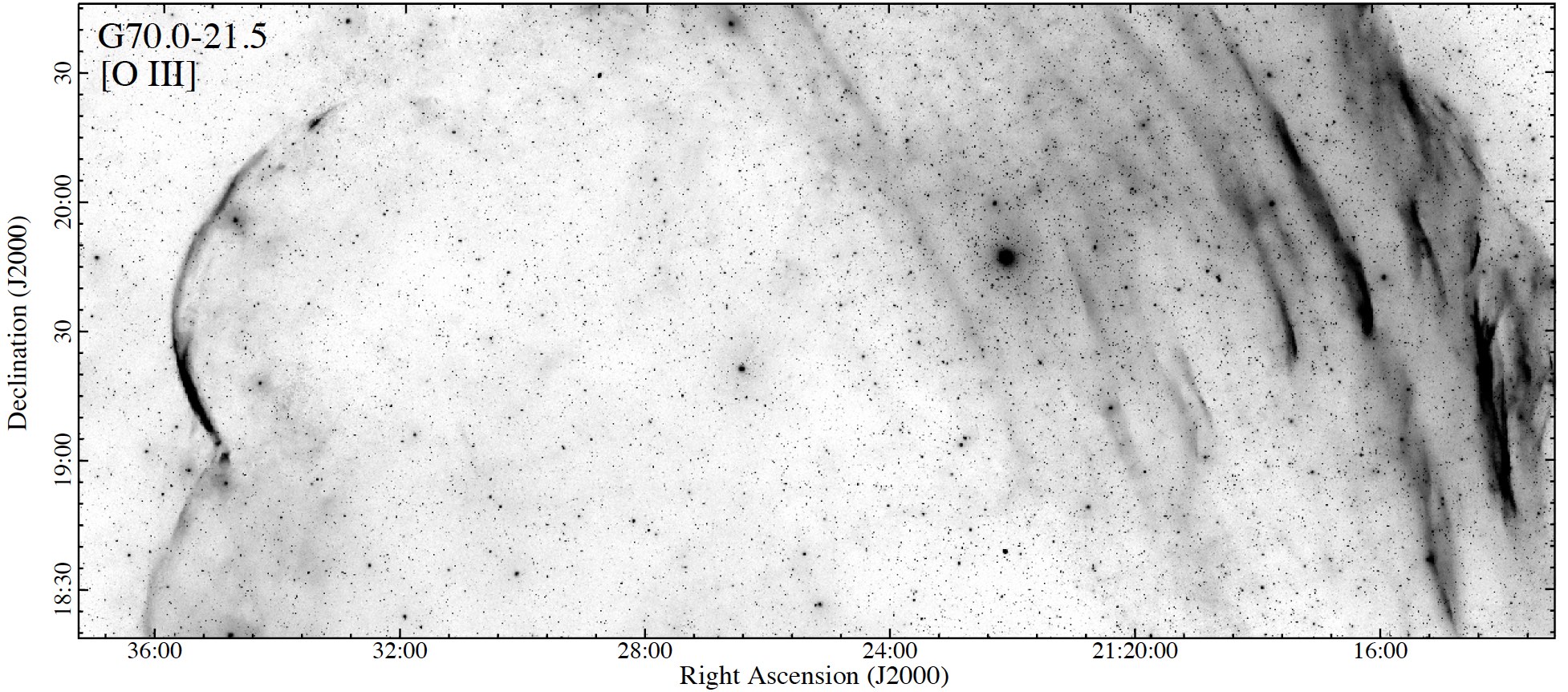}
\includegraphics[angle=0,width=0.70cm]{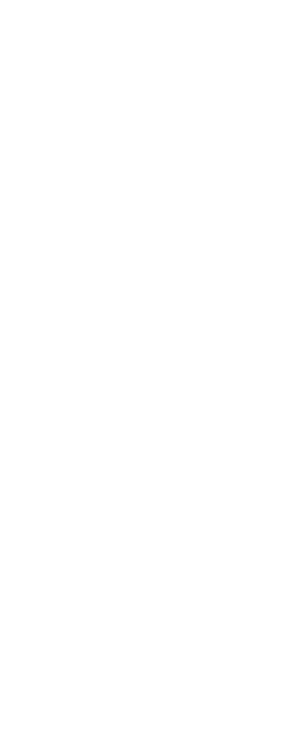}
\includegraphics[angle=0,width=16.4cm]{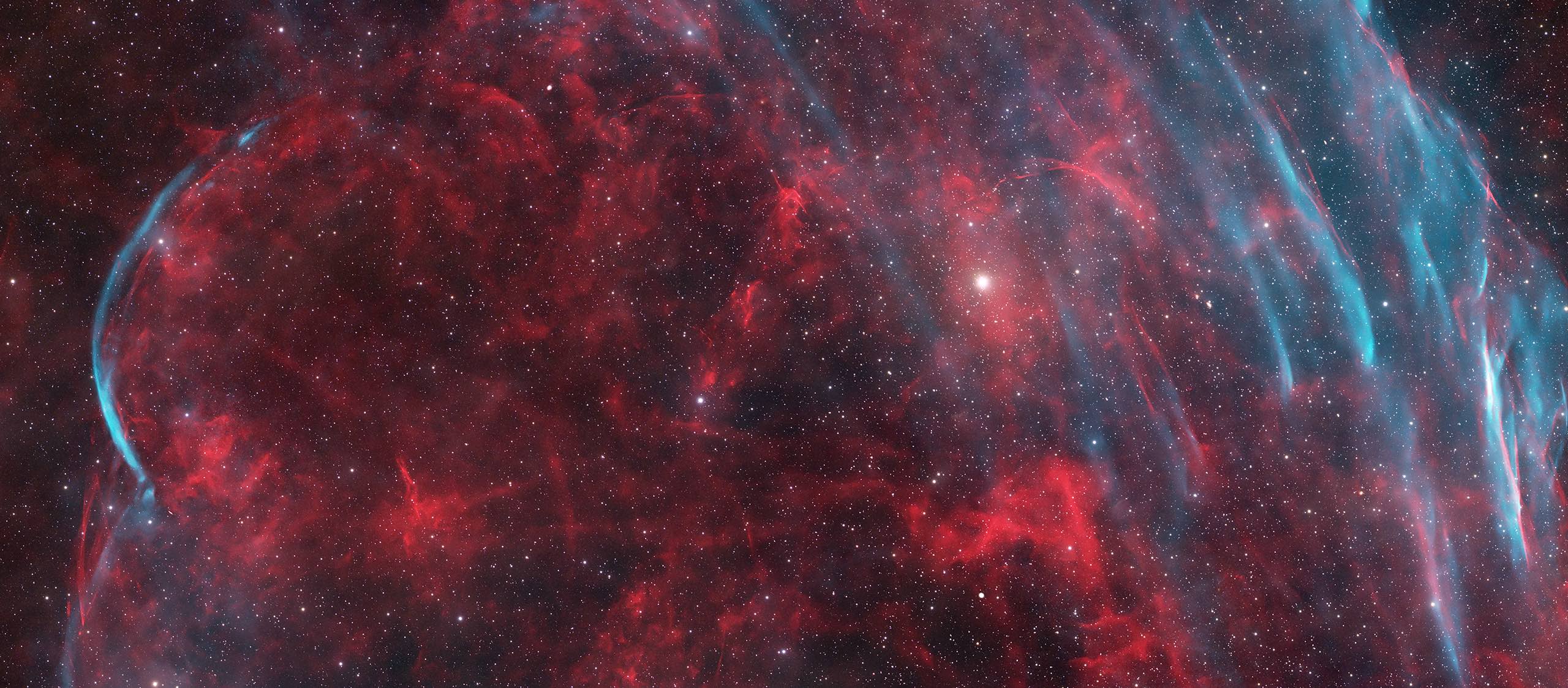}
\caption{H$\alpha$ and \O3 images 
of the central regions of G70.0-21.5 along with a matching color composite image.
\label{G70_enlarge}
} 
\end{center}
\end{figure*}

The stark difference in morphology between this remnant's  H$\alpha$ and \O3
emissions is shown in Figure~\ref{G70_enlarge}. While some H$\alpha$ emission
is seen throughout the remnant and is mainly diffuse in appearance, the
remnant's \O3 emission is highly filamentary and largely limited to the SNR's
eastern and western limbs. This means that spectral investigation of this
remnant sampling only a handful of individual regions and filaments bright in
either H$\alpha$ or \O3 emission may led to very different shock estimates and
results that may not be representative of the remnant as a whole.

\begin{figure*}[ht]
\begin{center}
\includegraphics[angle=0,width=17.0cm]{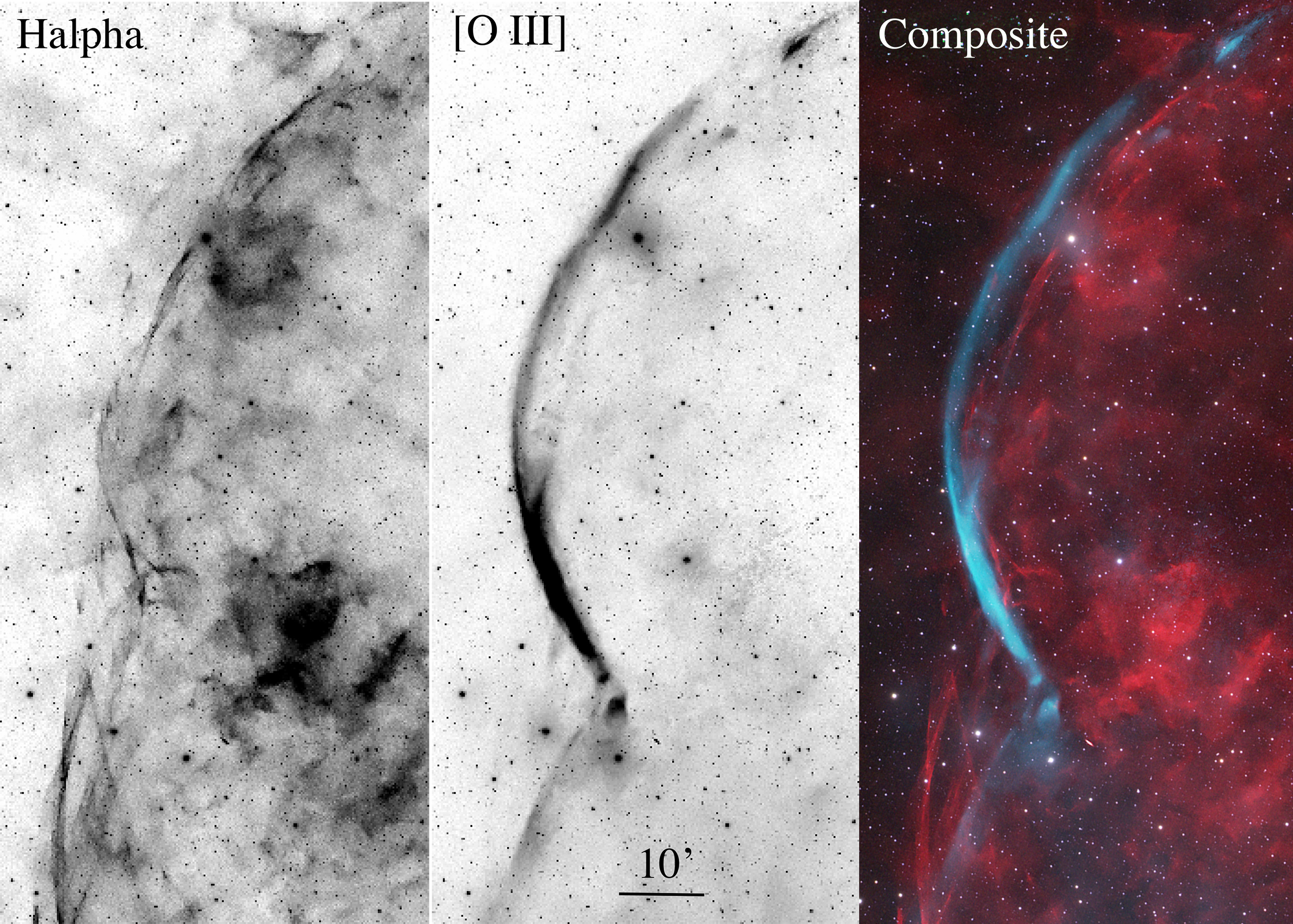}
\caption{Close-up of the H$\alpha$ and [\ion{O}{3}] $\lambda$5007 filter images plus color composite for the northeastern limb  
of G70.0-21.5.
\label{G70_triple}
} 
\end{center}
\end{figure*}

For example, the remnant's brightest H$\alpha$ and \O3 emissions are often
quite different.  This was
discussed in \citet{Raymond2020} where color composite images illustrated the
angular separation between H$\alpha$ and \O3 emissions where diffuse H$\alpha$
emission is often observed behind and interior to the shock front marked by
strong \O3 emission.

Even larger angular separations between H$\alpha$ and \O3 emission regions than
discussed by \citet{Raymond2020} are shown in the color composite in the lower
panel of Figure~\ref{G70_enlarge}.  An especially clear case of such
separations is presented in Figure~\ref{G70_triple} where we show enlargements of
H$\alpha$ and \O3 images of the remnant's eastern limb along with a color
composite where H$\alpha$ emission is red and \O3 is shown as blue. The bright  \O3
emission filament lies $2' - 5'$ out ahead of the H$\alpha$ emission, which
assuming a remnant distance of $\simeq$1 kpc \citep{Raymond2020}, translates
into a physical distance of $\sim$1 pc.

However, such an angular separation is not seen universally through the remnant. In the remnant's western section, \O3  
filaments are widely distributed and do not show the same pattern or separation for H$\alpha$ and \O3 emission features.

Finally, we note that the remnant's \O3 filaments appear unexpectedly broad, some $60''- 90''$ thick 
($\sim0.30 - 0.43$ pc at 1 kpc). This might give the initial impression that the \O3 image data are of much lower resolution than that of H$\alpha$, an impression dispelled upon examining the stellar point spread functions.
Such broad \O3 emission filaments are not commonly seen
in Galactic SNRs and may be a function of the remnant's age, the low density of the ambient ISM more than 20 degrees off the Galactic plane and the remnant's relatively close distance.

\begin{figure*}
\begin{center}
\includegraphics[angle=0,width=17.0cm]{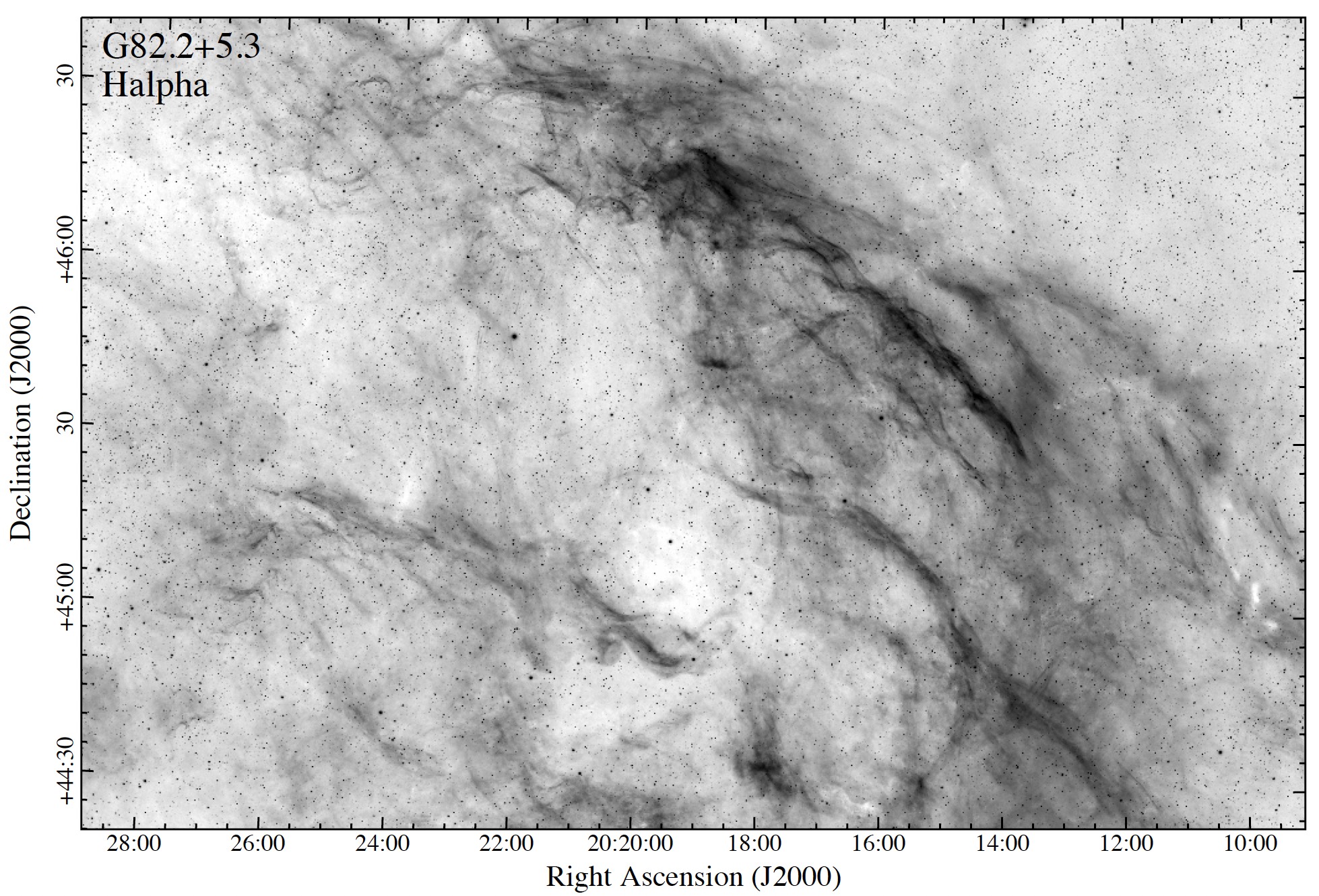} 
\includegraphics[angle=0,width=17.0cm]{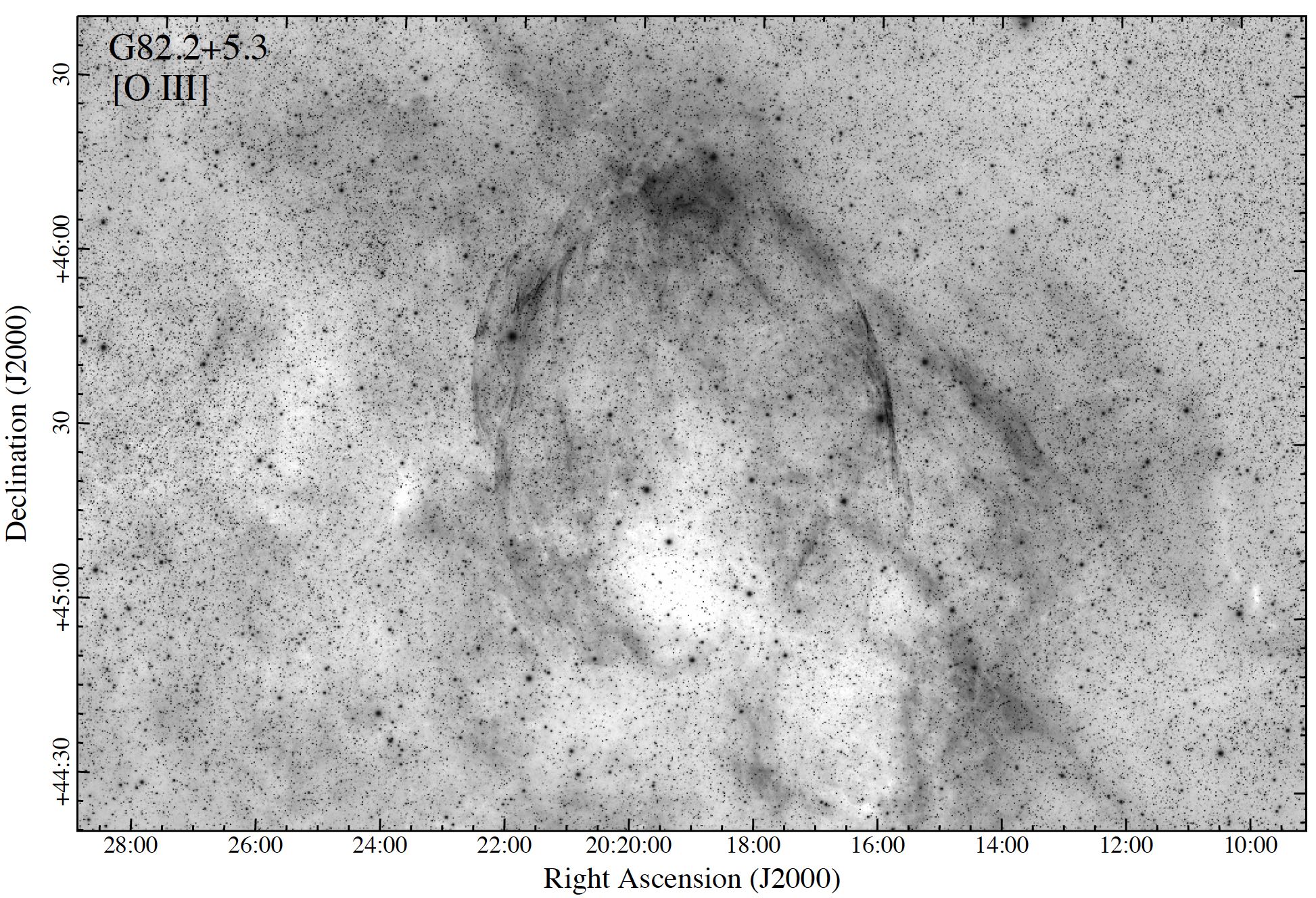} 
\caption{ {\bf{Top:}} H$\alpha$ image of G82.2+5.3 (W63) and surroundings. {\bf{Bottom:}} [\ion{O}{3}] $\lambda$5007 image of G82.2+5.3 (W63) and surroundings. \label{G82_Ha_n_O3}
} 
\end{center}
\end{figure*}

\begin{figure*}
\begin{center}
\includegraphics[angle=0,width=17.0cm]{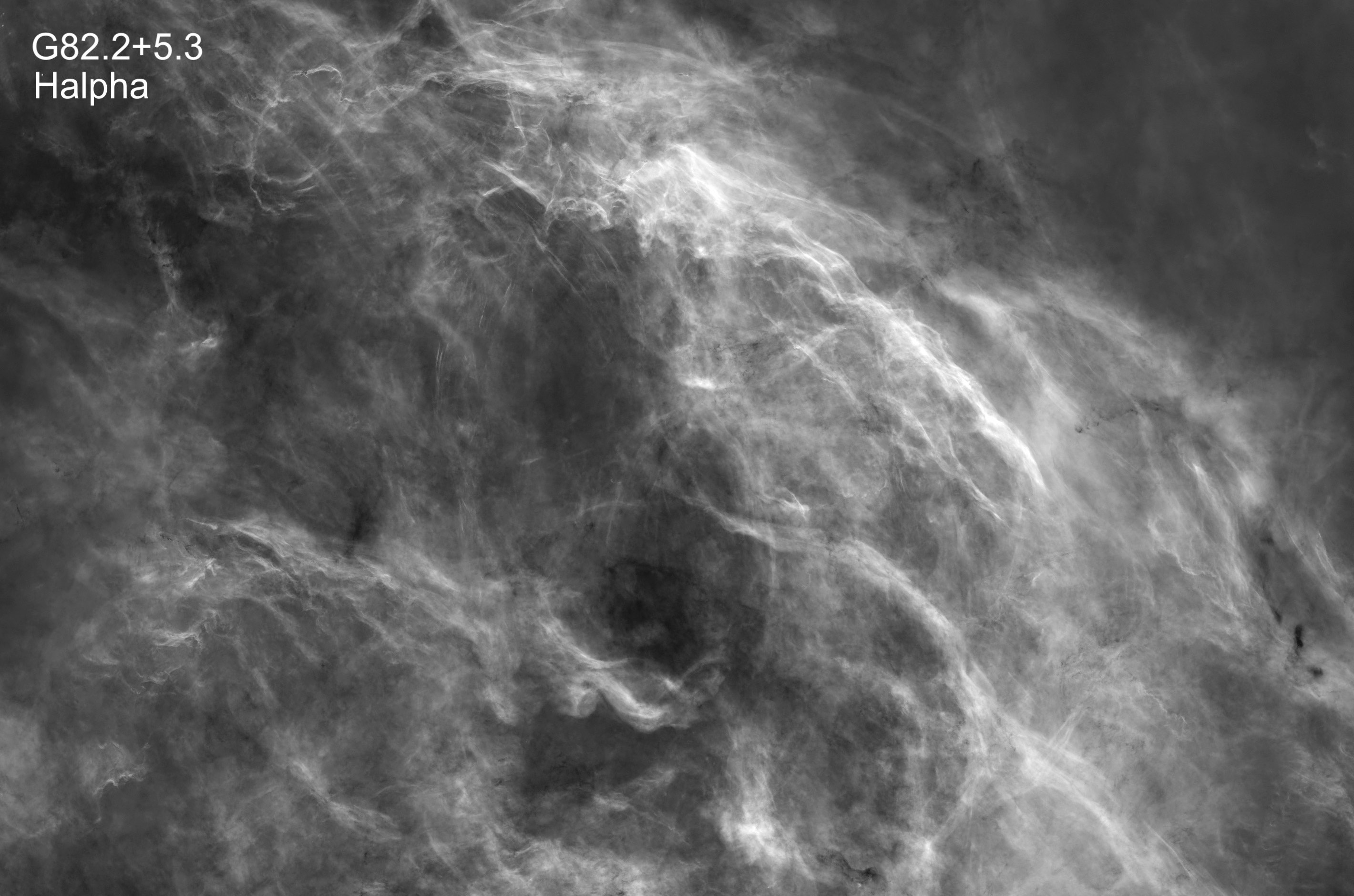}
\includegraphics[angle=0,width=17.0cm]{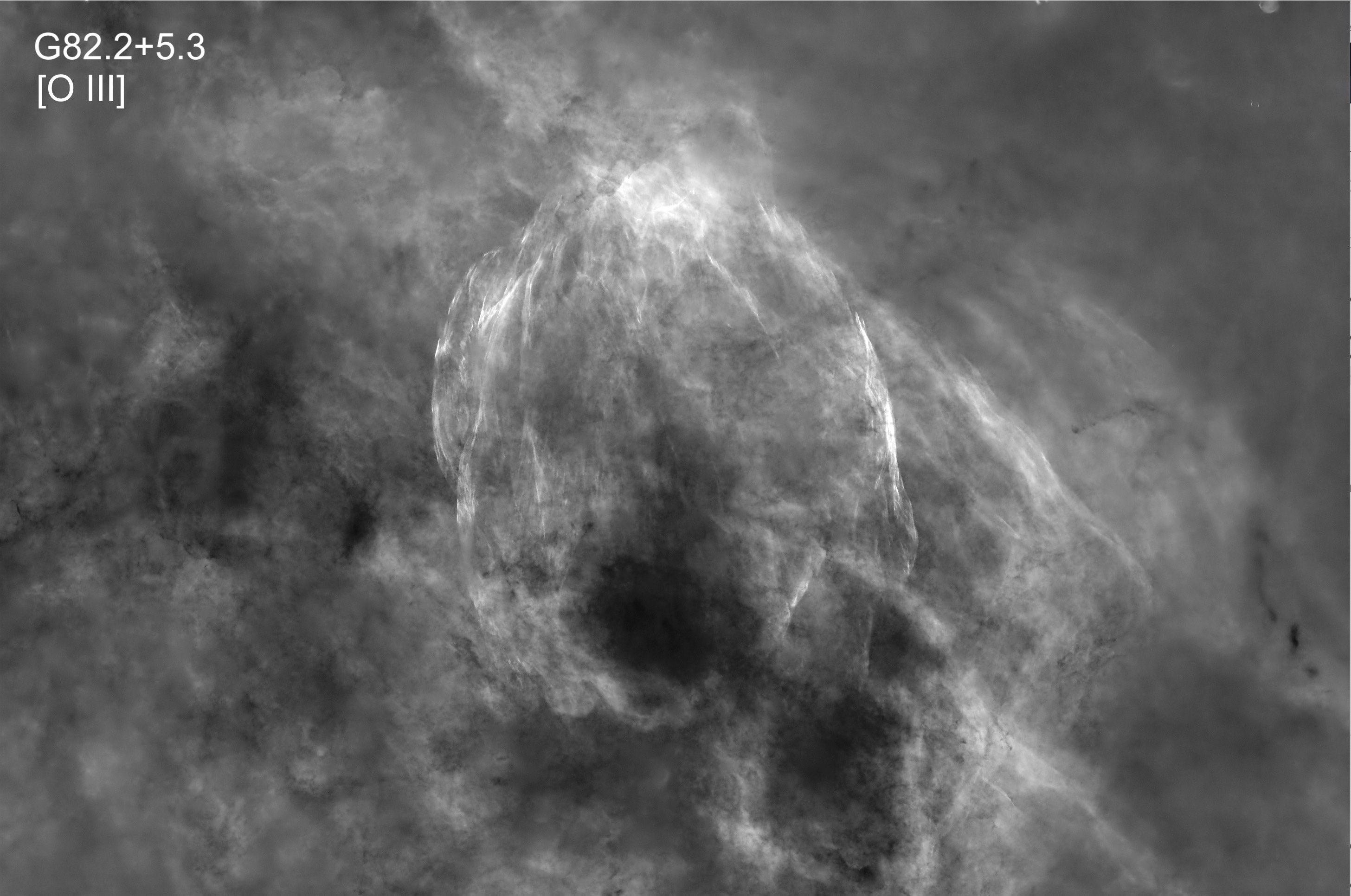}
\caption{Same data as in Figure~\ref{G82_Ha_n_O3}  for G82.2+5.3 (W63) but now shown as positive images with the stars removed by software to make the remnant's H$\alpha$ (top) and [\ion{O}{3}] (bottom) emissions and dust features more readily visible. Note the extensive [\ion{O}{3}] emission to the west and southwest of the remnant's prominent emission shell. 
\label{G82_positives}
} 
\end{center}
\end{figure*}

\begin{figure*}
\begin{center}
\includegraphics[angle=0,width=17.0cm]{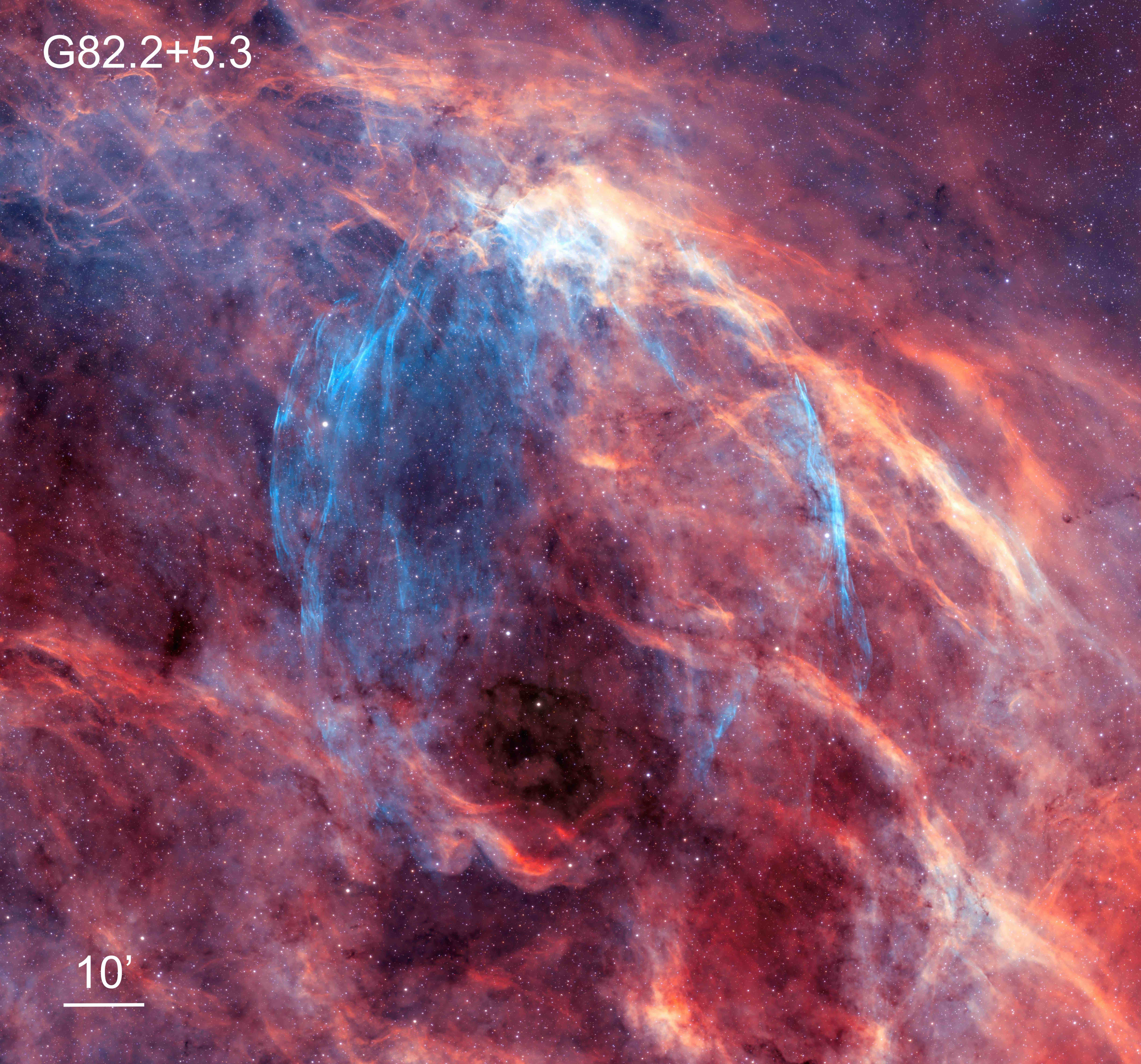} \\
\caption{ A color composite formed from H$\alpha$ (red) and [\ion{O}{3}] $\lambda$5007 (blue) images of G82.2+5.3 (W63). \label{G82_color}
} 
\end{center}
\end{figure*}

\subsection{G82.2+5.3 (W63)}

The strong Galactic radio source W63, found to be a 
likely SNR by \citet{Wendker1968}, was subsequently
discovered to have associated optical filaments by 
\citet{Dickel1969} and \citet{Wendker1971}.
Optical spectra of these filaments showed evidence for
interstellar shocks \citep{Sabbadin1976, Rosado1981}.

The only deep optical imaging of the G82.2+5.3/W63 remnant published to date 
are those taken by \citet{Mav2004}. They obtained flux calibrated 
$89' \times 89'$ wide images ($5''$ pixel$^{-1}$) of the remnant 
in several emission lines including
H$\alpha$ + [\ion{N}{2}], [\ion{S}{2}], \O3, and [\ion{O}{2}]. 
The W63 region of the Galactic plane in Cygnus is quite complex in terms of
H$\alpha$ emission features and the \O3 images of 
\citet{Mav2004} provided the first clear picture of the G82.2+5.3 remnant,
appearing markedly different to images
taken in the low ionization lines of [\ion{O}{2}] and [\ion{S}{2}].
The \O3 images revealed filamentary
emission along the remnant's west and east limbs
defining the opposite sides of an ellipsoidal shell 
similar with the morphology seen in the remnant's bright radio emission.

Our H$\alpha$ and \O3 images of G82.2+5.3 are shown in Figures~\ref{G82_Ha_n_O3} and \ref{G82_positives} where we present both the negative versions of the original
images and positive versions with the stars removed by software to enhance the visibility of the nebular features. A color composite of these images is shown in Figure~\ref{G82_color}.

\begin{figure*}[ht]
\begin{center}
\includegraphics[angle=0,width=8.0cm]{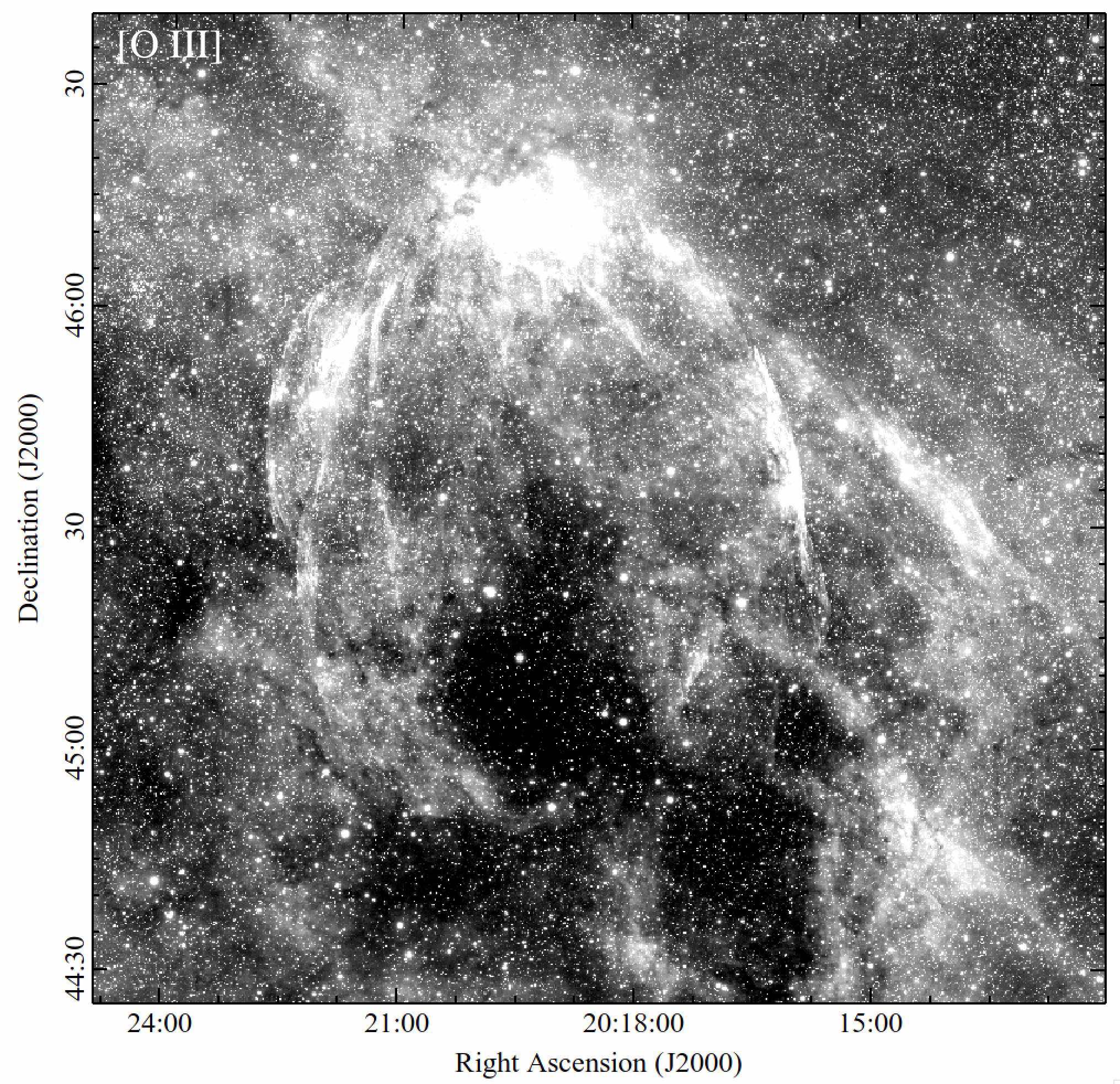}
\includegraphics[angle=0,width=8.0cm]{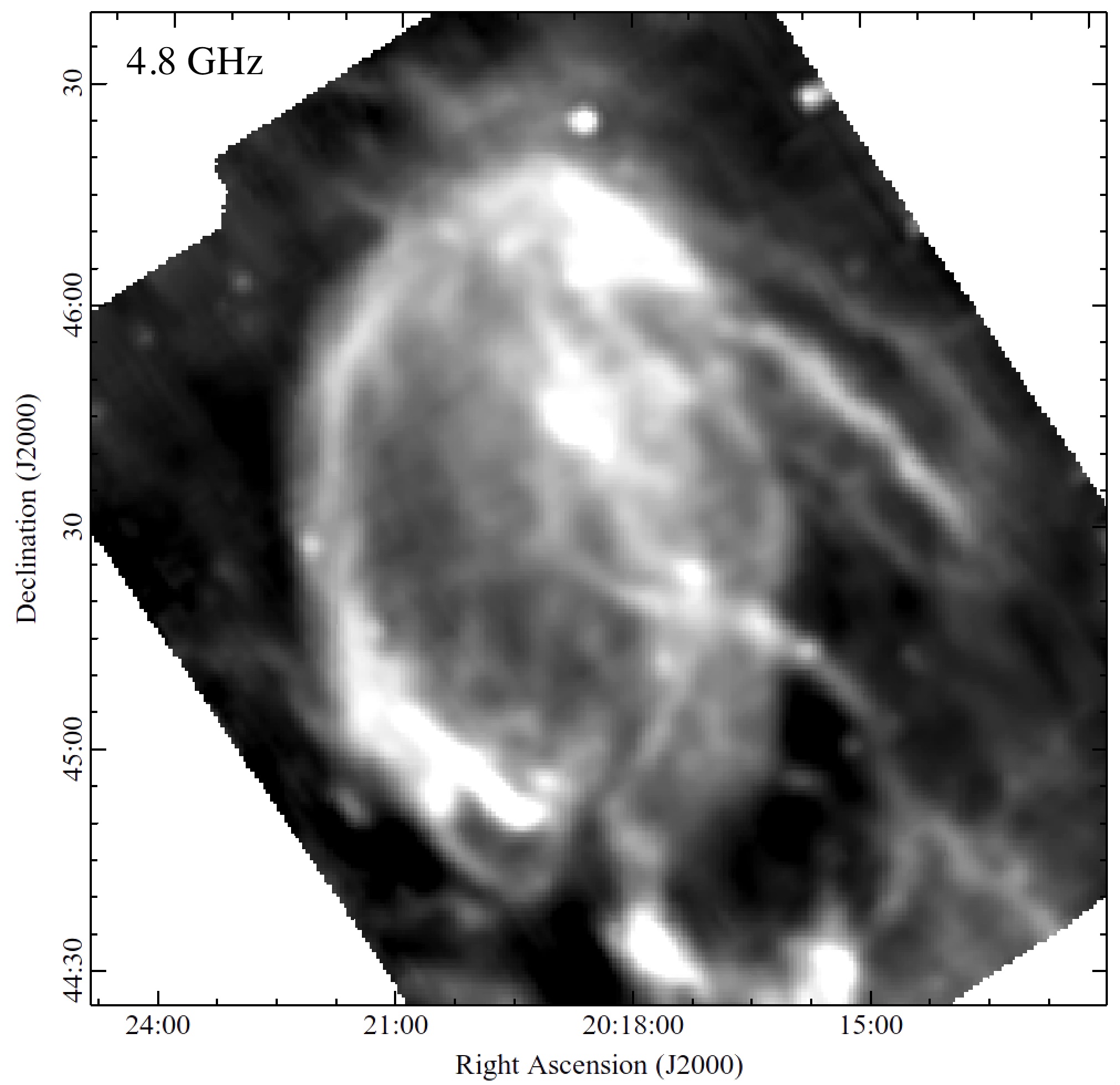} \\
\centerline{\includegraphics[angle=0,width=8.0cm]{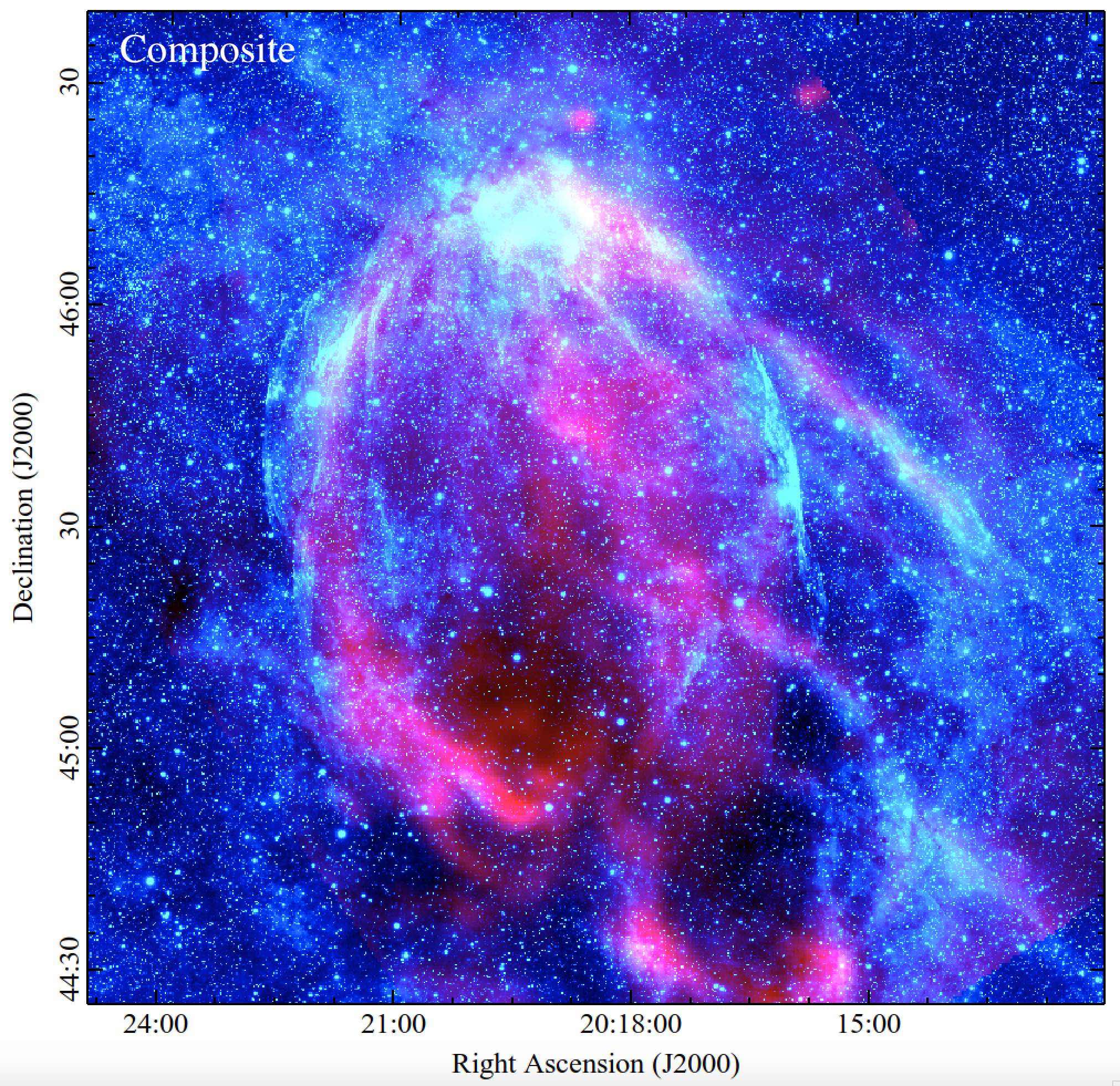}}
\caption{Optical \O3 $\lambda$5007, radio 4.8 GHz and color optical/radio composite images of G82.2+5.3 (W63) showing positional overlap of western and southern emissions features outside of main remnant shell.  \label{G82_radio}  }
\end{center}
\end{figure*}

In the H$\alpha$ image, numerous sharp filaments are visible but it is not clear exactly where the remnant lies without referring to radio or X-ray images. However even then, it is difficult to
determine its true extent of the remnant in the radio because, as noted by \citet{Kothes2006}, it is not clear which of the filaments surrounding the main source actually belongs to the SNR.

Part of the problem is that there is significant extensions (streaks) to the west and southwest as \citet{Mav2004} noted. 
While their \O3 images showed fainter emission off to the west and southwest
that seemed related to 4850 MHz radio emission in this region,  \citet{Mav2004} discounted
that possibility, viewing such emissions outside the remnant's emission shell as more likely associated with  H~II region emission. 

However, our deeper and wider FOV \O3 images show a more extensive \O3 emission filaments than shown in the images of \citet{Mav2004}. This can be seen in 
Figure~\ref{G82_positives} where we show our H$\alpha$ and \O3 images as positive versions without stars. The brighter \O3 filaments define a well structured, limb-brightened shell. But there are also significant narrow bands or streaks of emission off to the west that stretch farther to the west and southwest where one finds additional emission that joins with the southern portion of the shell. In addition, north of the top of the \O3 shell, our \O3 image detects a large region of faint diffuse \O3 emission.

These surrounding \O3 emission features appear to represent significant extensions of the remnant outside of its main shell and are not simply unrelated nebulae. Radio maps of the remnant support this conclusion by showing coincident nonthermal radio emissions with these \O3 emission features. This may explain the difficulty of defining the remnant's true extent \citep{Kothes2006}.

The correlation between the remnant's outlying \O3 features with its radio emission is shown in 
Figure~\ref{G82_radio} where we show a positive version of our \O3 image alongside an unpublished Effelsberg 4.8 GHz radio map (FHWM = 2.4$'$) of the remnant.
Both the \O3 and radio images show emission off to the west/southwest of W63's bright shell. Two bright radio
features/blob correspond to seemingly unremarkable
\O3 emission clouds. In addition, we note that faint \O3 emission located off to the shell's north is weakly detected in the radio.

The color panel in Figure~\ref{G82_radio} shows a composite image of our \O3 image (blue) with the
4.8 GHz radio map (red). As can be seen in the separate images
shown in the two  leftmost panels, the remnant's main shell
is wider than its radio emission. This difference is clear in the color composite where the radio emission filaments lie inside the optical \O3  emission shell. While there is little correspondence of radio emisson with the SNR's bright \O3 along the shell's
northern limb, there is a good match between radio and optical filaments along W63's eastern limb.

A summary of our optical imaging findings of the
G82.2+5.3 (W63) remnant is that while its
H$\alpha$ emission is hard to distinguish from the very complex H$\alpha$
emission in this Galactic direction, the remnant displays a clear and extensive \O3 emission structure. We also note that 
in addition to 
exhibiting a sharp and well defined emission shell in \O3, there is 
considerable related remnant emission outside of this shell in the form of long emission bands that extend
nearly a degree to the west and southwest.

\begin{figure*}[ht]
\begin{center}
\includegraphics[angle=0,width=14.7cm]{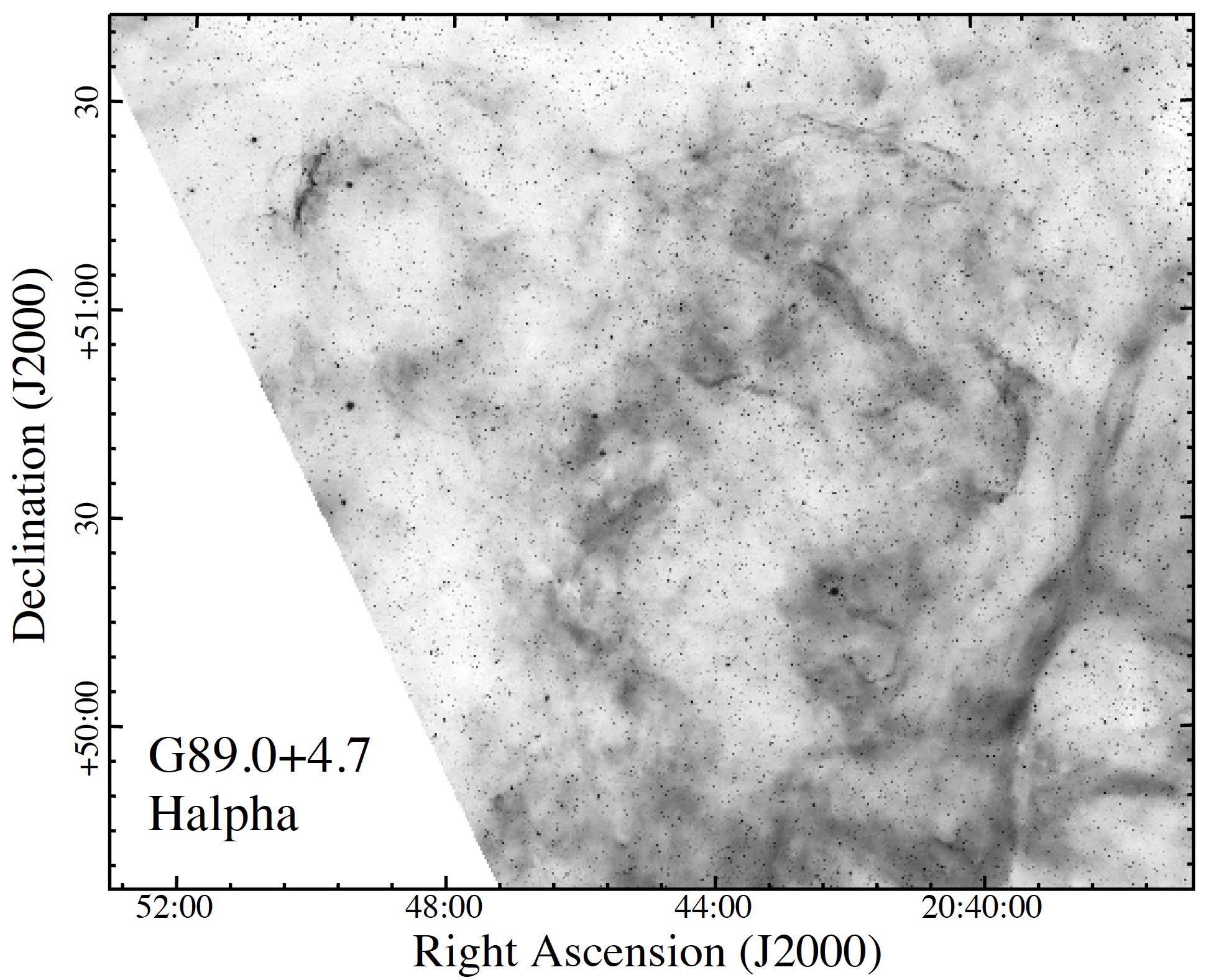} \\
\includegraphics[angle=0,width=14.7cm]{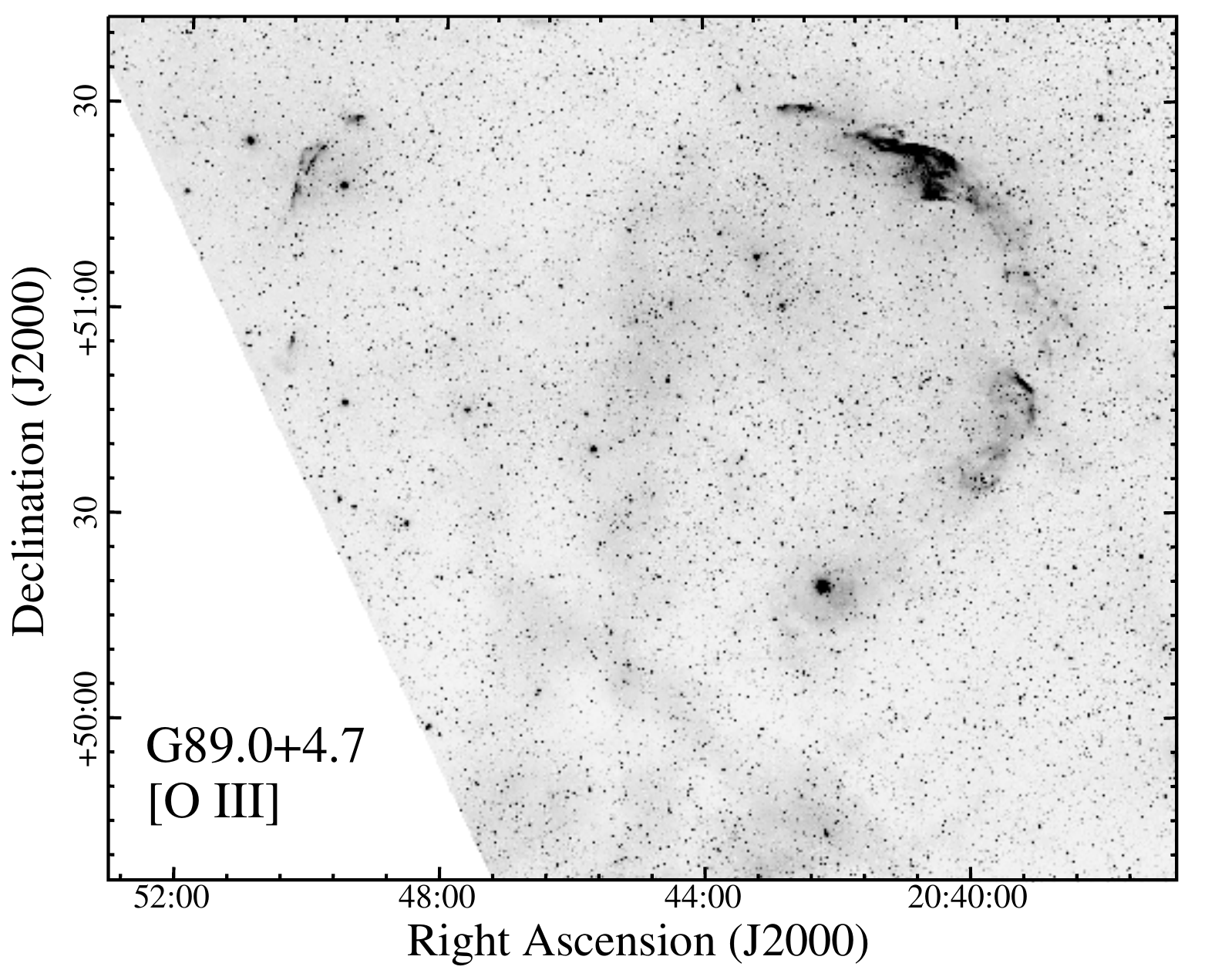} 
\caption{H$\alpha$ and \O3 images of G89.0+4.7 (HB~21). Bright star in lower right is 51 Cyg.
\label{G89_Ha_n_O3} 
} 
\end{center}
\end{figure*}

\begin{figure*}
\begin{center}
\includegraphics[angle=0,width=8.0cm]{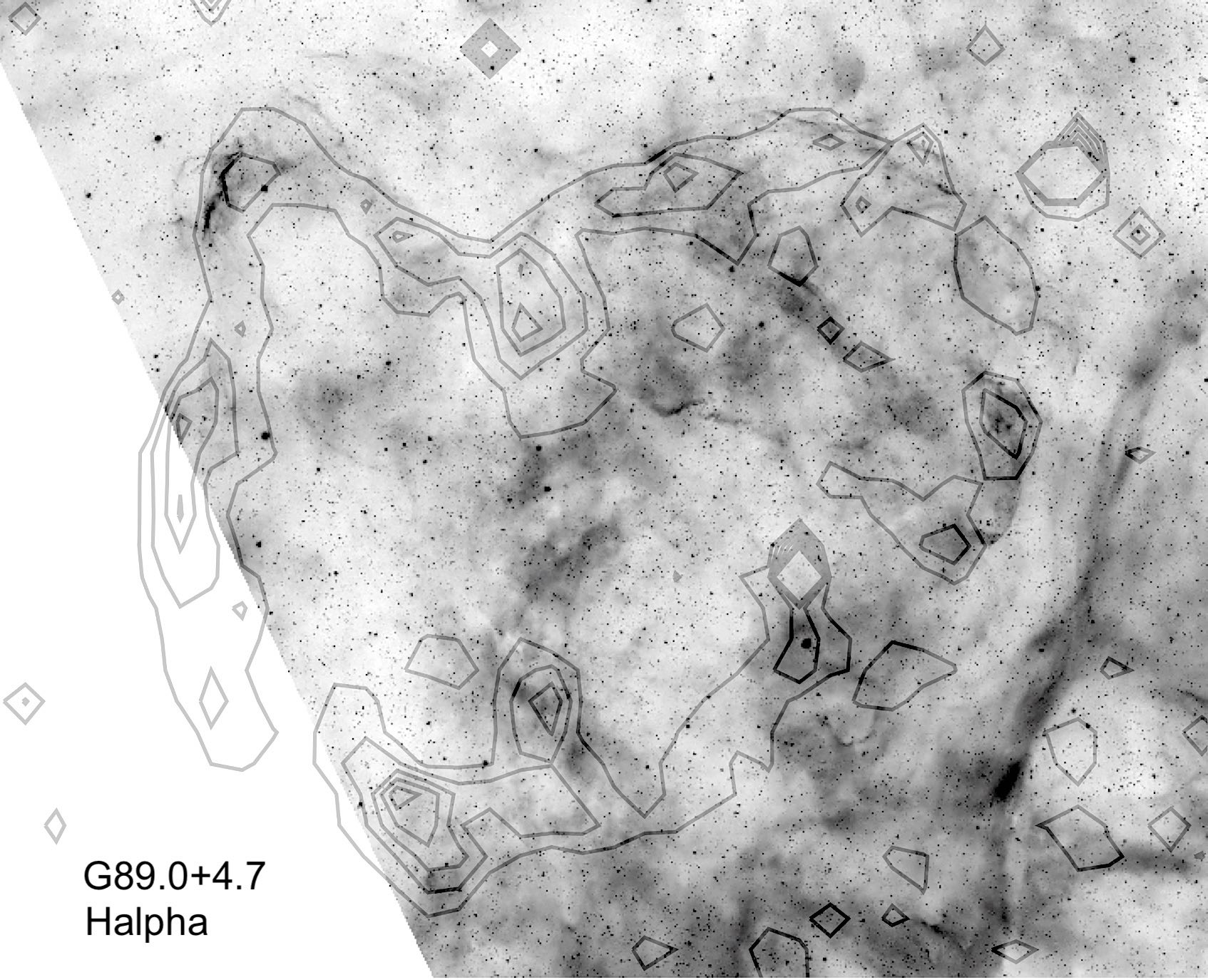} 
\includegraphics[angle=0,width=8.0cm] {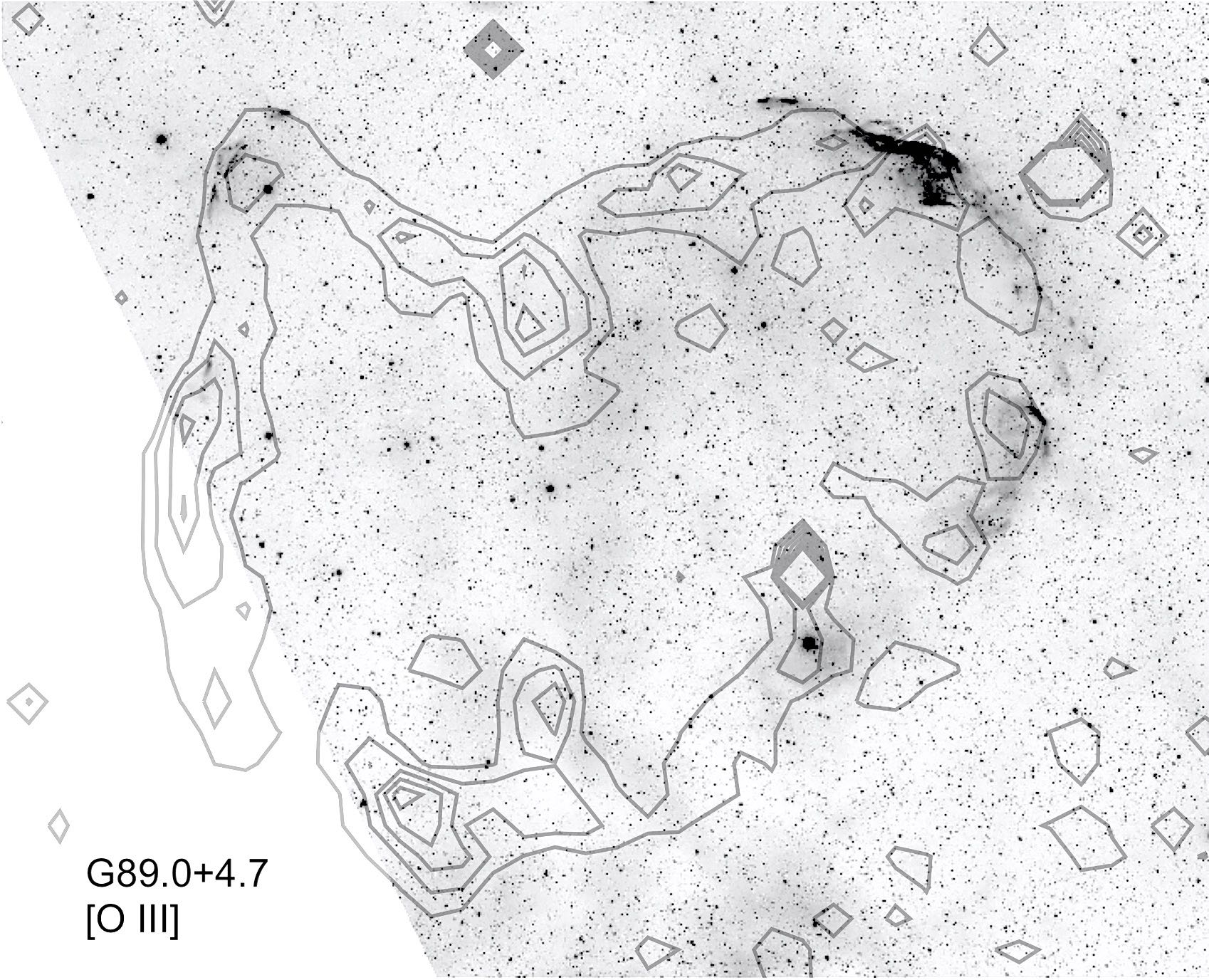}  \\
\vspace{0.20cm}
\includegraphics[angle=0,width=16.0cm]{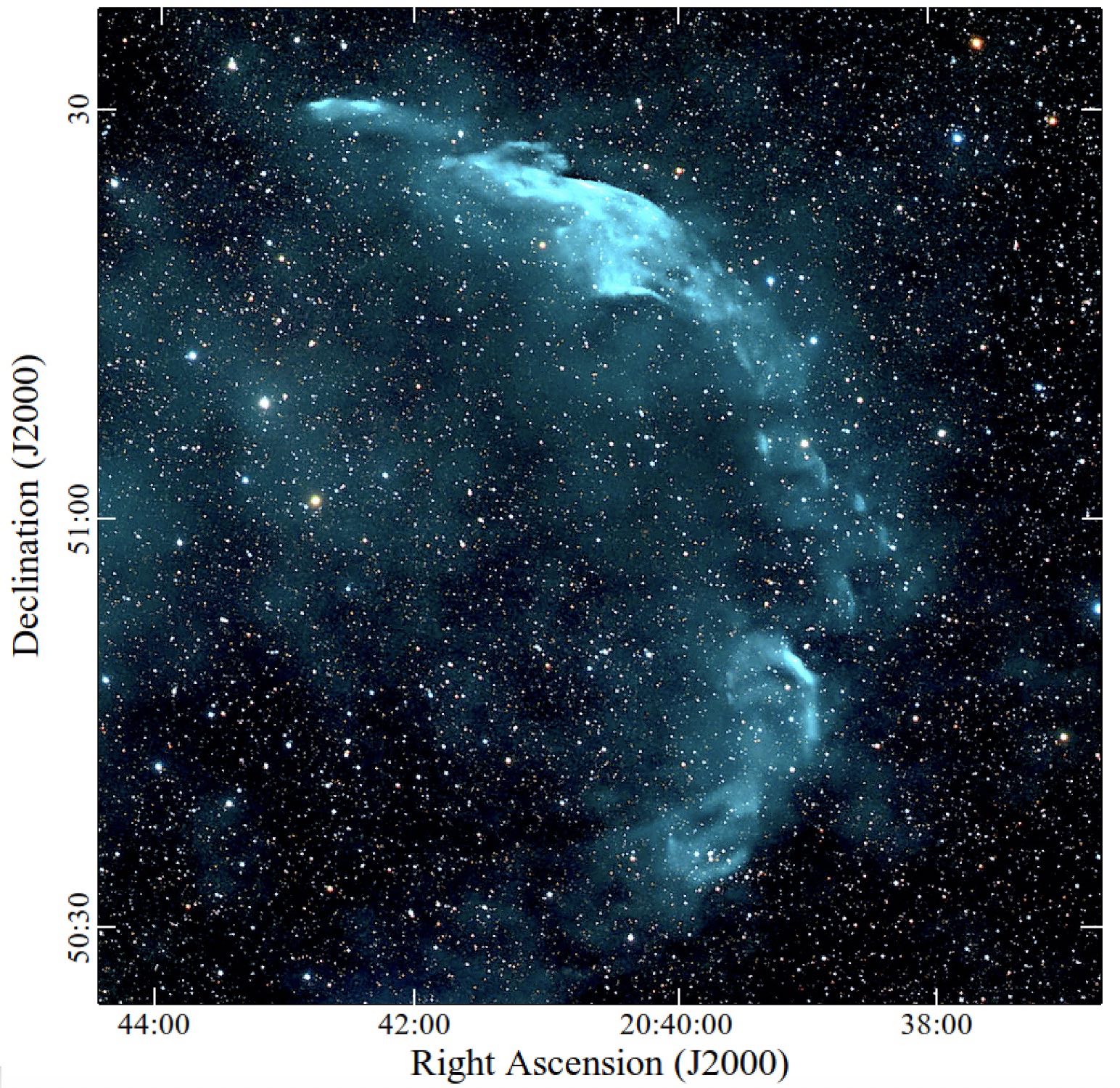}
\caption{Top panels: H$\alpha$ and \O3 images of G89.0+4.7 (HB~21) with overlay of WENSS 92 cm radio intensity contours.
Bottom: Our \O3 image plus 
RGB images
of G89.0+4.7's northern and western emission filaments.
\label{G89_radio} 
} 
\end{center}
\end{figure*}

\subsubsection{G89.0+4.7 (HB 21)}

Identified as a Galactic radio source some 70 years ago by \citet{HB1953},
G89.0+4.7 (HB 21)
has subsequently been a well studied Galactic remnant in the radio, infrared,
and X-rays in part due to its interaction with molecular clouds. However, its optical emission has been difficult to distinguish from other ISM emission features in the Galactc plane. Consequently, few  images have been published regarding its optical emission structure and properties.

\citet{Minkowski1958} noted filamentary nebulosity visible on Palomar Sky Survey images 
in the complicated region in Cygnus where HB~21 lies and suggested that some of this emission was likely associated with the remnant's radio emission
(see \citealt{Willis1973} for images). Using H$\alpha$ and \O3 
photographs of the HB~21 region with narrow 30--40 \AA \ filters,  
\citet{Lozinskaya1972} found faint optical filaments that were faintly
distinguishable from the background emission clutter, and were better seen in H$\alpha$ than \O3. 
However, the difficulty of clearly identifying the remnant's optical emission was demonstrated by
\citet{vdb1978a}'s failure to definitely find associated optical emission using
a 2 hr 098 emulsion + H$\alpha$ and a 110 minute 098 + [\ion{S}{2]]} Palomar 1.2m Schmidt images.

A clearer picture of HB~21's associated optical emission was made by \citet{Mav2007}
who obtained $70' \times 70'$ wide H$\alpha$ and [\ion{S}{2}] images of the remnant.
They found good correlation of [\ion{S}{2}] strong filaments with 4850 MHz radio emission
along the remnant's eastern area. Overall, the remnant's emission was relatively faint and
patchy in the central and west areas of the remnant.
These authors also obtained spectra of a half dozen regions which showed weak \O3 emission
suggesting relatively low shock velocities ($<$ 100 km s$^{-1}$).
This conclusion was viewed in-line  with the interferometric observations by
\citet{Lozinskaya1980} who observed velocities only as high as
60 to 80 km s$^{-1}$.

We obtained images of HB~21 centered on a region which preliminary images had indicated some \O3 emission, leading to our images missing the southeastern portion of the remnant. The results of our H$\alpha$ and \O3 images are
shown in Figure~\ref{G89_Ha_n_O3} where
our H$\alpha$ image is similar to that presented in \citet{Mav2007}. That is, 
we detected a handful of sharp H$\alpha$ filament along HB~21's northern and northeastern rims. 

However, our \O3 images revealed new emission features not previously seen.
Namely, they showed considerable \O3 emission along much of the remnant's northwestern limb, faint diffuse emission
toward the remnant's center, plus some
filamentary emission along its
northeastern edge coincident with
H$\alpha$ filaments. Although these \O3 emissions appears bright in our image possibly leading one to wonder why wasn't this emission detected previously, the remnant's \O3 emission is actually quite faint and was only detected through the co-addition and post-processing of five hundred 300 second (41.5 hr) \O3 images.

Figure~\ref{G89_radio} shows our H$\alpha$ and \O3 images along with radio emission contour intensity overlays from the Westerbork Northern Sky Survey (WENSS; \citealt{Rengelink1997}). In terms of H$\alpha$  vs radio flux one finds 
reasonably good agreement for the NE filament and for the few H$\alpha$ filaments along HB~21's northern rim.
However, the overall agreement is not strong and explains why firmly identifying associated optical emission to HB~21 was difficult.

A bit better correlation is seen for the remnant's \O3 emission, especially
along the northern and western limbs. Like seen for H$\alpha$, radio emission matches the northeastern \O3 filaments and a faint eastern limb \O3 filament near the edge of our FOV.

Although no prior optical study of HB~21 reported any detected 
\O3 emission, the remnant's \O3 emission along its NW and western edge 
is impressive. The bottom of Figure~\ref{G89_radio} also shows our \O3 image combined with short RGB exposures into a color composite. 
This image highlights the generally diffuse structure of the \O3 emission features. It also shows the very diffuse emission near the remnant's center. Although  this image indicates the presence of extremely faint emission out ahead of the band of \O3 emission, it is not clear if this is shock emission or dust reflected light.

\begin{figure*}[ht]
\begin{center}
\includegraphics[angle=0,width=16.5cm]{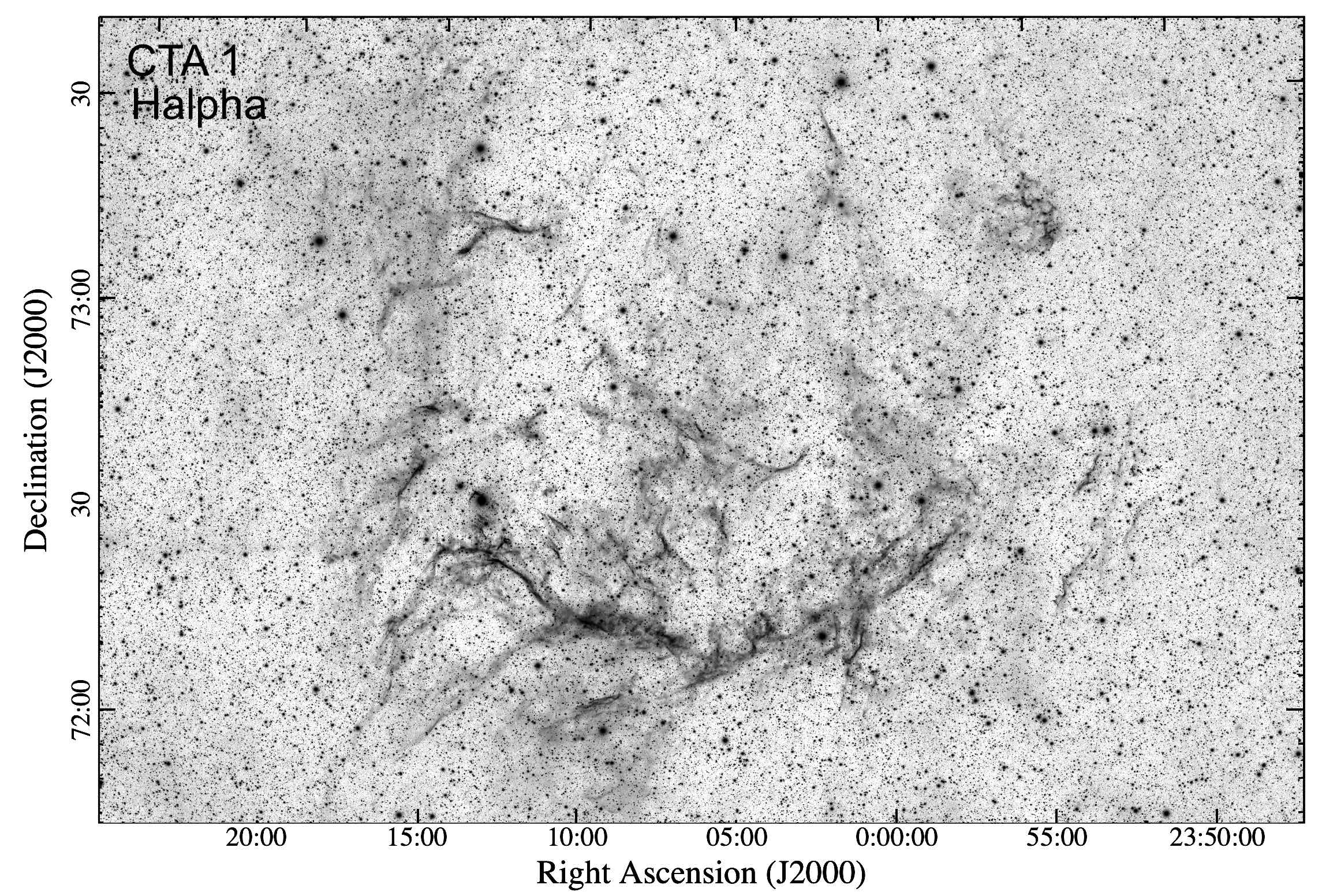}  \\
\includegraphics[angle=0,width=16.5cm]{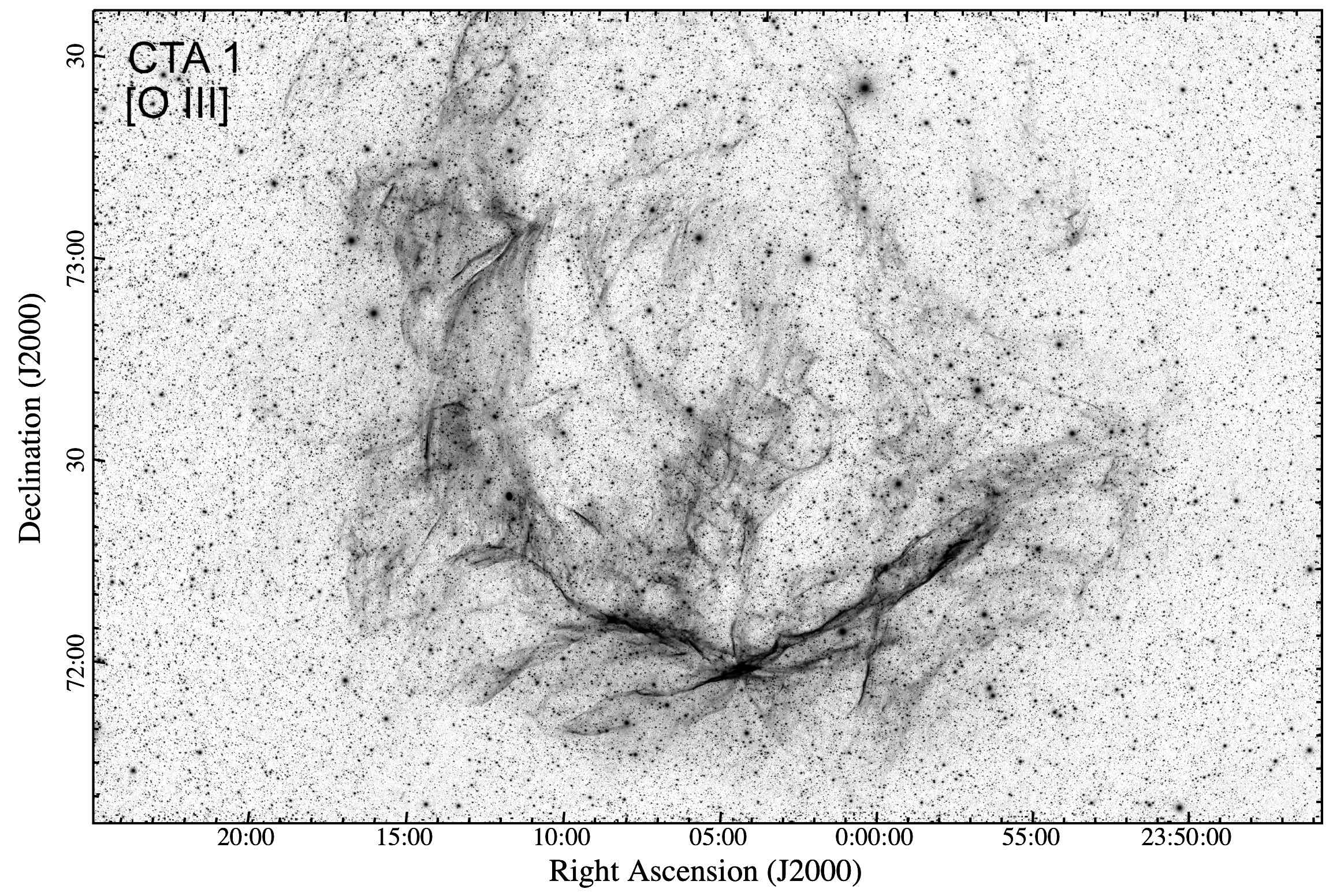} \\
\caption{H$\alpha$ and \O3 images of CTA 1 (G119.5+10.2).
\label{CTA1_O3_n_Ha} 
} 
\end{center}
\end{figure*}

\begin{figure*}
\begin{center}
\includegraphics[angle=0,width=17.0cm]{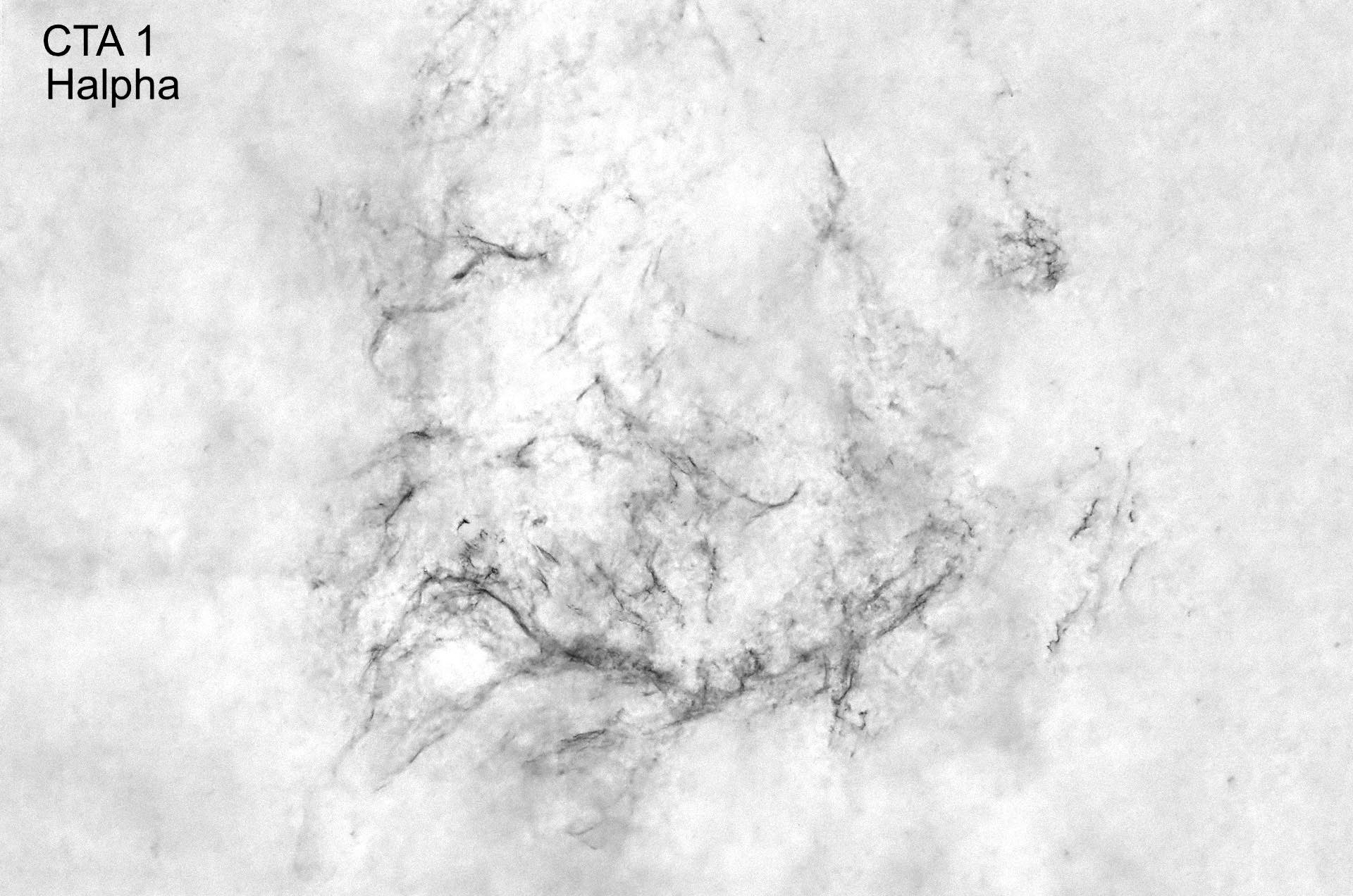}
\includegraphics[angle=0,width=17.0cm]{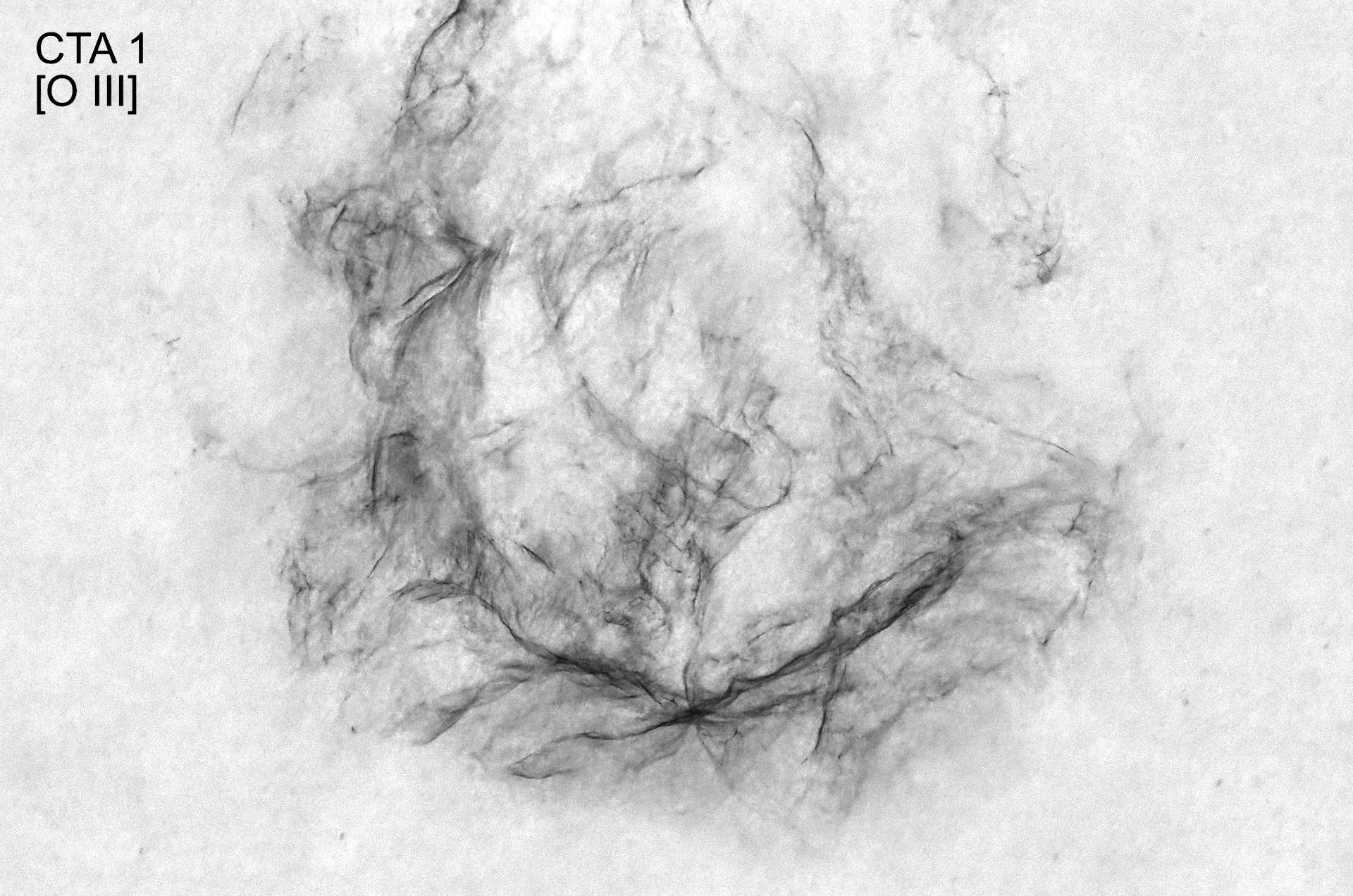}
\caption{Same data as in Figure~\ref{CTA1_O3_n_Ha} on CTA 1 (G119.5+10.2) but now shown as a positive images with the stars removed by software to make the remnant's H$\alpha$ (top) and [\ion{O}{3}] (bottom) emission features more readily visible.   \label{CTA1_pos}
} 
\end{center}
\end{figure*}

\begin{figure*}
\begin{center}
\includegraphics[angle=0,width=18.0cm]{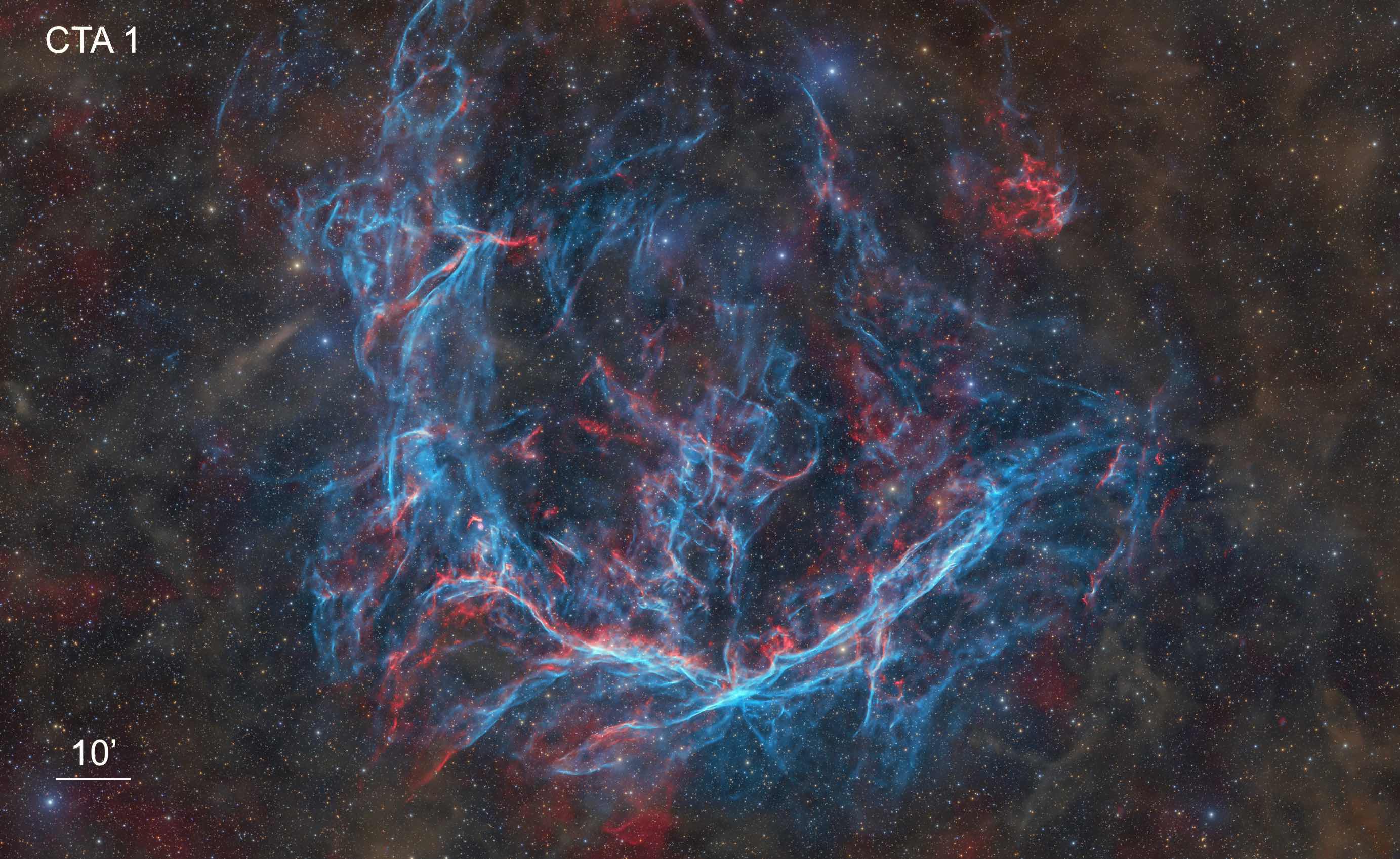} 
\includegraphics[angle=0,width=18.0cm]{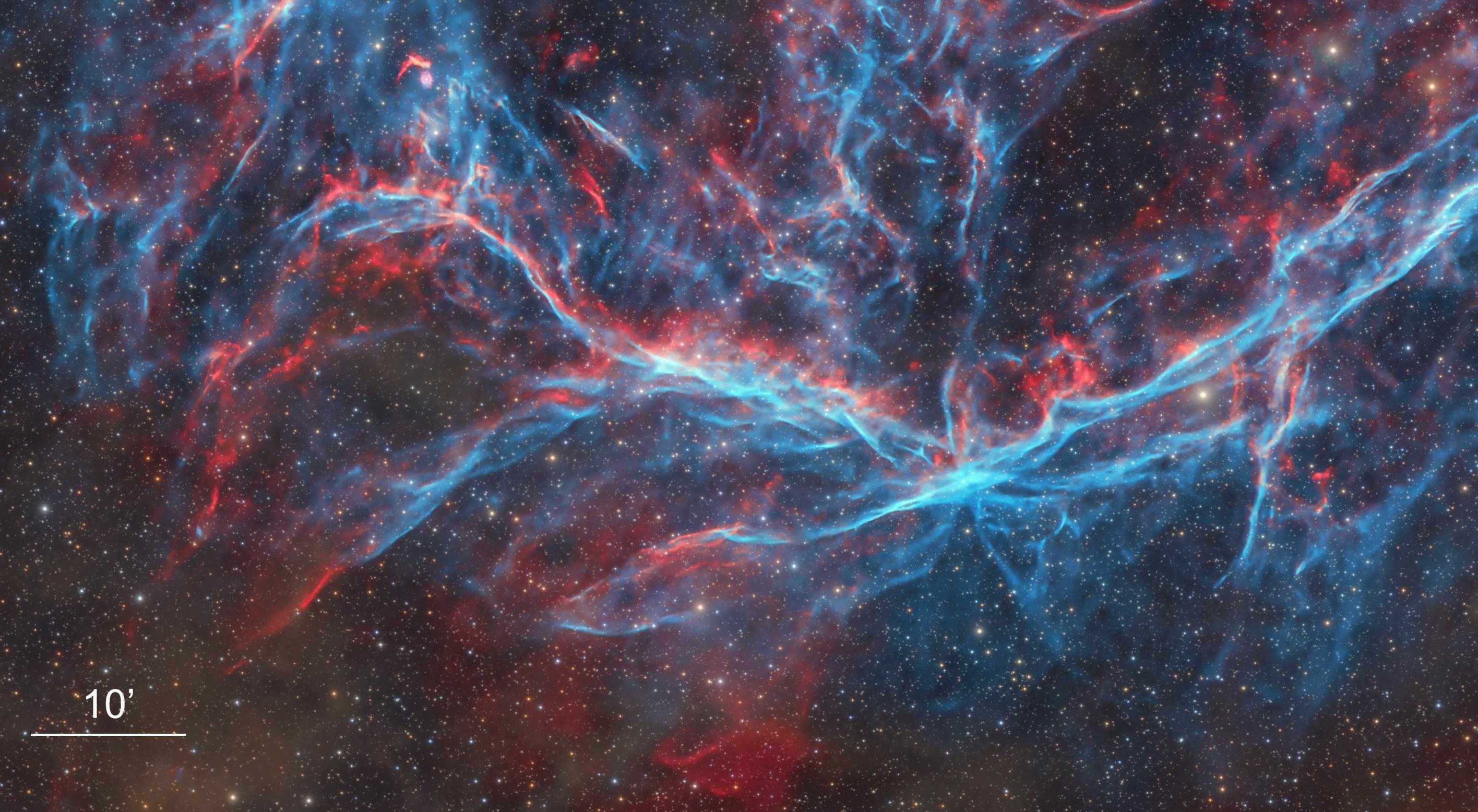}
\caption{Color composite of CTA 1 (G119.5+102) with H$\alpha$ (red) and \O3 (blue) emissions along with RGB images. Top panel shows the entire remnant as imaged. Bottom panel is a blowup of its southern region. North is up, East to the left.
\label{CTA1_color} 
} 
\end{center}
\end{figure*}

\subsubsection{G119.5+10.5 (CTA 1)}

\citet{Harris1960} detected a large, 
$2\degr$ diameter radio source at 960 MHz located 
well off the Galactic plane at $l$ = 119.5$\degr$, $b$ = 10.2$\degr$.
Because a few faint wisps of optical emission visible
on the red print of the Palomar Observatory Sky Survey
(POSS I) agreed in location with the radio
emission, \citet{Harris1962} suggested
that this radio source, named CTA 1 due to it being the first in Caltech's Catalog A of radio emission sources, was a possible supernova remnant. This
conclusion was later confirmed through subsequent 2695 MHz observations made by \citet{Sieber1979}.

The remnant's radio emission shell is incomplete along its north and northwest
and more recent radio studies of the remnant \citep{Pineault1997,Sun2011} have concluded that this gap is due to the
remnant expanding into a much lower density ISM along these sections of its northern limb.
X-ray data point to a center-filled SNR whose spectrum suggests the presence of a pulsar, a synchrotron nebula, plus a thermal component associated with the remnant's expanding ISM shocks \citep{Slane1997}.
The presence of a pulsar inside CTA~1 was subsequently confirmed by the Fermi Gamma-Ray Space Telescope which found a pulsar located near the center of the compact synchrotron nebula in CTA 1 \citep{Abdo2008,Li2016}. The pulsar has a period of 316.86 ms and a characteristic age of 10$^{4}$ yr roughly comparable to that estimated for the SNR.

\begin{figure*}[ht]
\begin{center}
\includegraphics[angle=0,width=18.0cm]{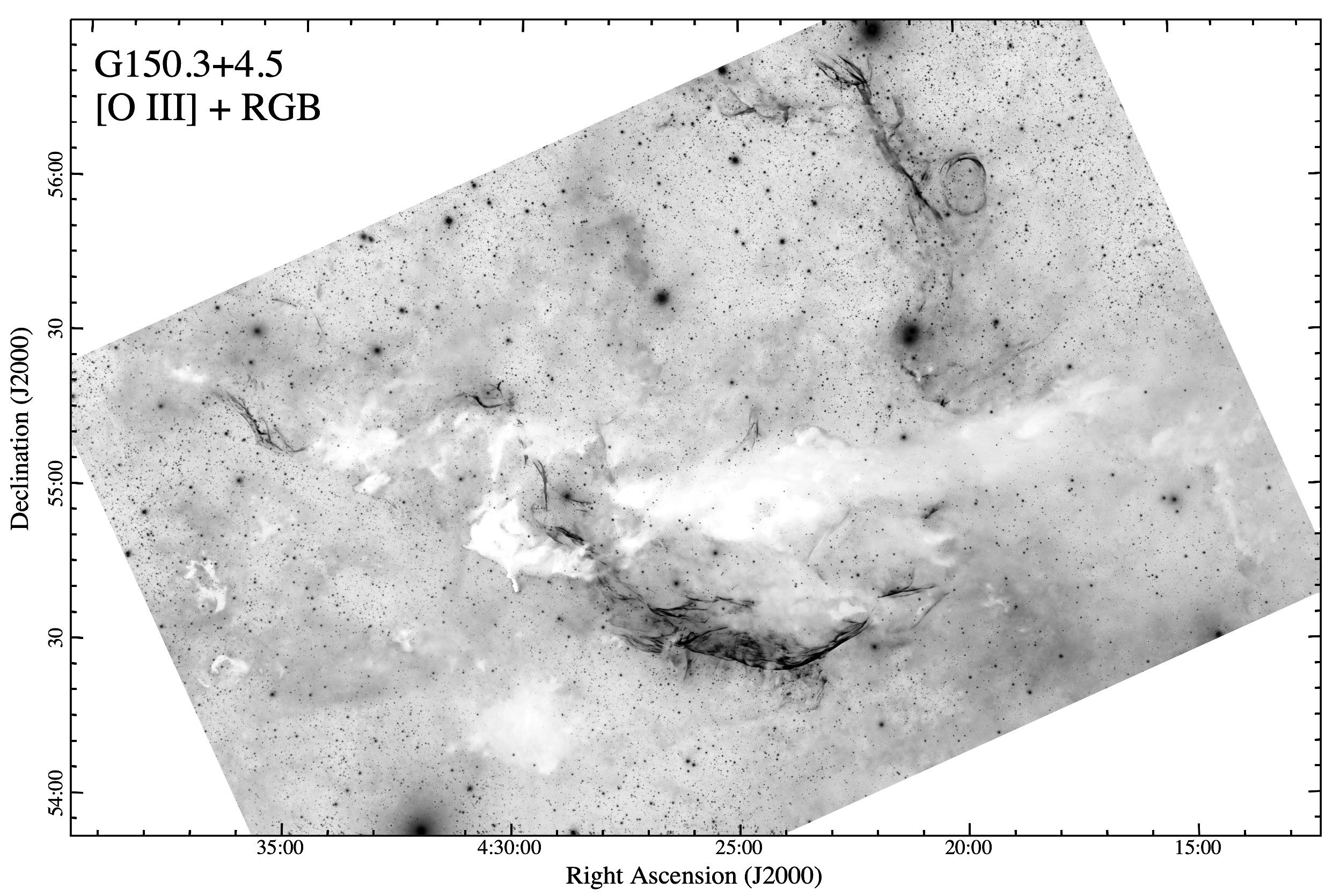}  \\
\caption{Composite \O3 plus RGB images of G150.3+4.5. Note the appearance of several dust clouds in the region. 
\label{G150_image} 
} 
\end{center}
\end{figure*}

The optical emission associated with G119.5+10.2 (CTA~1) is relatively faint, so much so 
that in the atlas of optical SNRs by \citet{vdb1973},
it was the only one of 24 objects that was 
considered too faint to be
reproduced in print. However, images taken as part of the ELSMW \citep{Parker1979} yielded clear images of the remnant, revealing it to be 
stronger in \O3 compared to its H$\alpha$ emission, with most of its northern and northwestern portions missing consistent with that seen in the radio. Higher resolution \O3 images of four small
regions indicated an overall mainly filamentary emission structure \citep{Fesen1983}. 
Deep H$\alpha$, [\ion{S}{2}], and \O3 images of the entire remnant along with optical spectra were obtained by \citet{Mav2000} who confirmed the remnant's optical emission to be mainly filamentary in \O3, and noticeably more diffuse in H$\alpha$.

\begin{figure*}[ht]
\begin{center}
\includegraphics[angle=0,width=8.2cm]{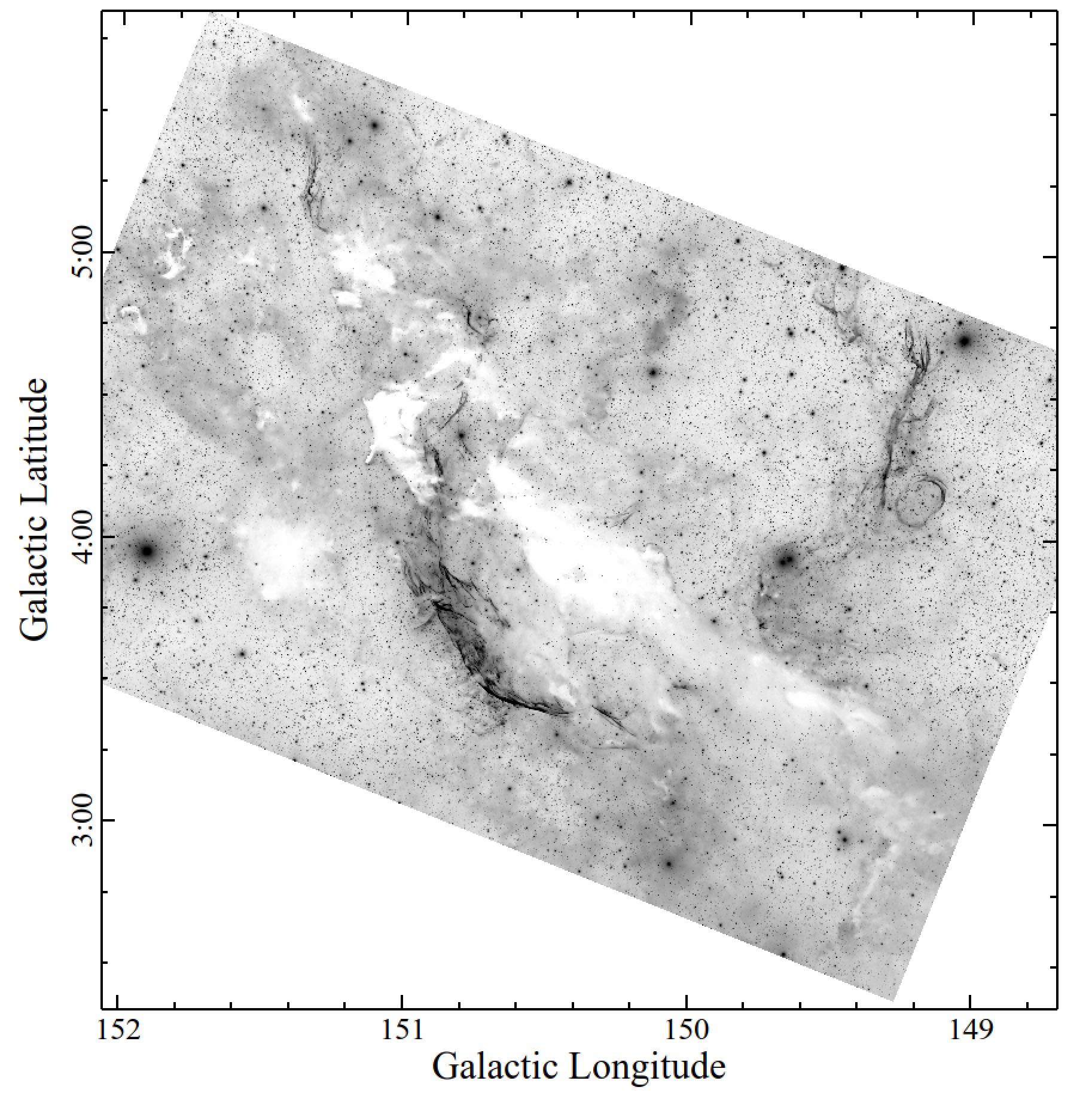} 
\includegraphics[angle=0,width=8.0cm]{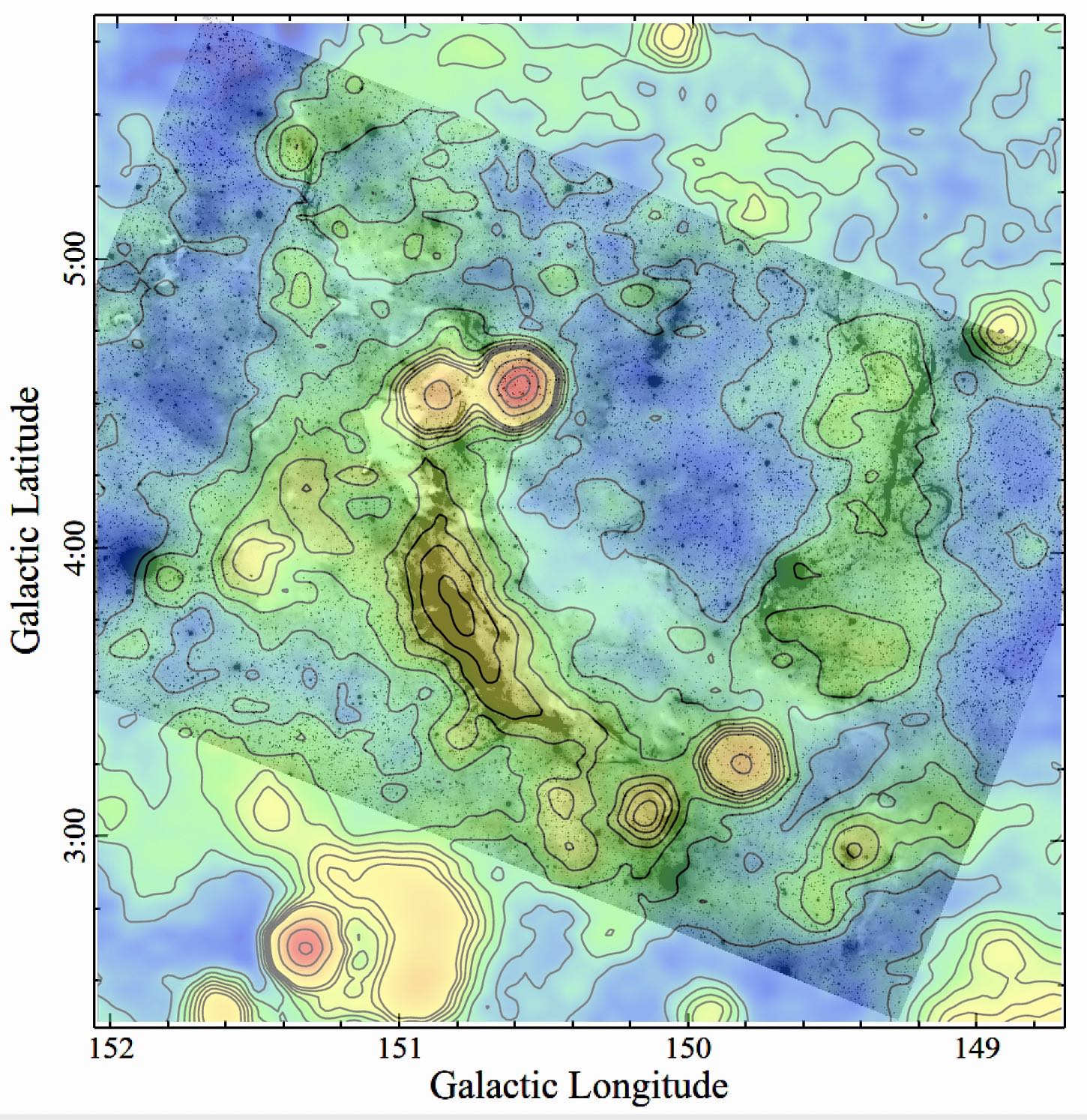}
\caption{
The left panel our \O3 + RBG image of G150.3+4.5 displayed
in Galactic coordinates, with the right panel
showing an overlay of the remnant's color coded 6 cm radio emission and brightness contours 
from \citet{Gao2014}.
\label{G150_radio} 
} 
\end{center}
\end{figure*}

\begin{figure*}
\begin{center}
\includegraphics[angle=0,width=16.5cm]{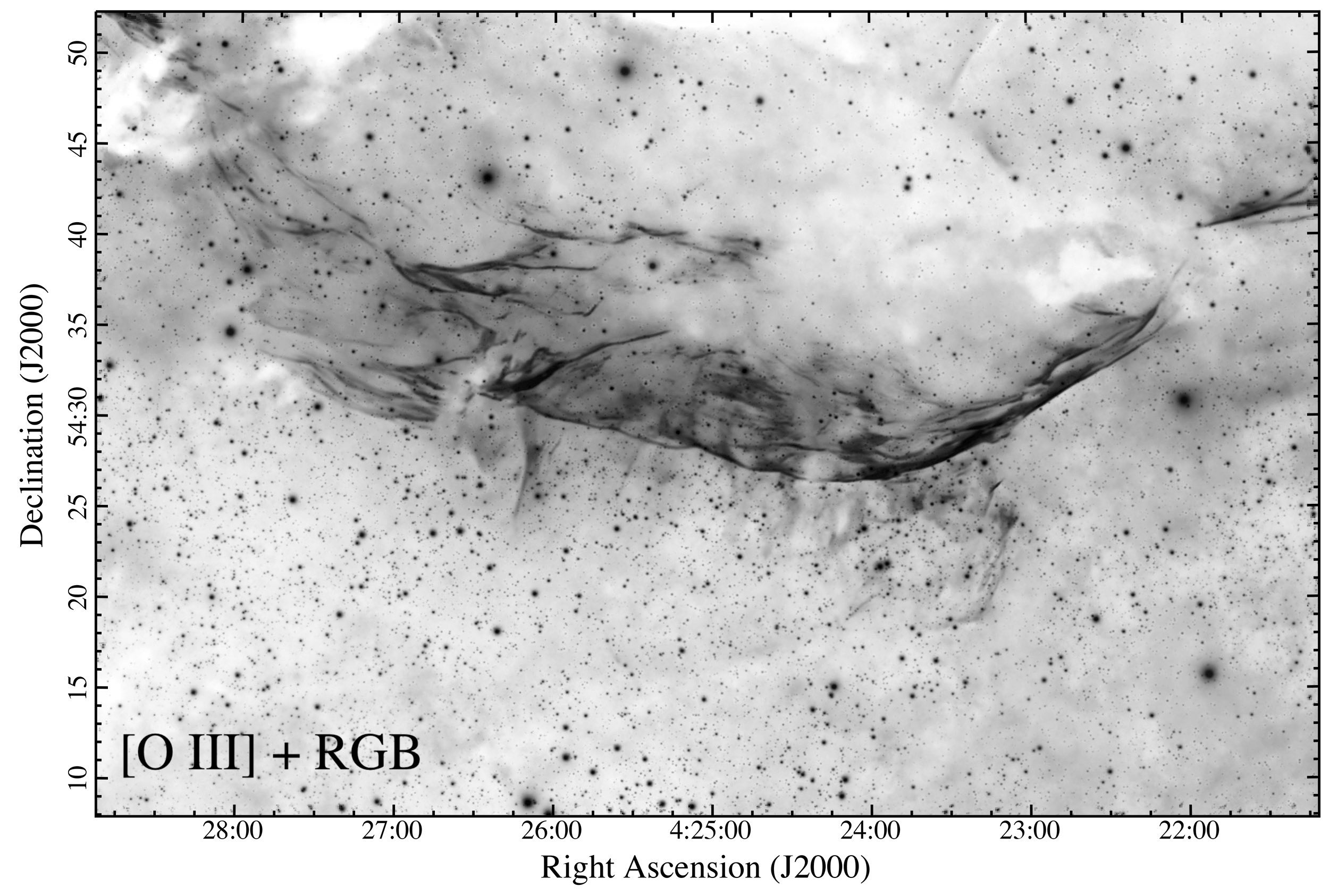}  \\
\includegraphics[angle=0,width=16.5cm]{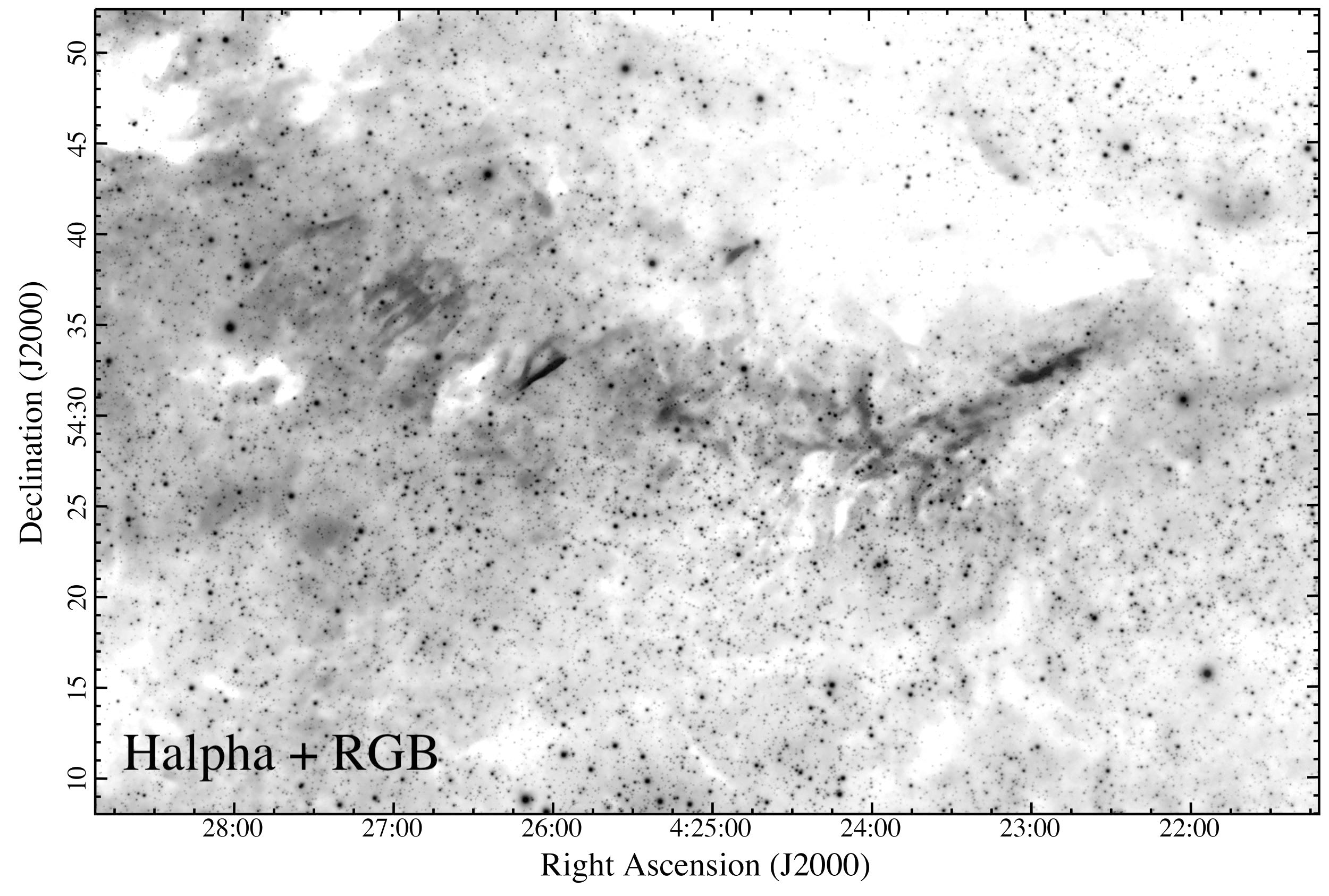} 
\caption{Composite \O3 and H$\alpha$ plus RGB images of the southern limb of G150.3+4.5 showing both the presence of sharp \O3 filaments along with scattered emission extending farther to the south, and the lack of associated filamentary H$\alpha$.
\label{G150_O3_n_Ha} 
} 
\end{center}
\end{figure*}

\begin{figure*}
\begin{center}
\includegraphics[angle=0,width=16.5cm]{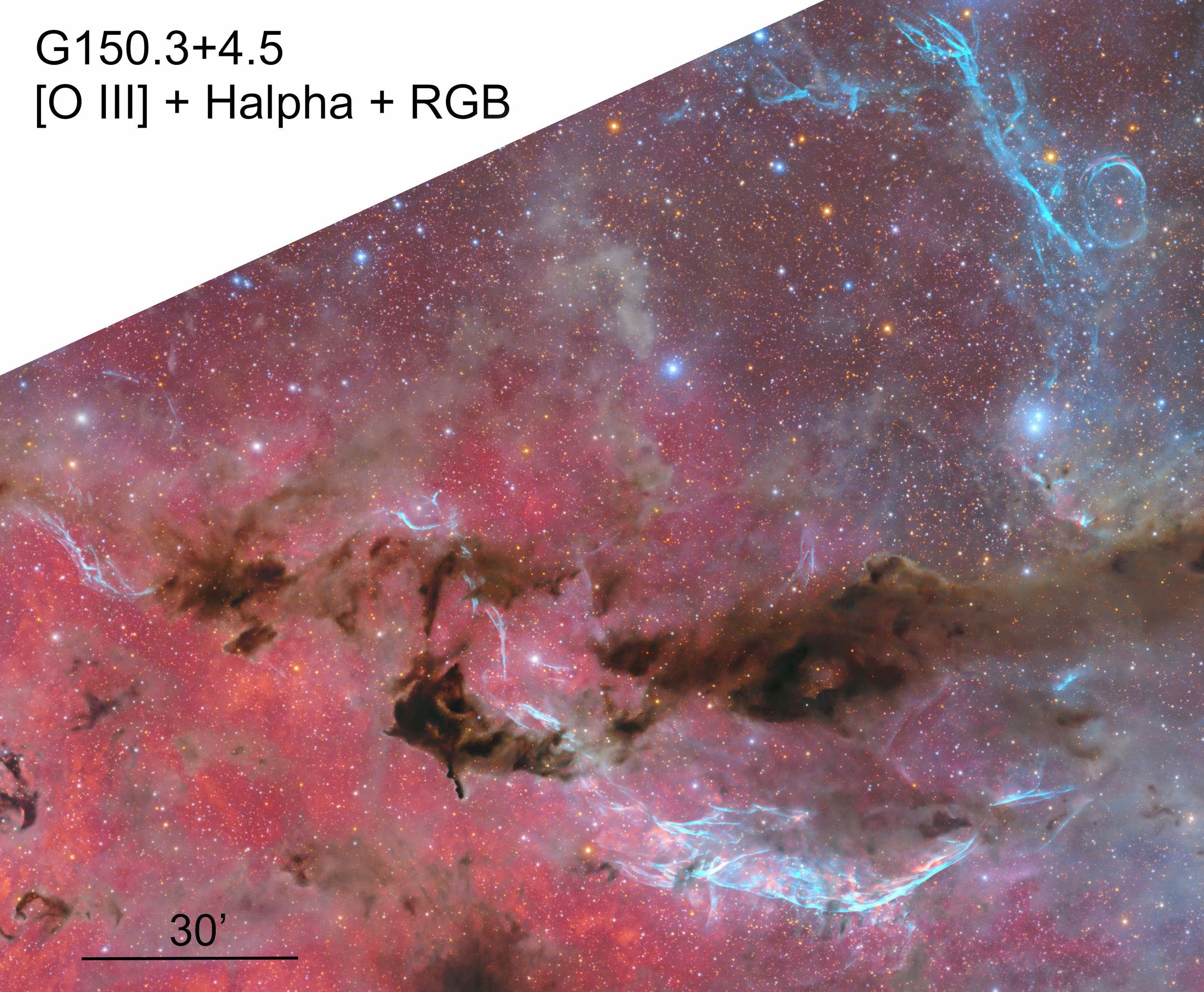}  \\
\includegraphics[angle=0,width=16.5cm]{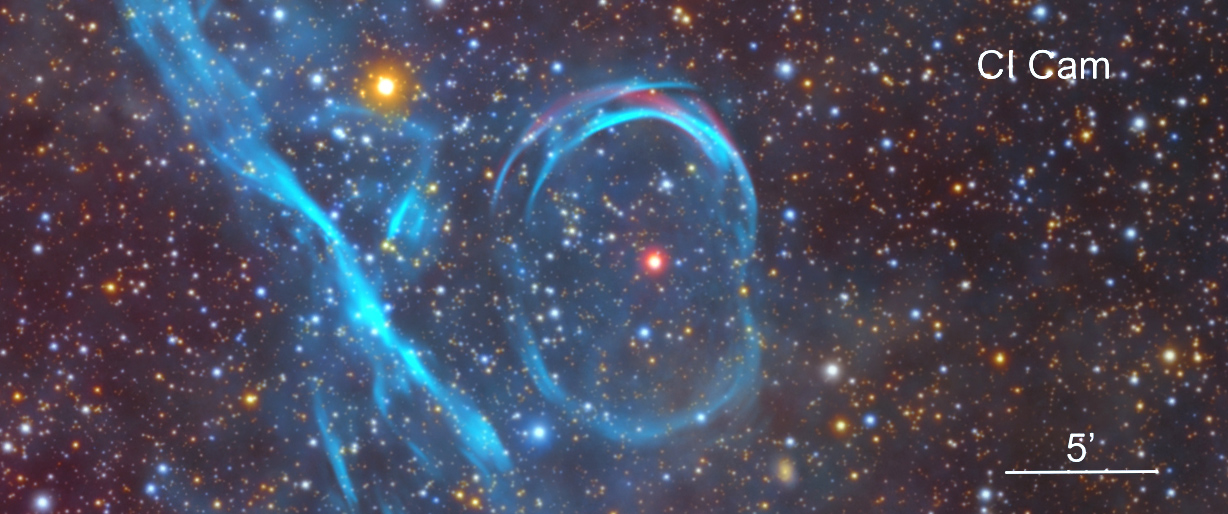} 
\caption{Top: Color composite made from \O3 (blue) and H$\alpha$ (red) plus broad RGB images of G150.3+4.5.
North is up, east to the left. Bottom: Enlargement of a section along remnant's northwestern limb showing an \O3 emission shell centered on the variable X-ray binary B[e] star, 
CI Cameleopardalis (CI Cam).
\label{G150color} 
} 
\end{center}
\end{figure*}

The CTA 1 remnant is roughly $2.0\degr \times 2.5\degr$ in size placing it among the largest known Galactic SNRs.
Our H$\alpha$ and \O3 images did not cover the entire remnant, missing a small part of its northeastern section, i.e., above Dec = 
+73.6$\degr$. Nonetheless, our images presented in 
Figure~\ref{CTA1_O3_n_Ha} cover the bulk of the remnant's optical emission.  Our images are both deeper and of higher resolution than those of \citet{Mav2000} and clearly define the structure of the remnant's optical emission. They also better highlight the substantial differences in the remnant's H$\alpha$ and \O3 emission structure.  

As has been long known from the ELSMW images of \citet{Parker1979},
CTA~1's optical emission is dominated by \O3 emission.
The remnant's brightest \O3 filaments are found in the south where one finds a complex of bright long, curved and overlapping filaments. Other bright \O3 filaments are seen along the SNR's eastern limb,
plus a surprising number of short filaments located near the remnant's projected center. However, \O3 filaments
 are almost entirely absent along CTA~1's west and northwestern limbs.

Although the remnant's \O3 emission exhibits many bright and sharp filaments, our images also show the presence of extensive diffuse emission regions across the remnant. While such diffuse emission often coincides with the remnant's bright filaments, it is widespread albeit very faint in places.
There is also noticeable diffuse, very faint \O3 emission surrounding most of the remnant's entire southern limb,
extending as far south as Dec = 71.8$\degr$. 

CTA~1's Halpha emission, in contrast, is far more limited, being brightest and
largely confined to the SNR's southern limb. 
Although there are a few sharp 
H$\alpha$ filaments scattered across the whole of the remnant, much of the H$\alpha$ emission is diffuse and not always coincident with diffuse \O3 emission. Faint, diffuse H$\alpha$ emission is also seen to extend farther south than that of the \O3 emission, especially to the southeast.

There are also a number of small clouds which are bright
in H$\alpha$ scattered across the remnant. A small $\sim15'$ emission patch along the remnant's northwest
limb is noticeable for its dominated H$\alpha$ emission.

Fine-scale differences between CTA~1's H$\alpha$ and \O3 emission morphologies 
can be more clearly seen in Figure~\ref{CTA1_pos} where we present
both H$\alpha$ and \O3 images with the stars removed.
Whereas H$\alpha$ emission consists mainly of short filaments
mainly confined to the CTA~1's central region and southern and eastern rims, its \O3 emission is quite extensive consisting equally
of filaments and diffuse emission regions.

Color composite images of CTA~1 made from combining H$\alpha$, \O3 and RGB exposures are presented in 
Figure~\ref{CTA1_color}. The upper panel shows a wide view of the remnant while the lower panel presents a blowup of just the southern
limb region. These figures showcase the complexity of the remnant's optical emission-line and thus the danger of spectral analysis based on single slit data taken of just a few regions.

Notable emission features include the parallel, ripple-like emission filaments seen near the center of the blow-up image,
and the small patches of H$\alpha$ emission lining the inner edge
of the nearly east-west line of bright \O3 filaments.
These color images also highlight the presence of
neighboring dust clouds, seen as grayish features in the 
wide top panel.

\subsubsection{G150.3+4.5}

From measurements made at 4.8 GHz, 2.7 GHz, 1420 MHz, 408 MHz,
and 327 MHz,
\citet{Gerbrandt2014} found two crescent shaped nonthermal emission
features that stretched from $l = 150.5\degr$, $b = 3.4\degr$ to
$l = 150.9\degr$, $b = 3.9\degr$. While they were unsure if 
these crescents form part of an eastern shell limb of a
single large object, a subsequent study by \citet{Gao2014} taken
at 6 cm showed they were in fact part of a large $2.5\degr$ wide
and $3.0\degr$ tall oval-shaped
remnant centered at $l = 150.3$, $b = 4.5\degr$.
The emission shell was incomplete, with a large gap in the northwest.

Although no associated optical emission has been published on this remnant,
\citet{Gerbrandt2014} noted that the red DSS2 image showed
faint red filaments spatially coincident with the crescent radio feature. Using WHAM data, \citet{Gao2014} also found broad H$\alpha$ emission overlapped with the eastern shell of the G150.3+4.5. 

Because our initial images showed little H$\alpha$ emission associated
with the G150.3+4.5 remnant, we concentrated on taking deep
\O3 imaging. Figure~\ref{G150_image} shows a composite of our \O3 plus broad RGB images, highlighting both the extent of
the remnant's filamentary \O3 emission but also the significant amount of obscuring ISM clouds in this direction. We find many \O3 bright filaments throughout the region of the remnant we imaged, especially along the remnant's southern limb.

As a check on whether these \O3 filaments are really connected to the G150.3+4.5 SNR, the two panels of Figure~\ref{G150_radio}
show the \O3 + RGB composite image now in Galactic
coordinates along side with an overlay of color coded 6 cm radio emission and brightness contours taken from \citet{Gao2014}. 
We find the positional agreement between the optical \O3 filaments and the radio emission to be excellent; specifically, along the
remnant's southeastern shell where the radio emission is brightest and where \citet{Gerbrandt2014} found 
crescent shaped emission we find the brightest \O3 filaments. Also, 
the \O3 emission filaments at
$l = 149.5\degr$, $b = 4.5\degr$ appear coincident with the
western side of the remnant's radio shell.
These and the overall agreement between the nonthermal radio shell discussed by \citet{Gao2014} and our detected \O3 emission filaments make it clear that this optical emission is associated with the SNR and marks the first fully confirmed optical emission from G150.3+4.5.

Figure~\ref{G150_O3_n_Ha} shows close-up views of our  
\O3 and H$\alpha$ images for the remnant's southern limb.
Like that seen in other remnants in our imaging program, we find substantial differences in the optical emission
morphology between these two emission lines. We note that
although some faint emission visible on the red DSS2 images that was interpreted by \citet{Gerbrandt2014} as being
filamentary, our H$\alpha$ images show little in the way of filamentary structure.
Instead, we mainly find short thick patches of H$\alpha$ emission with only a small handful of
short filaments around RA: 04:27, Dec: +54:32.
We also find little correspondence between the emission
features seen in \O3 and that of H$\alpha$.
However, one notable feature of the \O3 emission
here is the presence of faint \O3 emission extending farther to the south from the bright, sharp \O3 filaments.

\begin{figure}[t]
\centerline{\includegraphics[angle=0,width=8.5cm]{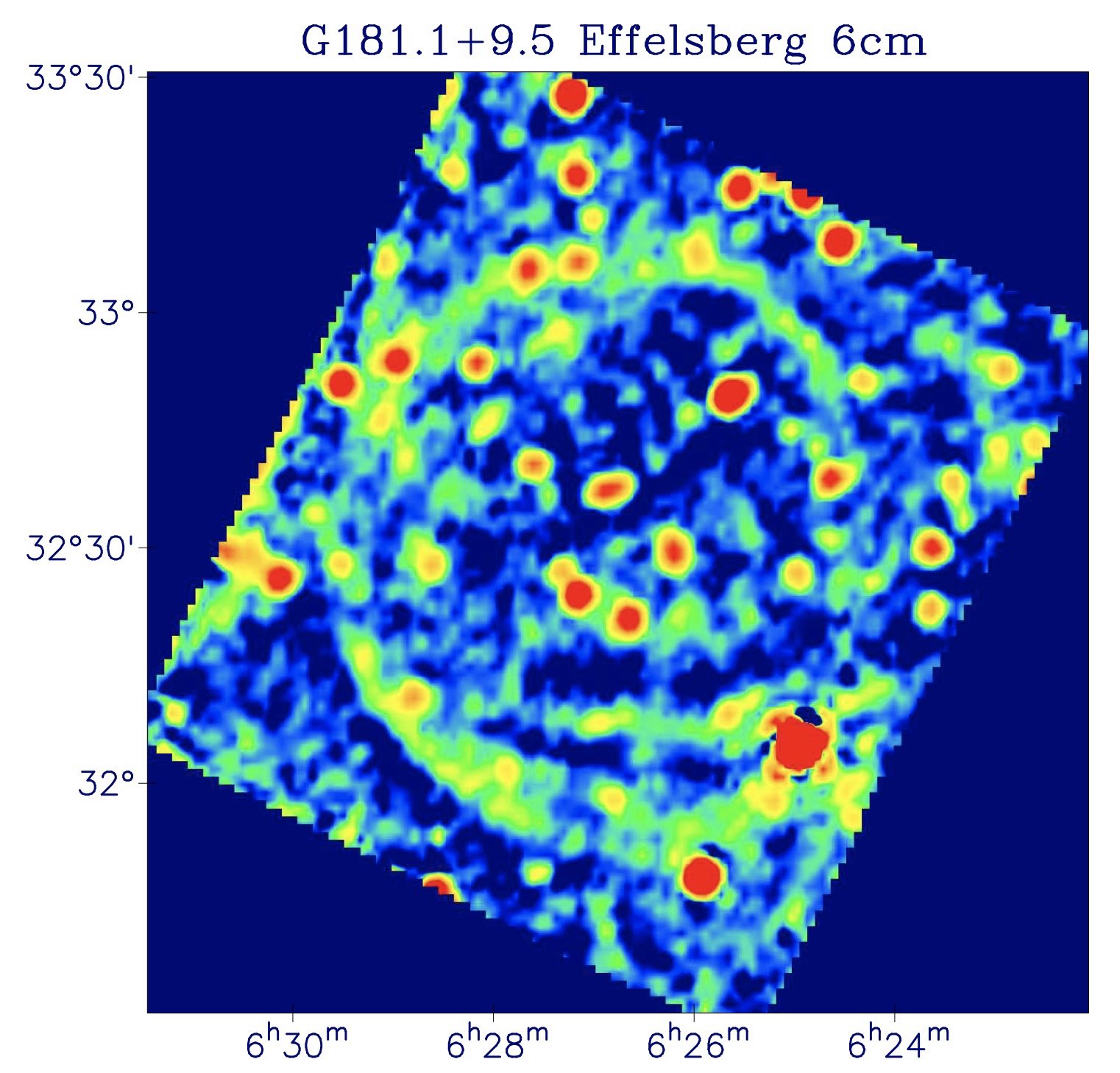}}
\caption{Effelsberg 6 cm
image of G181.1+9.5 \citep{Kothes2017}.
  \label{G181_radio}  }
\end{figure}

\begin{figure*}[ht]
\begin{center}
\includegraphics[angle=0,width=8.0cm]{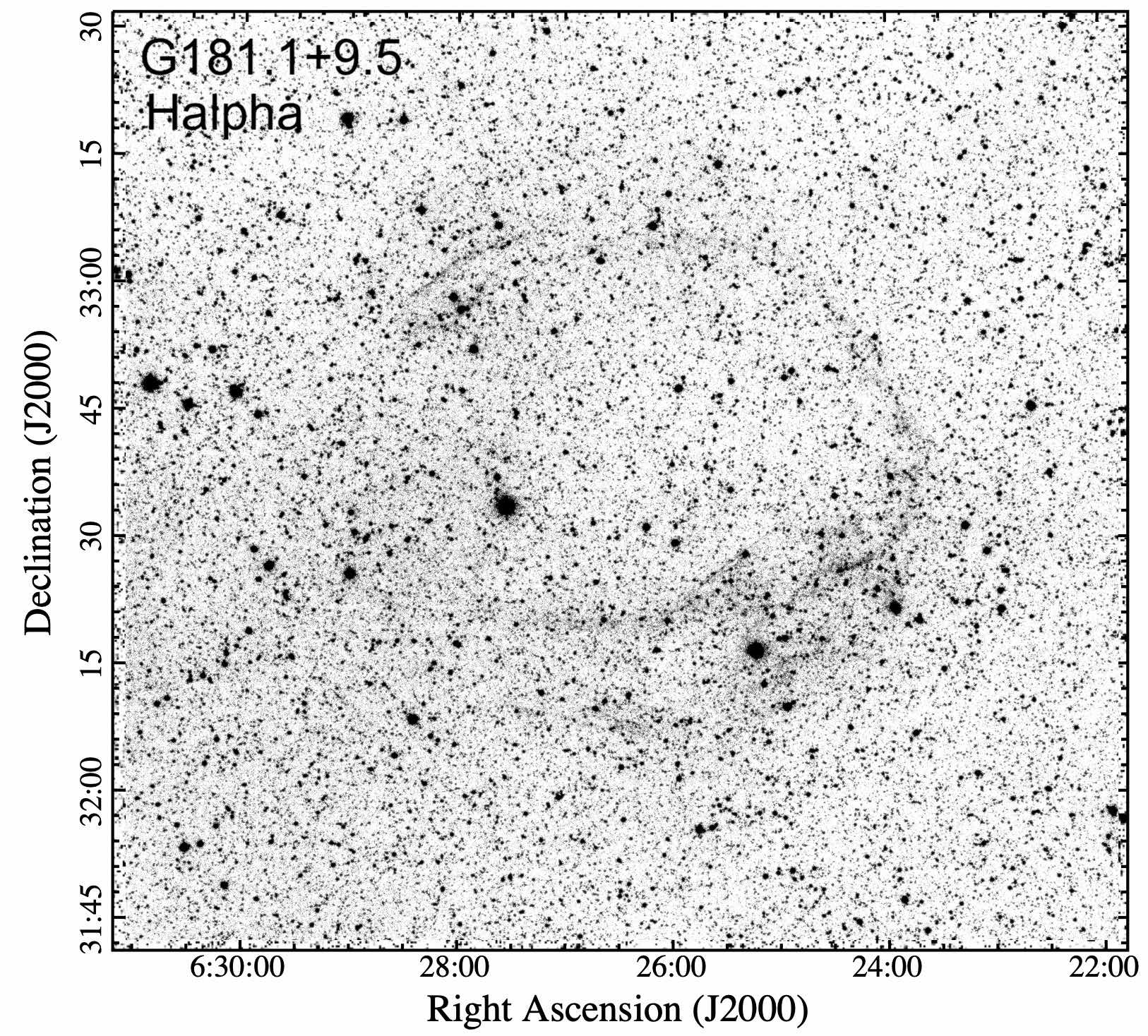} 
\includegraphics[angle=0,width=8.0cm]{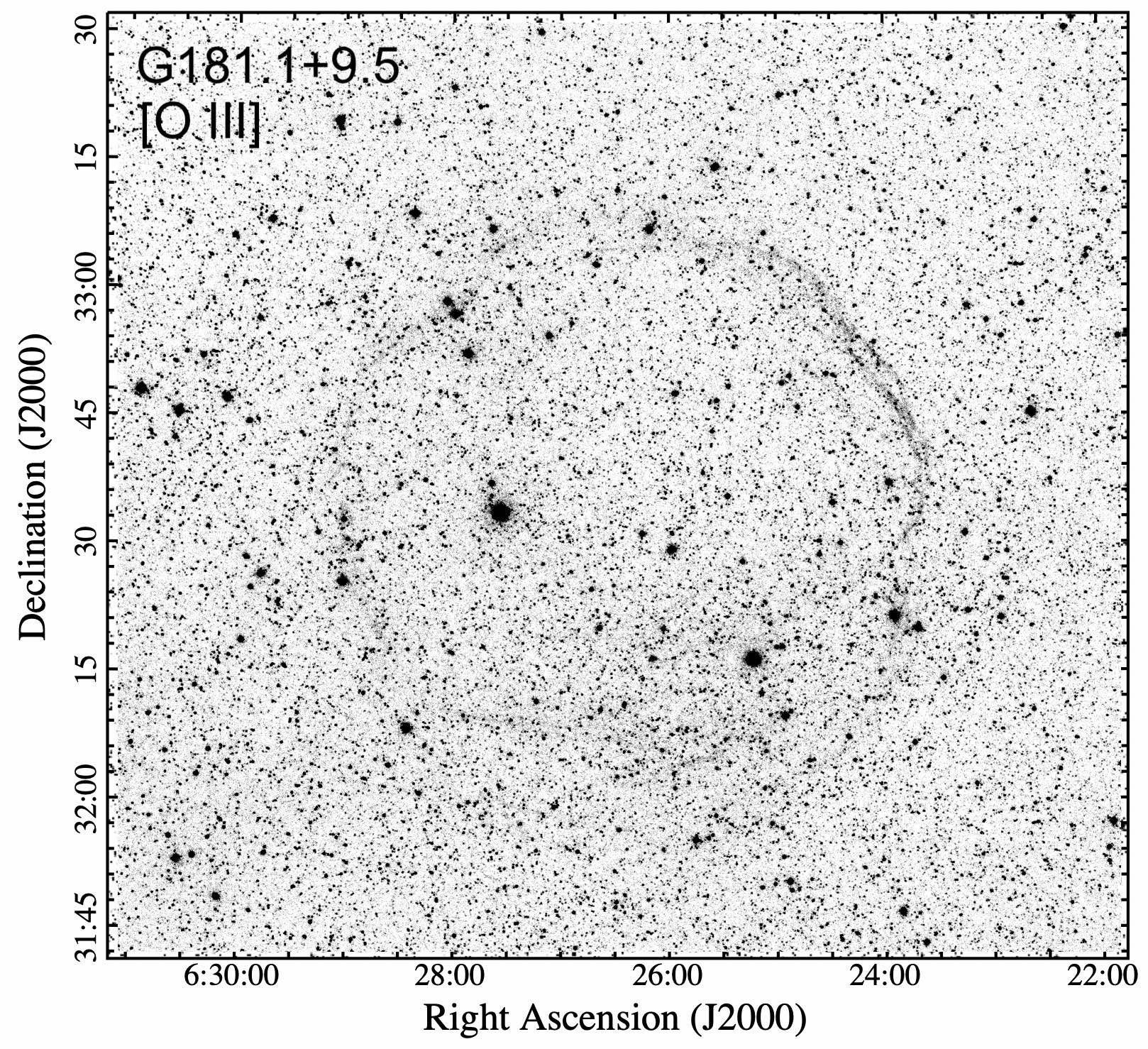} \\
\includegraphics[angle=0,width=0.5cm]{blank.jpg}
\includegraphics[angle=0,width=7.1cm]{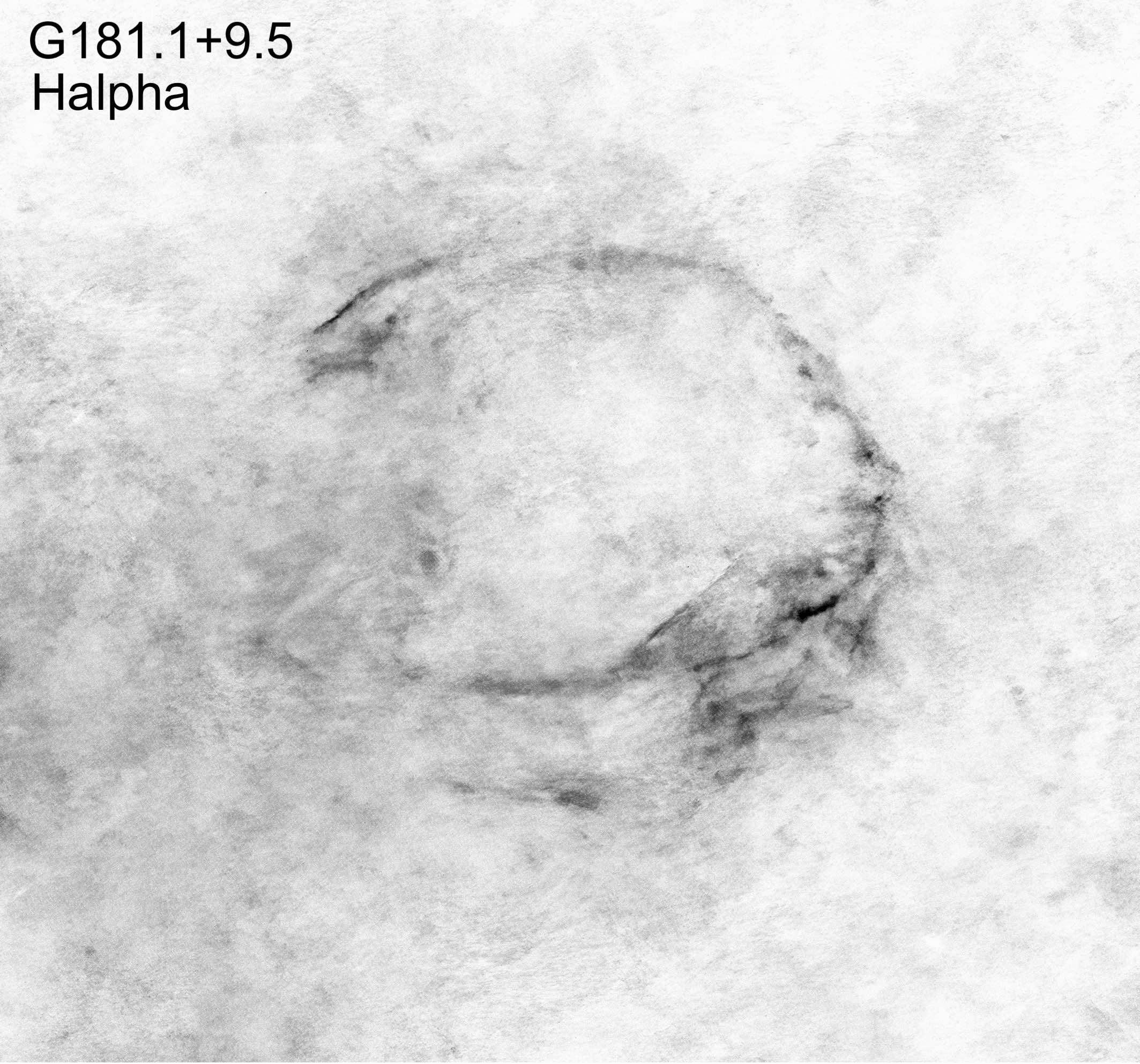} 
\includegraphics[angle=0,width=0.8cm]{blank.jpg}
\includegraphics[angle=0,width=7.1cm]{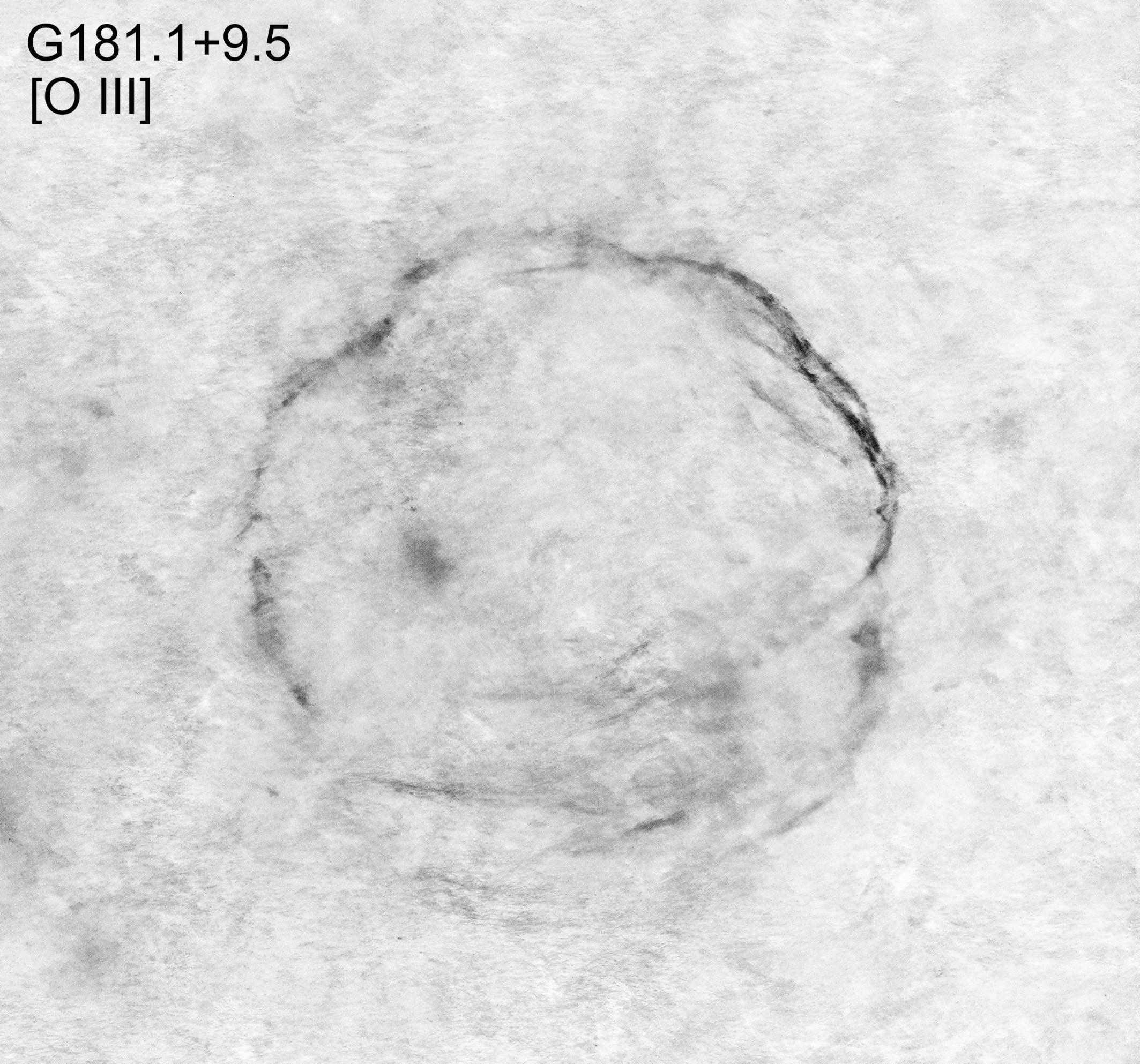}
\caption{Top Panels: Deep H$\alpha$ and \O3 images of the 
faint radio SNR G181.1+9.5. Bottom Panels: Same images but now 
with stars removed using software.
\label{G181_plots} 
} 
\end{center}
\end{figure*}

\begin{figure*}[ht]
\begin{center}
\includegraphics[angle=0,width=8.5cm]{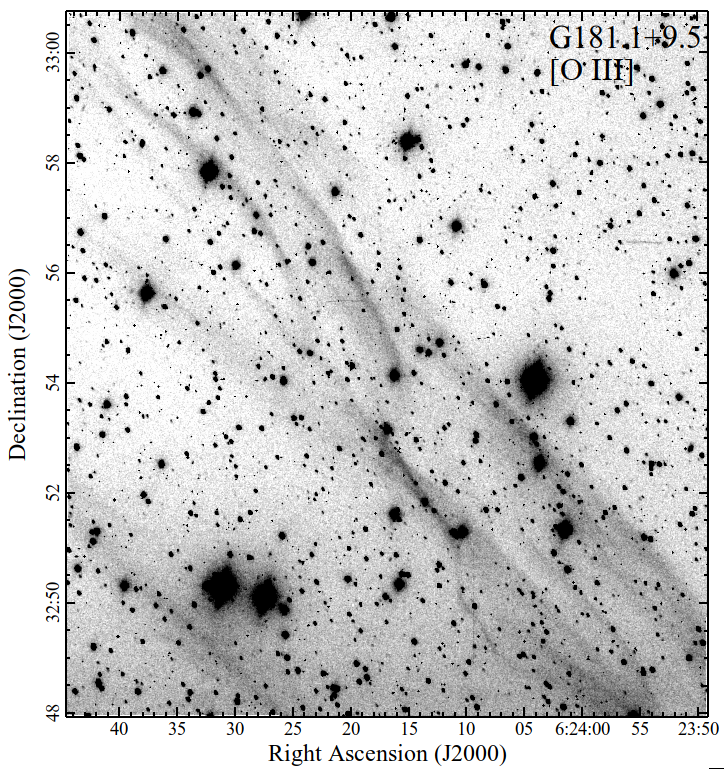} 
\includegraphics[angle=0,width=8.5cm]{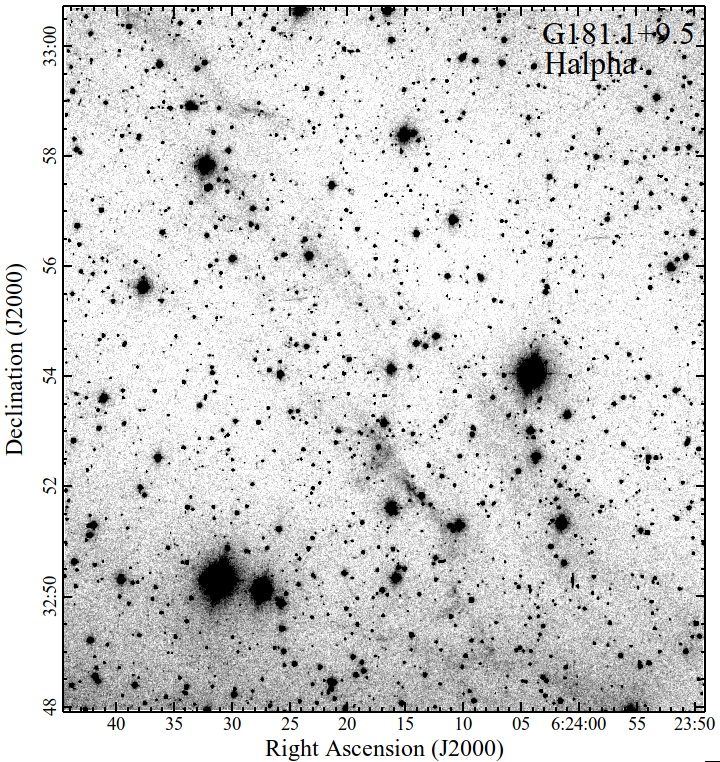} \\
\includegraphics[angle=0,width=8.5cm]{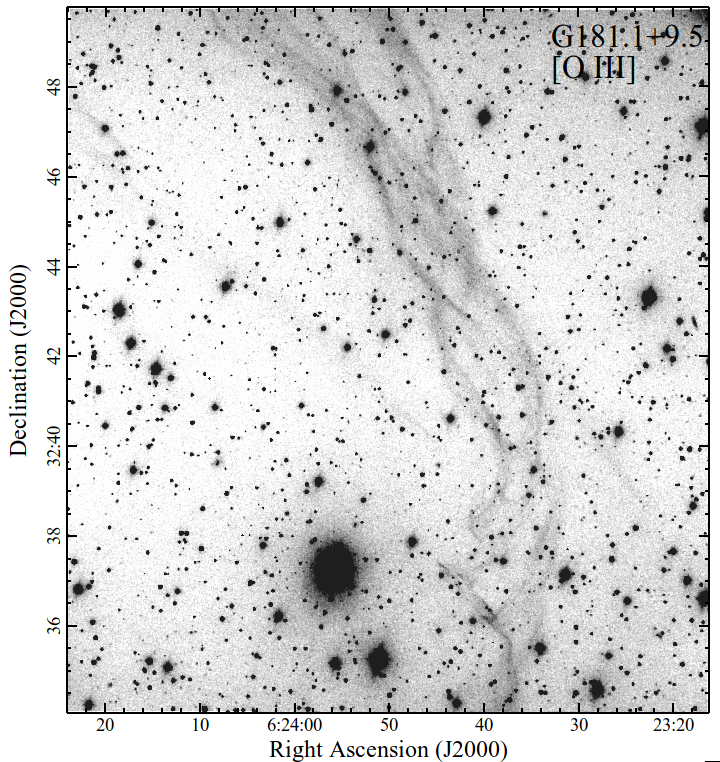} 
\includegraphics[angle=0,width=8.5cm]{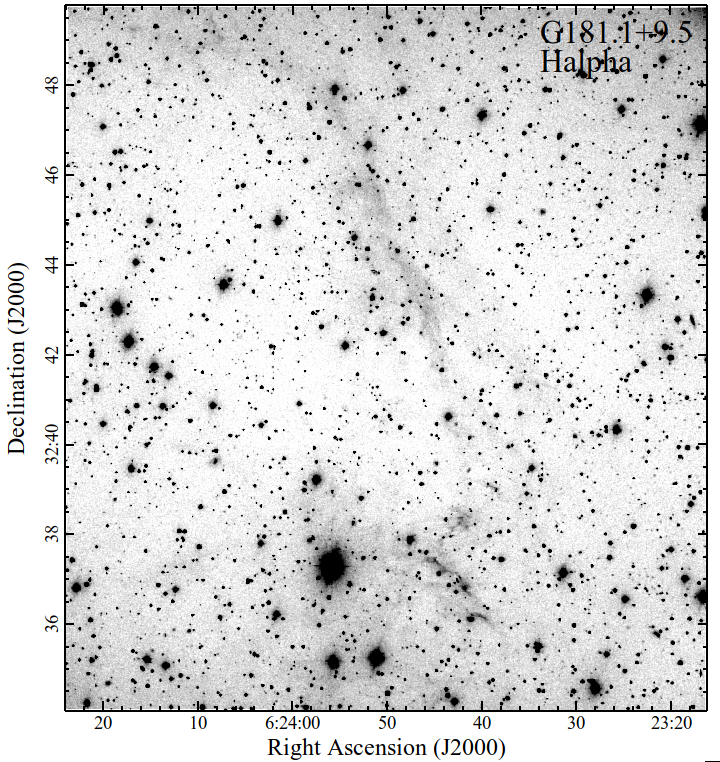}
\caption{MDM \O3 and H$\alpha$ images of G181.1+9.5's northwestern (top) 
and western (bottom) regions.
\label{G181_MDM} 
} 
\end{center}
\end{figure*}

\begin{figure*}[ht]
\begin{center}
\includegraphics[angle=0,width=17.0cm]{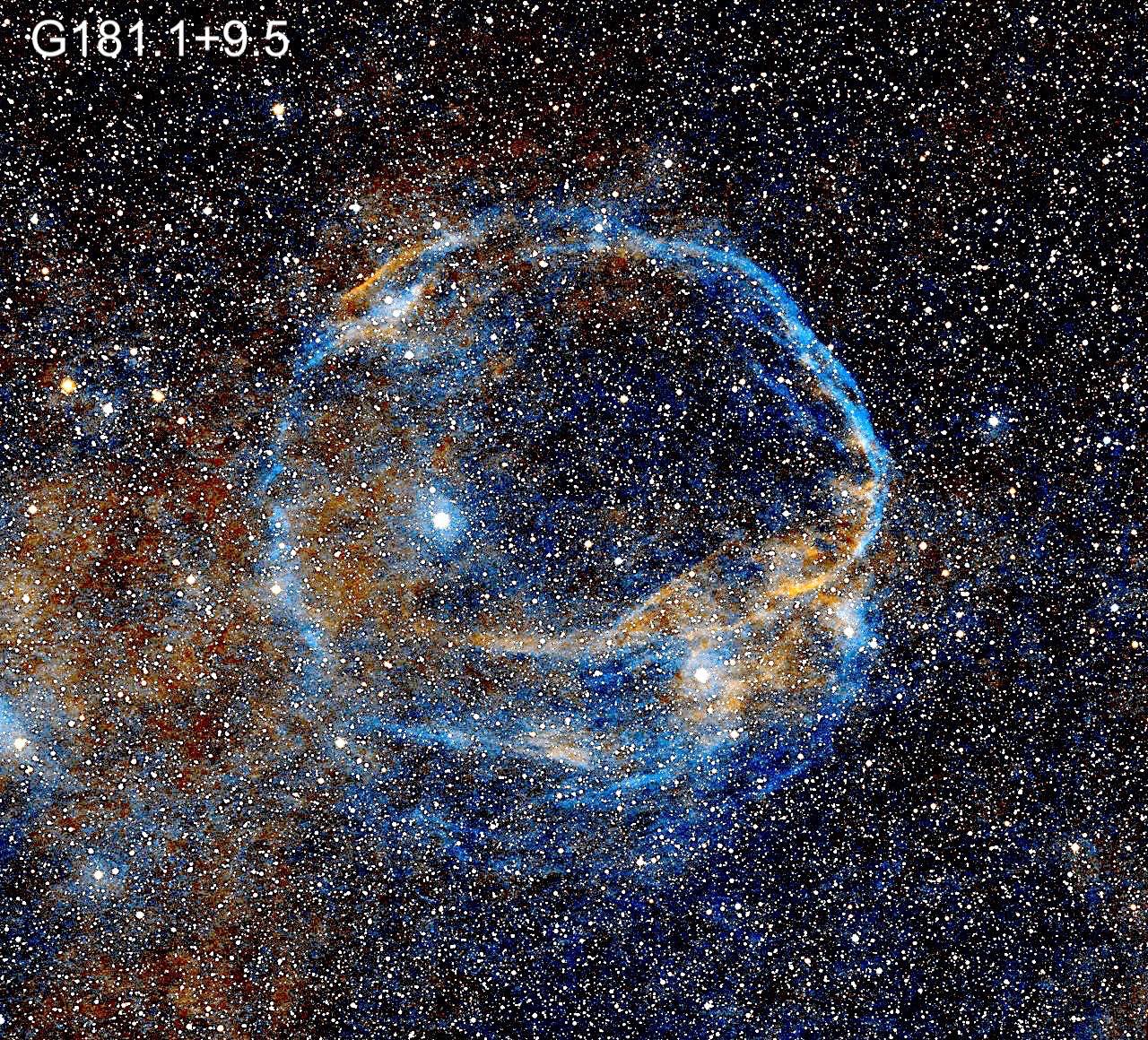}
\caption{Color H$\alpha$, \O3 and RGB composite image of SNR G181.1+9.5. 
\label{G181_color} 
} 
\end{center}
\end{figure*}

We show a color \O3 + H$\alpha$ + RGB composite image in Figure~\ref{G150color}. The upper panel of this image illustrates
the complex emission structure of this Galactic region: broad and diffuse red H$\alpha$ emission in the
southeast, a broken line of dusty ISM clouds cutting
across the remnant east to west, 
and a north-south
column of \O3  filaments marking the remnant's western edge. 

Interestingly, near the top right of Figure~\ref{G150color}  we discovered a small complete shell of \O3 emission lying
just west of the main line of remnant \O3 filaments.
The lower panel of Figure~\ref{G150color} shows this elliptical \O3 emission shell in more detail. It is centered on the variable
X-ray binary source CI Cam containing a 
B[e] star plus a compact companion 
\citep{Bartlett2019,Aret2020,Barsukova2023}. 
The shell, which has angular dimensions 
around $9' \times 11'$,
has to our knowledge not been previously reported
(see \citealt{Thureau2009, Barsukova2023}).

The fact this shell is located along the western boundary of the G150.3+4.5 remnant and exhibits similar \O3 emission line strengths, suggests the two objects might lie at
a similar distance. But this appears unlikely.
CI Cam's Gaia DR3 parallax  is 
0.210 mas indicating a statistical corrected
distance 4.10$^{+0.36}_{-0.32}$ kpc, a value
consistent with estimated based on other data
\citep{Aret2020}. However, if the remnant's distance is also  $\sim 4.1$ kpc, its angular diameter of $2.5\degr \times 3.0\degr$
would indicate unrealistic physical dimensions around $180 \times 215$ pc, thereby ranking it as the largest (and oldest) SNR known despite displaying an abundance of \O3 bright
filaments indicating a majority of its shocks have velocities well above 100 km s$^{-1}$.
Instead, we suggest CI Cam is unrelated to G150.3+4.5 with the
SNR lying at a distance much closer than that of CI Cam \citep{Devin2020, Feng2024}).
Consequently, the location of CI Cam so close to the western limb of G150.3+4.5 is likely to be a chance coincidence.

\begin{figure*}[ht]
\begin{center}
\includegraphics[angle=0,width=17.71cm]{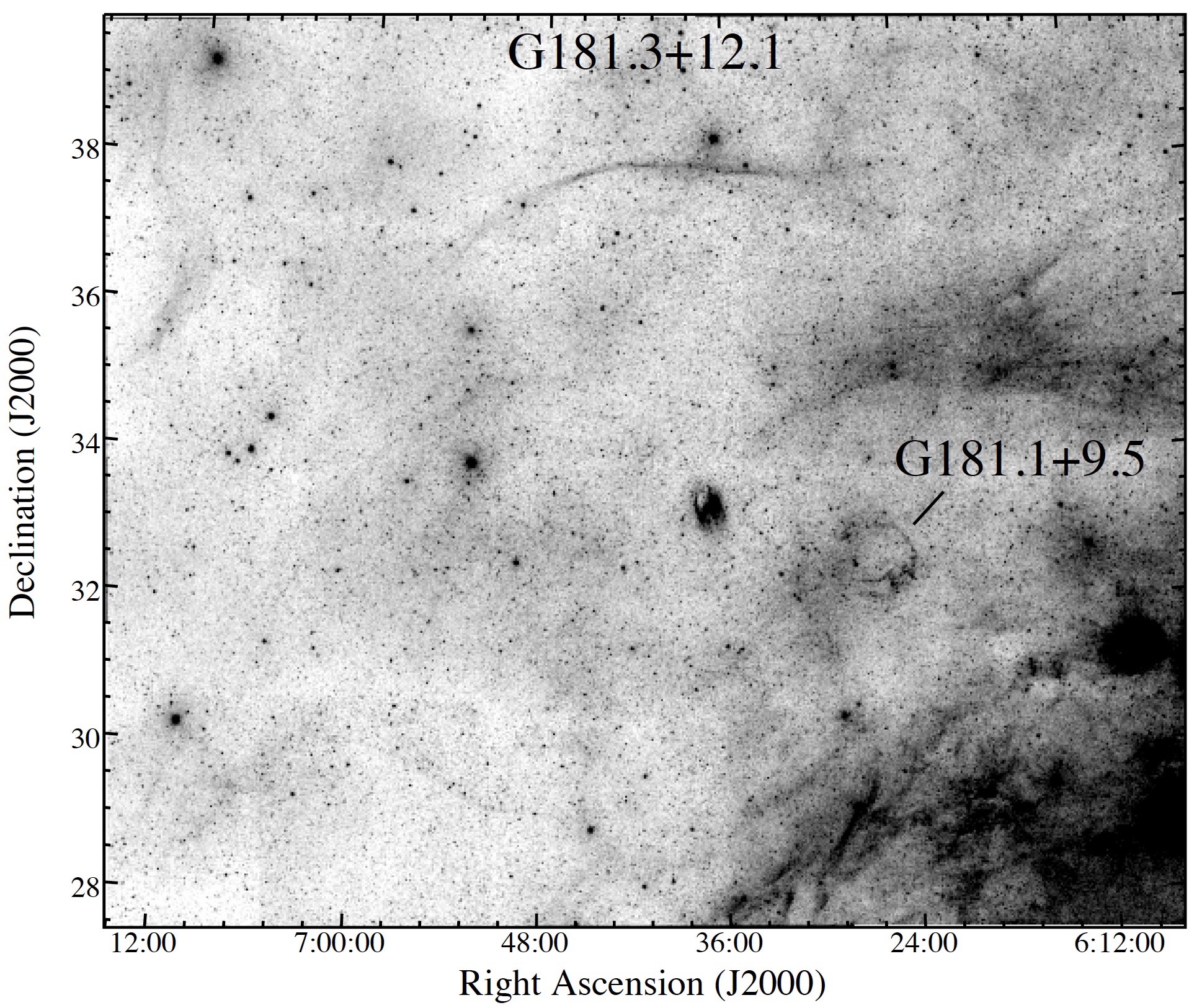} 
\caption{A $14.7\degr \times 12.4\degr$  MDW H$\alpha$ mosaic image covering the area around SNR G181.1+9.5 showing the appearance of a faint and incomplete H$\alpha$ emission shell
roughly 9.0$\degr$ in diameter. If an unrecognized SNR it would be G181.3+12.1.
\label{G181_XTRA} 
} 
\end{center}
\end{figure*}

\subsubsection{G181.1+9.5}

This very high Galactic SNR was only recently discovered in the radio by
\citet{Kothes2017} who found it at 1420 MHz to have nearly a circular,
shell-like morphology with an angular diameter of $\sim1.2\degr$ and a physical
size of about 30 pc at an estimated distance of 1.5 kpc. \citet{Kothes2017}
also found it to exhibit an unusually high degree of linearly polarized
emission with a non-thermal spectrum and noted ROSAT X-ray emission was present
from interior it. 

In Figure~\ref{G181_radio}, we show the Effelsberg  6 cm intensity radio map of
G181.1+9.5 \citep{Kothes2017}. The remnant's structure is mainly circular with a noticeable bright
extension to the south.

\citet{Kothes2017} noted that G181.1+9.5 was the faintest SNR detected up to that time in terms of its radio emission with a surface brightness of $1.1 \times 10^{-23}$ W$^{-2}$ Hz$^{-1}$ sr$^{-1}$ at 1 GHz
which put it some three times fainter than the previously faintest known SNR in 2017.

Despite its  extreme faintness in the radio, G181.1+9.5 is not really all that faint
optically. Figure~\ref{G181_plots} presents our H$\alpha$ and 
\O3 images (top panels) along with software removal of stars (lower panels). Similar to the remnant's 6 cm emission structure, we find
the remnant's H$\alpha$ emission is strongest and most extensive
in the south and southwest. In the southern region, the H$\alpha$
emission shell appears doubled, a feature that is also seen in both our \O3 and the 6 cm radio image.

The remnant's H$\alpha$ emission is weakest is along the SNR's eastern limb. The remnant also exhibits considerable \O3 emission, especially along the shell's west and northwest edge where it is quite filamentary.
This is shown in Figure~\ref{G181_MDM} where
we show MDM 2.4m images of these regions.
In both cases, the \O3 emission lies outside
(to the west) of the bulk of the H$\alpha$ emission.

Differences between the remnant's bright H$\alpha$ and \O3 emissions
can be readily seen in the color image of Fig.~\ref{G181_color}.
This remnant, like several others in our survey, displayed strong \O3 emission across most of its regions, but
especially along its southern boundary where the radio 
emission is also strong.
In addition to the remnant's emissions,
our images detected a broad region of diffuse H$\alpha$ emission
off to and partially coincident with the remnant's eastern limb.

During the process of analyzing the deep MDW H$\alpha$ images of G181.1+9.5 and its surroundings, we constructed a very wide FOV H$\alpha$ mosaic image near the SNR. This deep
H$\alpha$ image led to the discovery of what appears to be an
incomplete $\sim 9\degr$ diameter emission shell centered about a degree east of G181.1+9.5.  This possible shell, we are calling
G181.3+12.1, is shown
in Figure~\ref{G181_XTRA}. Higher resolution MDM H$\alpha$ images of the shell's bright northern edge showed
no sharp filaments, only broad, very faint diffuse emission. If this shell is an unrecognized
SNR shell, it would be among the largest known Galactic
SNRs and at $b = 12.1\degr$, would rank as one of the farthest off the Galactic plane.

\begin{figure*}[ht]
\begin{center}
\includegraphics[angle=0,width=8.5cm]{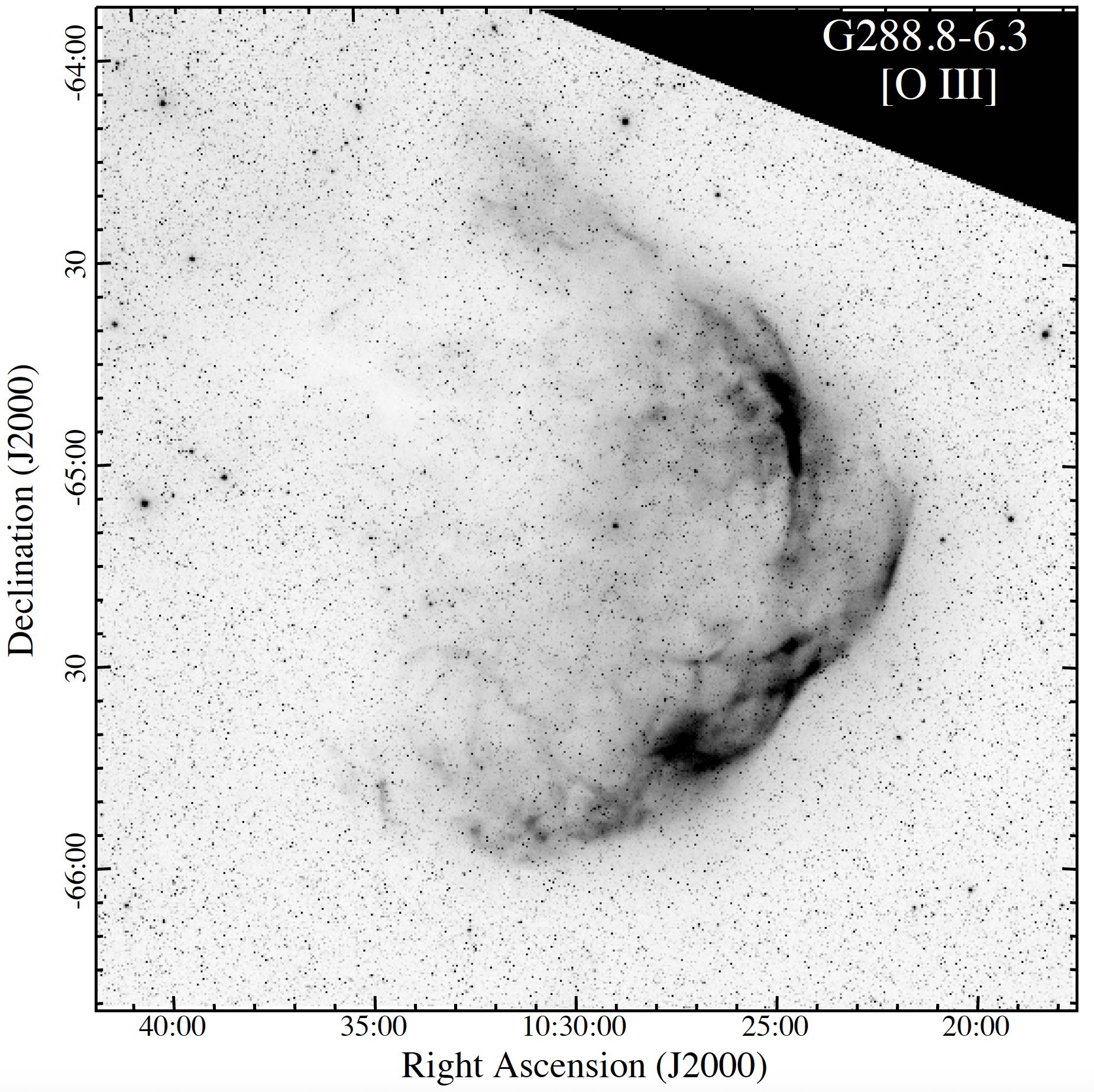}
\includegraphics[angle=0,width=8.5cm]{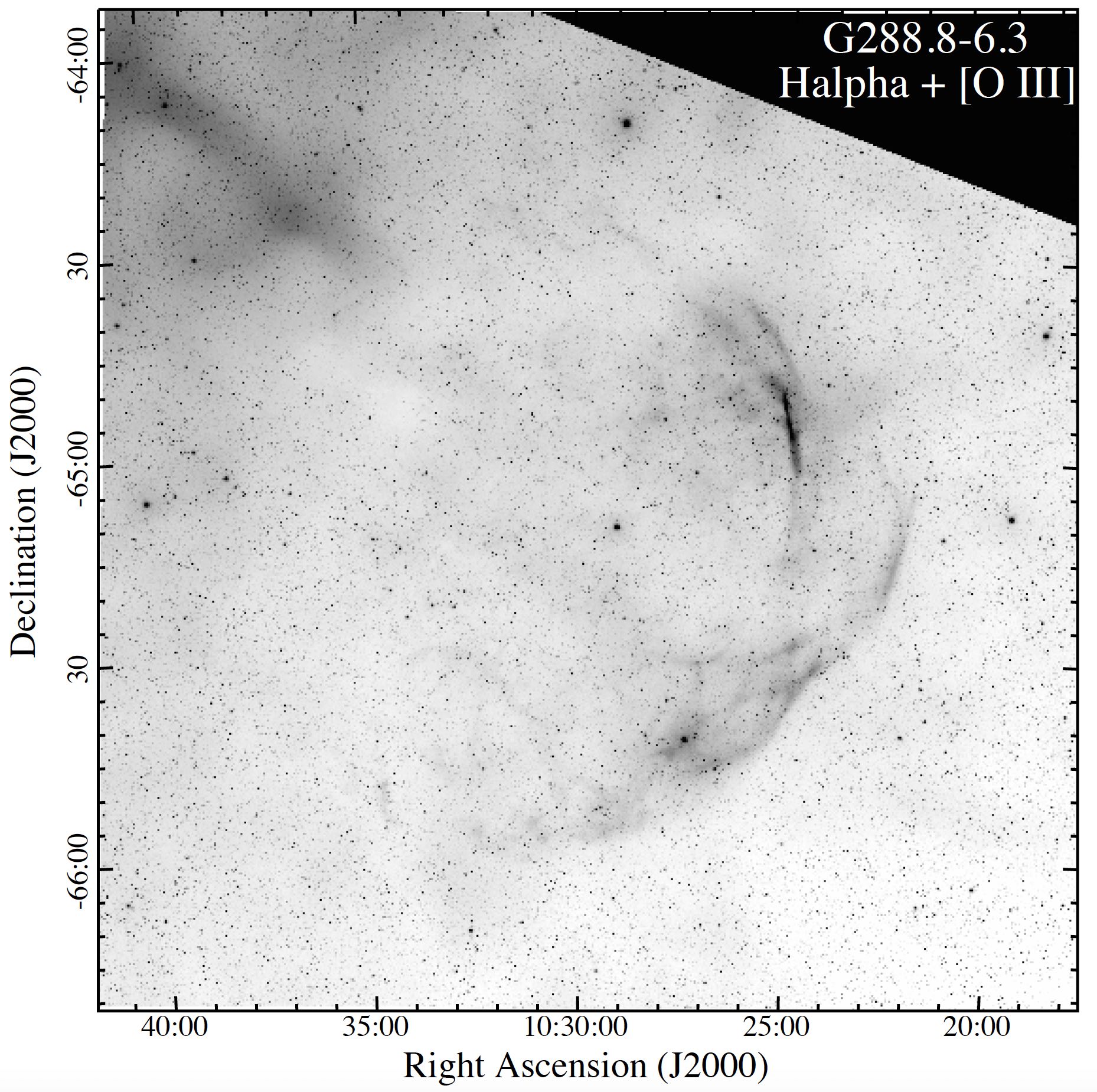} \\
\includegraphics[angle=0,width=14.0cm]{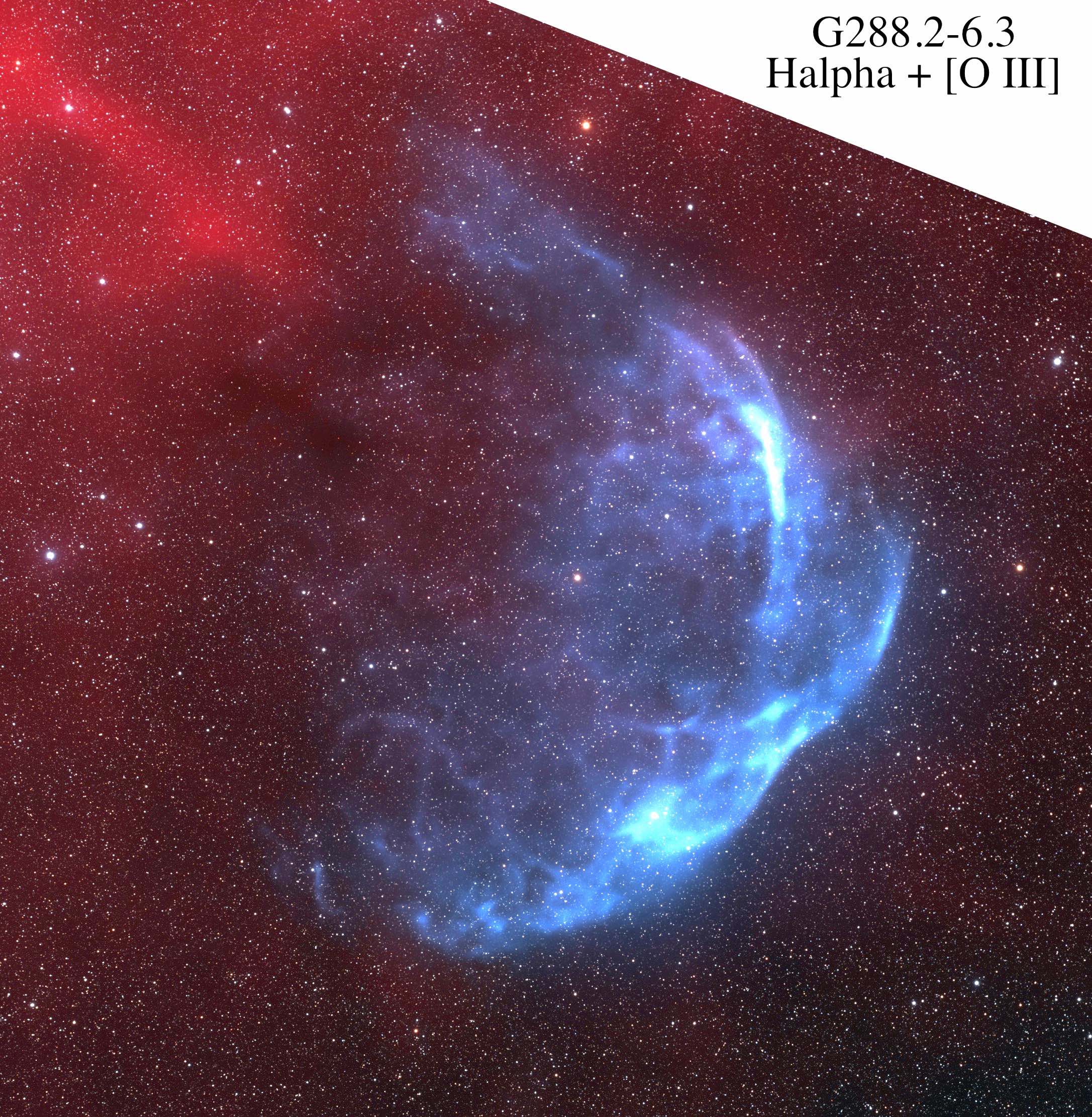}
\caption{Top: Deep \O3 image (left) and composite \O3 + H$\alpha$ (right) images of the SNR G288.8-6.3. Bottom: Color composite of \O3, H$\alpha$ + RGB images of the G288.8-6.3 SNR.
\label{G288_plots_n_color} 
} 
\end{center}
\end{figure*}

\begin{figure}
\begin{center}
\includegraphics[angle=0,width=8.2cm]{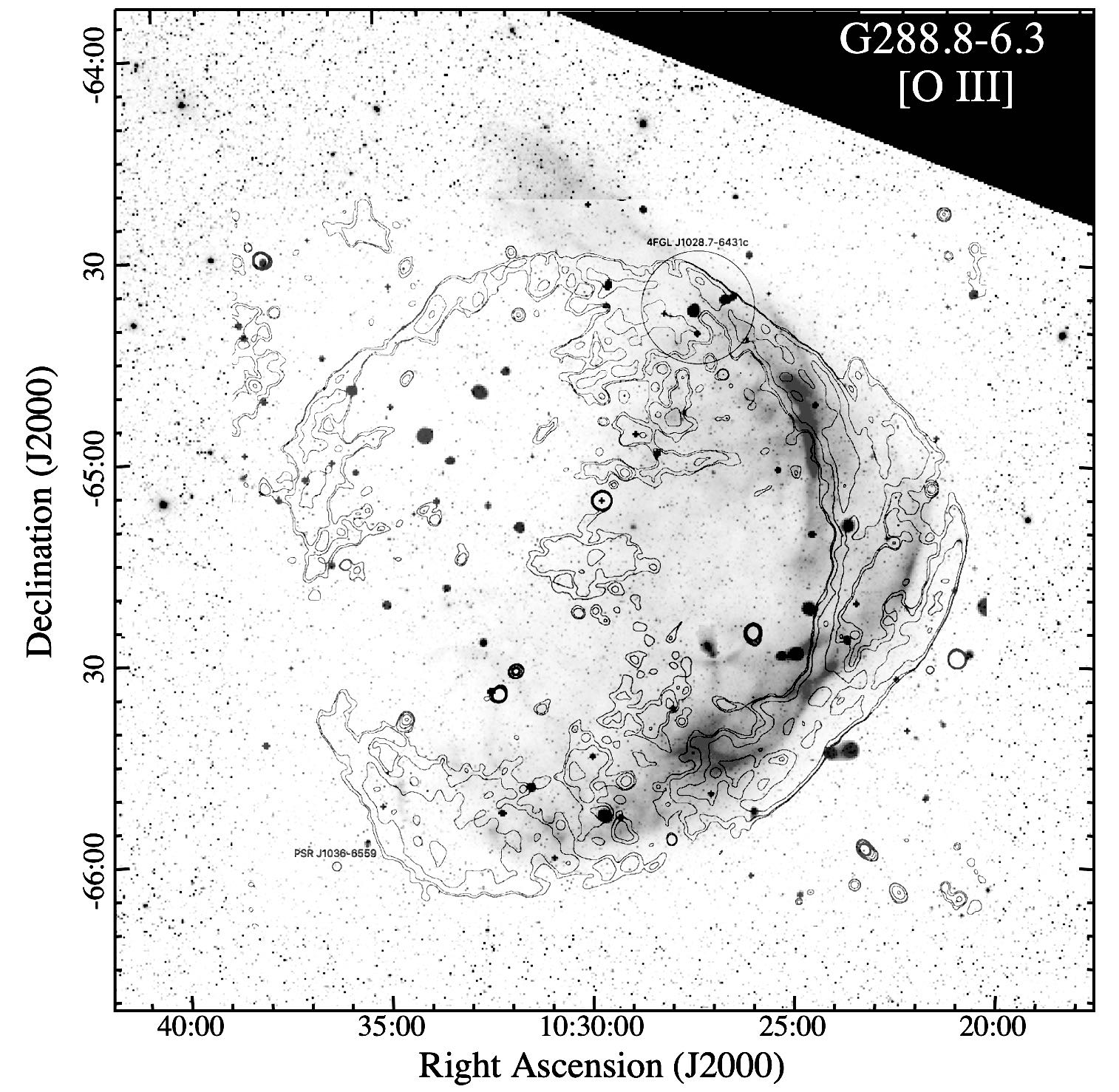} 
\caption{\O3 image of G288.8-6.3 with 943 MHz radio emission contours 
adapted from \citet{Filipovic2023}.
\label{G288_radio}
} 
\end{center}
\end{figure}

\subsection{G288.8-6.3} 

This large remnant ($1.8\degr \times 1.6\degr$) was first identified
as a SNR candidate by \citet{Duncan1997} from 
4.85 GHz Parkes-MIT-NRAO (PMN) survey data.
\citet{Stupar2008} examining H$\alpha$ images taken as part of the
Anglo-Australian
Observatory/United Kingdom Schmidt Telescope (AAO/UKST) survey of the southern Galactic plane, found a small 2.5 arcmin patch of H$\alpha$ nebulosity coincident with the northwest part of the
\citet{Duncan1997} SNR candidate. Follow-up optical spectra
showed evidence of shocks, namely strong  [\ion{O}{2}] $\lambda$3727, [\ion{O}{3}] 
$\lambda$5007, and [\ion{S}{2}] $\lambda\lambda$6716,6731 emission lines with a [\ion{S}{2}]/H$\alpha$ ratio of 0.54.

This remnant was subsequently more thoroughly studied through  data taken as
part of Australian Square Kilometre Array Pathfinder (ASKAP) Evolutionary Map
of the Universe (EMU) survey \citep{Filipovic2023}.  A multifrequency analysis
radio data of the remnant showed it to have a thin shell morphology with a
typical SNR spectral index of $\alpha$ = $0.41 \pm 0.12$ and estimated its
distance to be $\sim$1.3 kpc which would imply a physical diameter of around 40
pc. The remnant has also been detected in $\gamma$-rays with the $Fermi$-Large Area Telescope \citep{BS2023}.

\citet{Filipovic2023} noted the presence of a faint H$\alpha$ filament detected in
the AAO/UKST H$\alpha$ SuperCOSMOS survey \citep{Parker2005} which overlapped with radio contours along the remnant's
south-western limb. However, it
appears the remnant's H$\alpha$ emission is rather meager compared to its
\O3 emission. 

Instead of a small 2.5 arc minute isolated
H$\alpha$ emission cloud in the north and a tiny streak
of H$\alpha$ along the remnant's southern boundary,
G288.8-6.3's optical is both relatively bright and quite extensive when imaged in \O3. We reach this
conclusion even though we did 
not obtain a pure
H$\alpha$ image. Instead, we employed a dual bandpass filter that is sensitive to both H$\alpha$ and \O3 line emissions.

The right hand panel shown in Figure~\ref{G288_plots_n_color} is the result
of our dual bandpass H$\alpha$ + \O3 filter image.
This should be compared with the pure
\O3 image shown in the left hand panel of the same figure. The main difference in the two
images is the extended diffuse emission cloud off to the northeast in the H$\alpha$ + \O3 image which is not present in the pure \O3 image. There is also
some faint diffuse H$\alpha$ emission along the northwest limb
which extends a bit farther west than seen in \O3. 
This comparison makes it clear that the remnant's optical emission is dominated by its \O3 line emission.

We show a color composite of our images in Figure~\ref{G288_plots_n_color}
which emphasizes the remnant's dominated \O3 emission shown here as blue emission.
We note that we obtained the \O3 image first and then an exploratory dual emission-line filter which suggested that a pure H$\alpha$ image would not show many additional features.
We therefore did not pursue obtaining a separate set of deep H$\alpha$ images. As seen in other SNRs in our survey, the
lack of  \O3 images of this remnant prevented previous researchers from
realizing the remnant's fuller optical emission structure.

Lastly, Figure~\ref{G288_radio} shows 
our optical \O3 image 
overlayed with 943 MHz radio emission contours
adapted from \citet{Filipovic2023} (their Fig.\ 7).
While the remnant's \O3 optical and its 933 MHz radio emissions are fairly well correlated along the remnant's western limb, optical \O3 emission
is virtually absent along the northeast, east, and southeast limbs where substantial radio emission is detected. Other differences include
along G288.8-6.3's western limb where radio emission
appears out ahead (farther west) of the \O3 emission by several arcminutes. In contrast, the radio map shows no correspondence to the remnant's northernmost
\O3 emission which extends well past the radio emission boundary. Overall, however, we find enough agreement to confirm the \O3 emission seen is from the G288.8-6.3 SNR.

\begin{figure*}[ht]
\begin{center}
\includegraphics[angle=0,width=11.5cm]{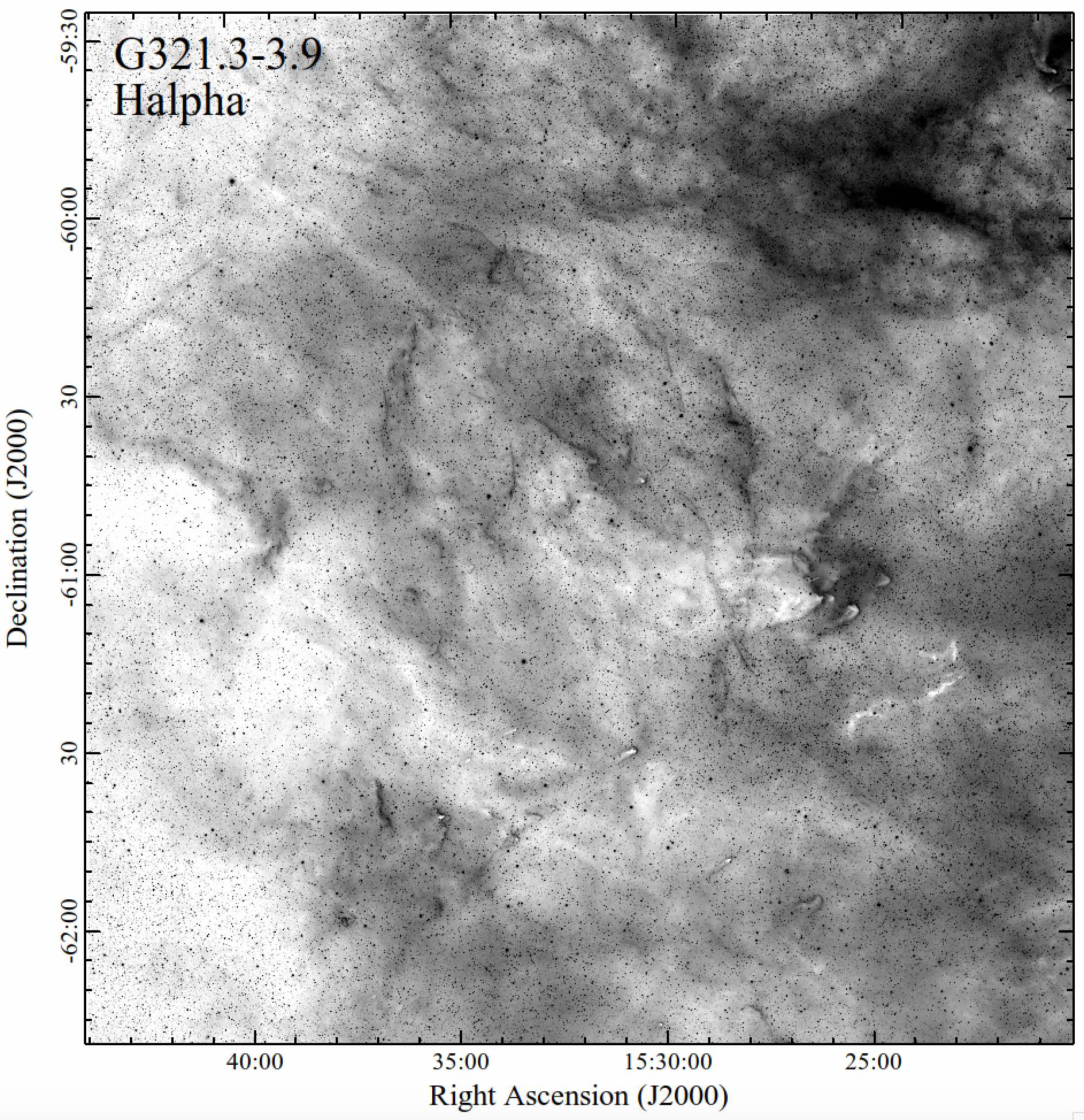} \\
\vspace{0.15cm}
\includegraphics[angle=0,width=11.5cm]{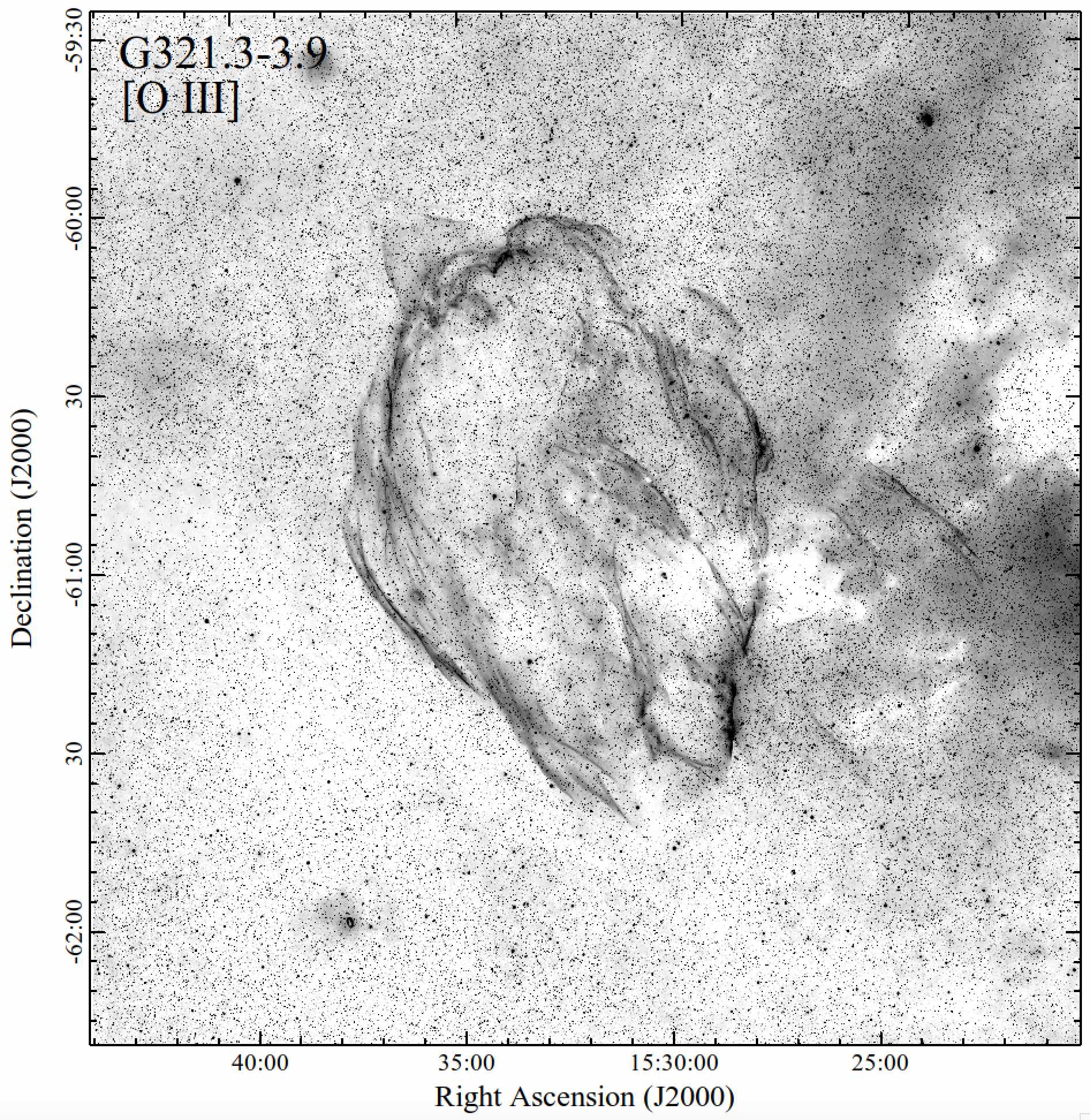}
\caption{Matching H$\alpha$ and \O3 images of the SNR G321.3-3.9.
\label{G321_plots} 
} 
\end{center}
\end{figure*}

\begin{figure*}
\begin{center}
\includegraphics[angle=0,width=18.5cm]{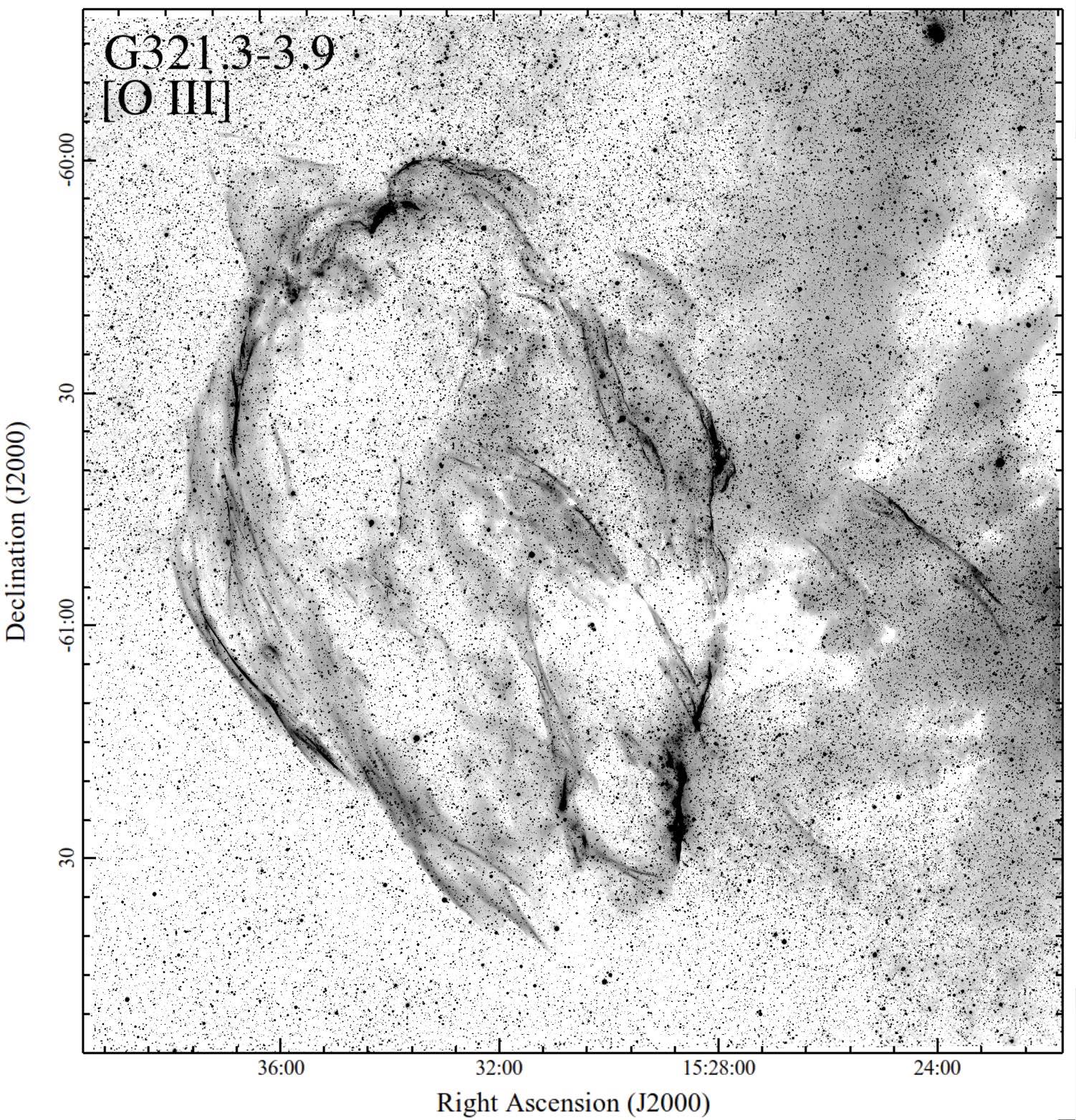}
\caption{Enlargement of \O3 image of G321.3-3.9 highlighting a 
northern and a large western shock `breakout'.
\label{G321_blowups} 
} 
\end{center}
\end{figure*}

\begin{figure*}
\begin{center}
\includegraphics[angle=0,width=16.0cm]{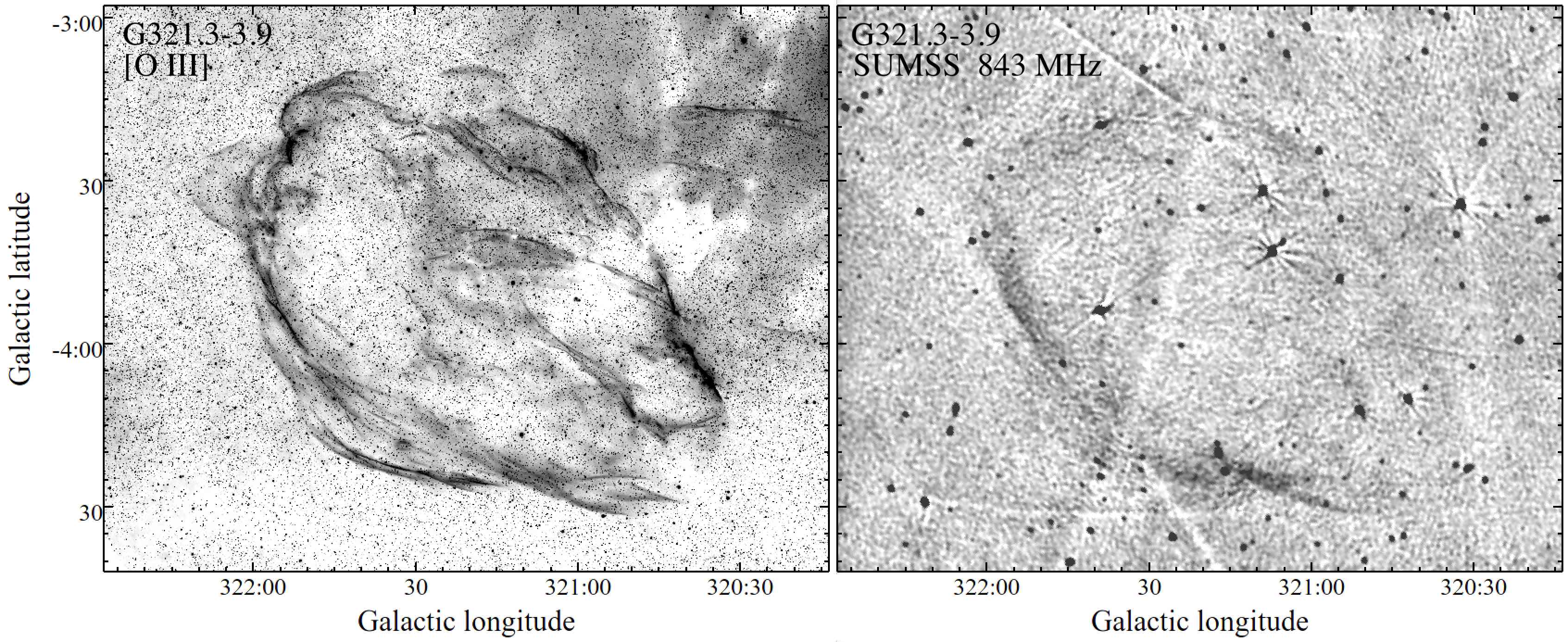}
\includegraphics[angle=0,width=16.0cm]{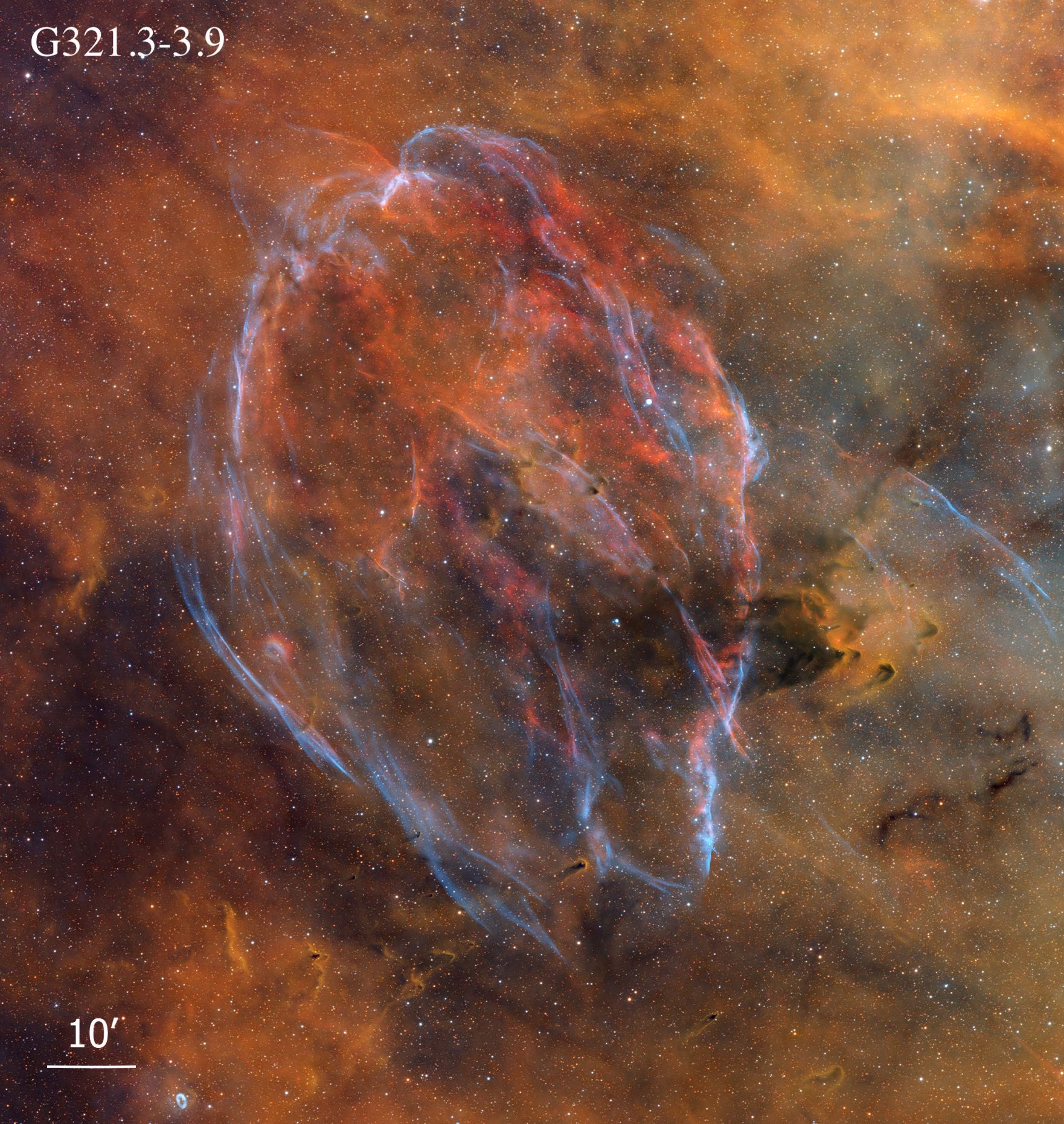} 
\caption{Top: Comparison of optical \O3 emission vs.\ radio SUMSS 843 MHz emission in Galactic coordinates. \\
Bottom: Color composite of the SNR G321.3-3.9.
Blue: \O3, red: [\ion{S}{2}], yellow: H$\alpha$. North is up, east to the left.
\label{G321_color} 
} 
\end{center}
\end{figure*}

\subsubsection{G321.3-3.9}

This  elliptical shaped $1.3\degr \times 1.8\degr$ remnant was discovered by \citet{Duncan1997} at 2.4 GHz who called it G321.3-3.8. The remnant was later better mapped 
in the 2nd epoch Molongo Galactic Plan Survey
by \citet{Green2014} who renamed it G321.3-3.9.
Recently, \citet{Mant2024} presented additional radio data
along with eROSITA X-ray data confirming its SNR identification.

No optical emission has been reported for this
SNR by \citet{Duncan1997} or
\citet{Green2014}  and no mention of it in the southern hemisphere AAO/UKST H$\alpha$ survey for SNRs by \citet{Stupar2008}. Nonetheless, our imaging revealed the remnant to exhibit significant optical line emissions.

\begin{figure*}[ht]
\begin{center}
\includegraphics[angle=0,width=12.5cm]{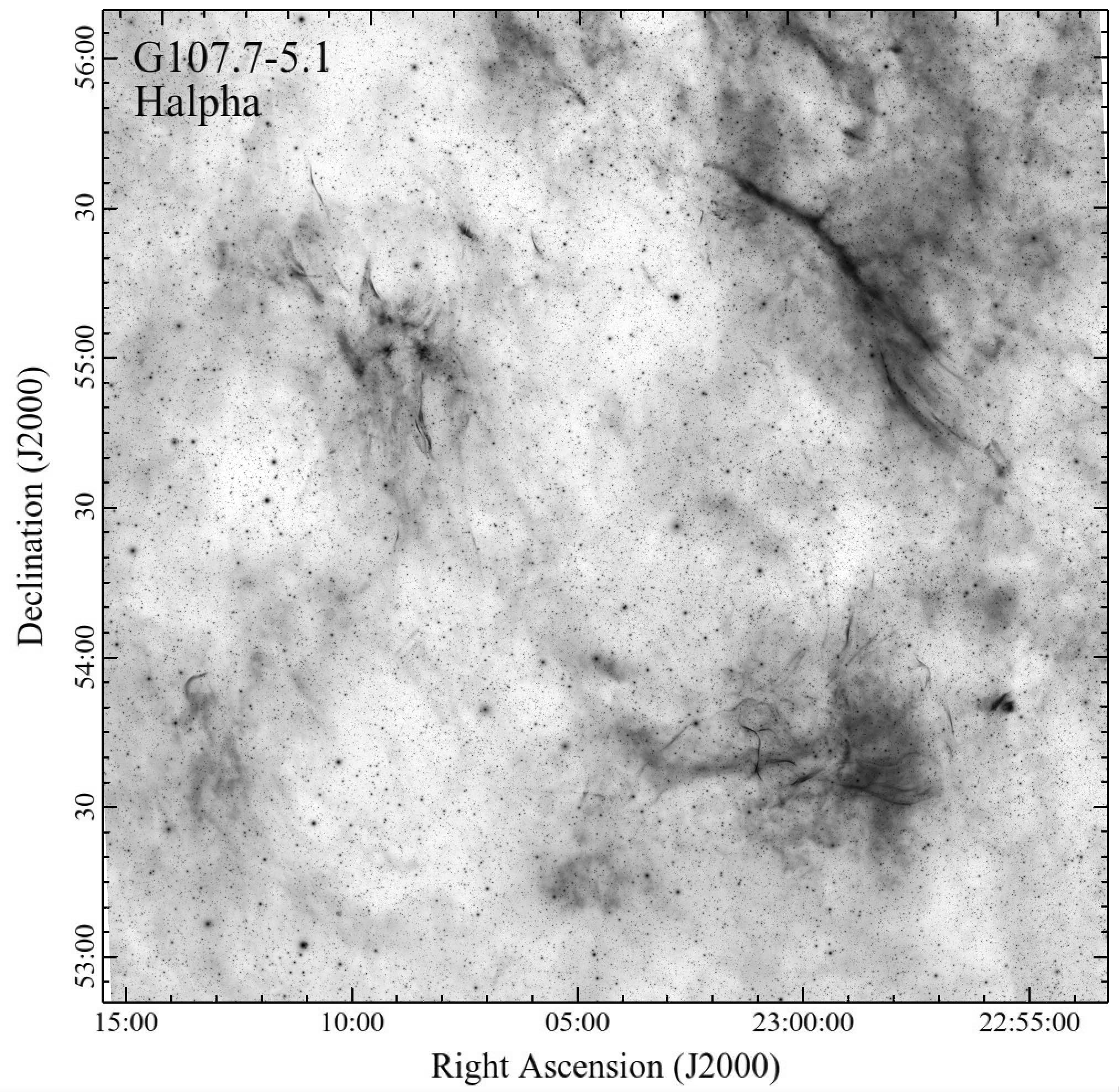}  \\
\includegraphics[angle=0,width=12.5cm]{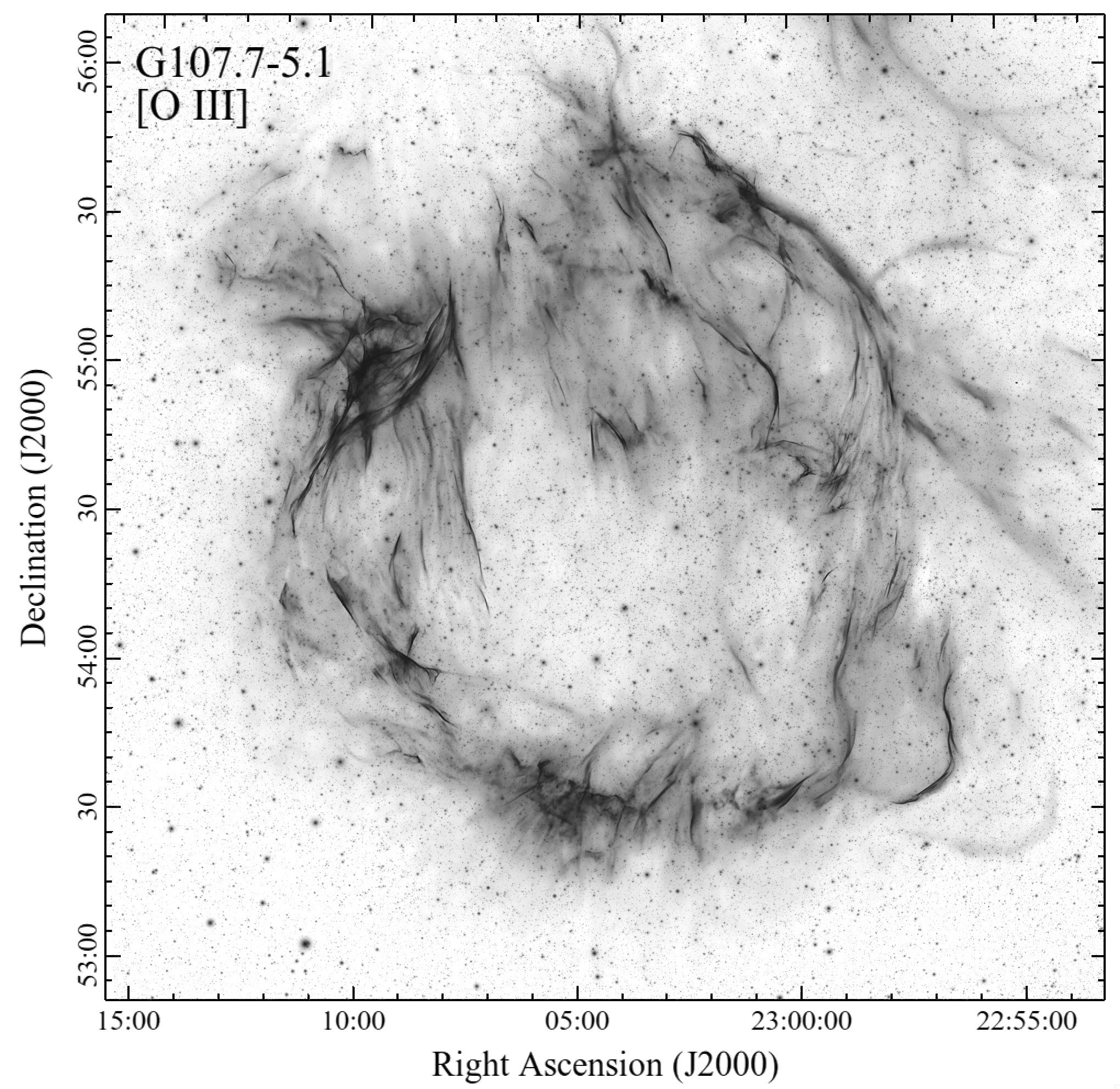}
\caption{Matching $3.2\degr \times 3.2\degr$ H$\alpha$ and \O3 images 
of the large Nereides SNR G107.7-5.1.
\label{G107_plots} 
} 
\end{center}
\end{figure*}

\begin{figure*}
\begin{center}
\includegraphics[angle=0,width=18.0cm]{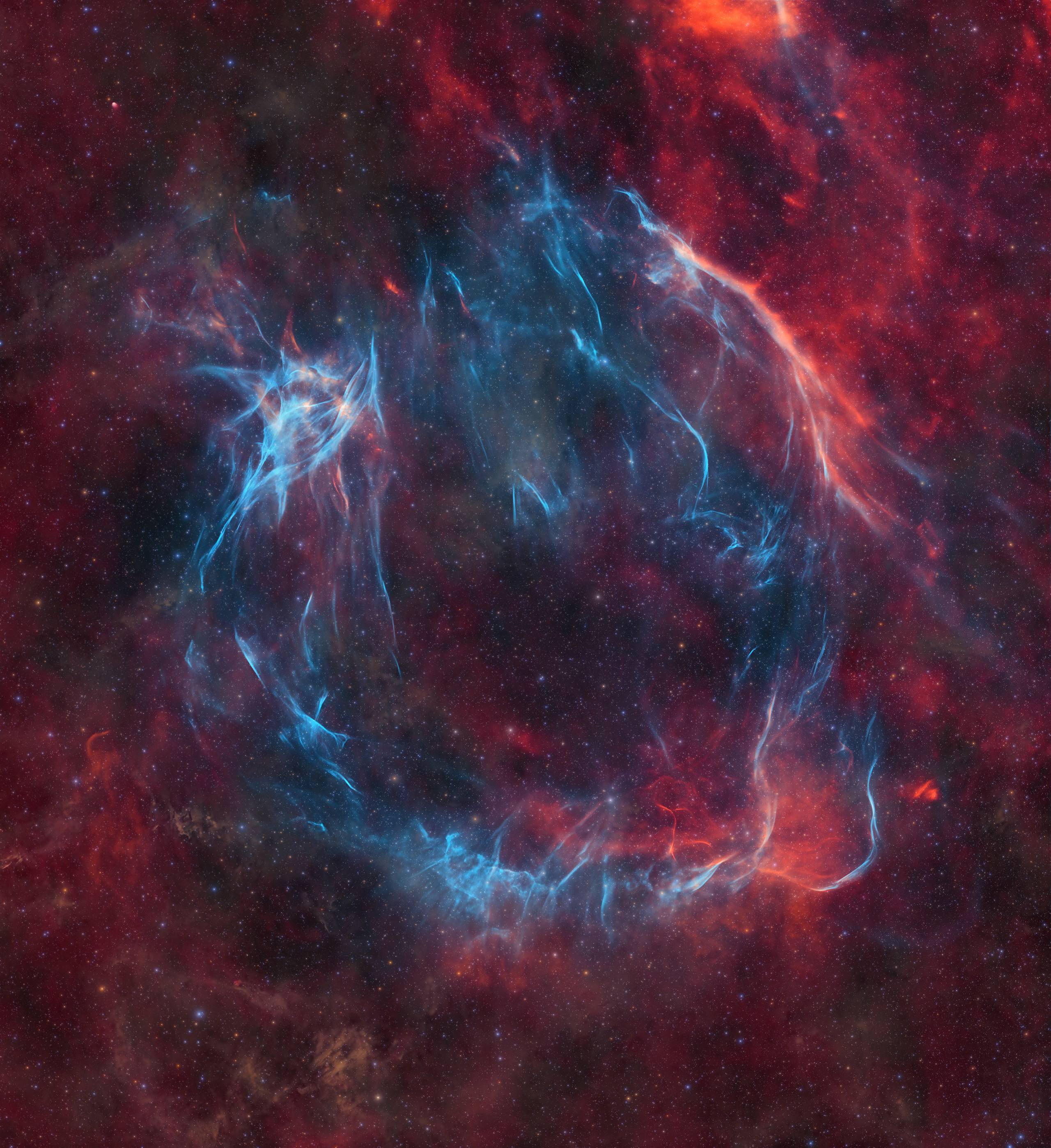} \\
\caption{Color composite of H$\alpha$, \O3, and RGB images of the Nereides SNR G107.7-5.1. 
\label{G107_color} 
} 
\end{center}
\end{figure*}

\begin{figure*}
\begin{center}
\includegraphics[angle=0,width=18.0cm]{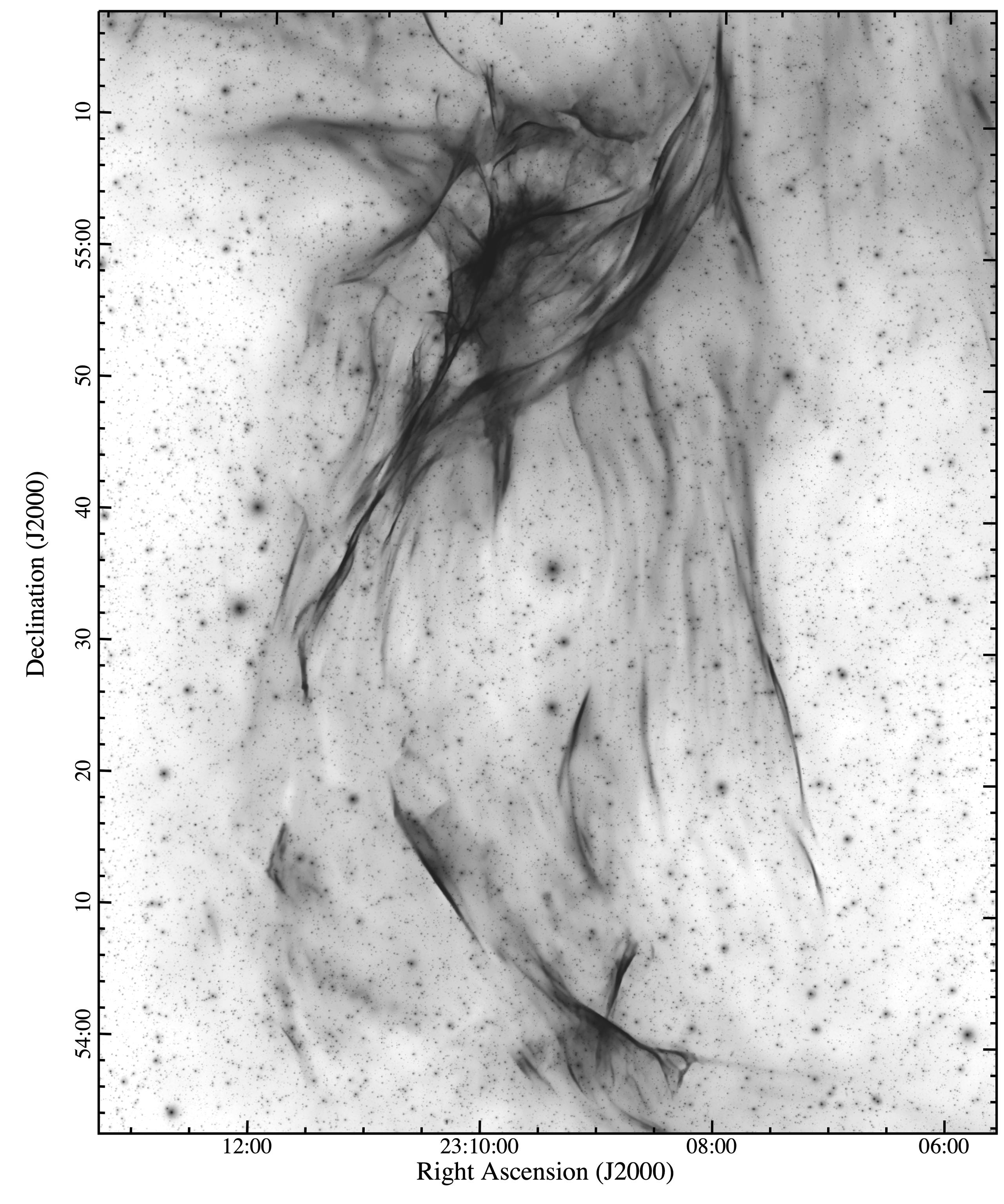} \\
\caption{Enlargement of our \O3 image for the eastern region of G107.7-5.1. 
\label{G107_east} 
} 
\end{center}
\end{figure*}

\begin{figure*}
\begin{center}
\includegraphics[angle=0,width=18.0cm]{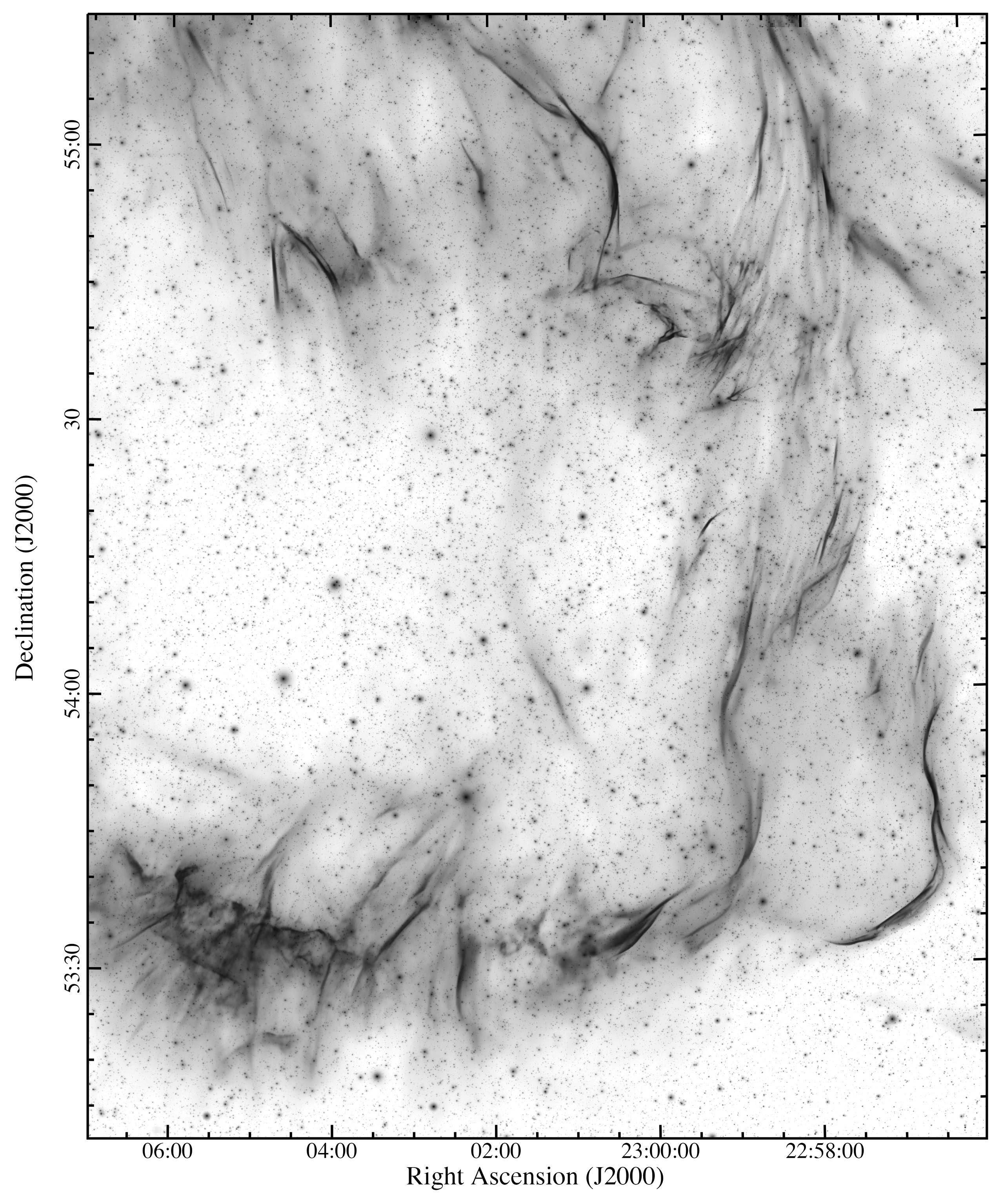} \\
\caption{Enlargement of our \O3 image for the western region of G107.7-5.1. 
\label{G107_west} 
} 
\end{center}
\end{figure*}

Figure~\ref{G321_plots} presents our 
H$\alpha$ and \O3 images. As can be seen in the H$\alpha$ image, the Galactic region
toward G321.3-3.9 is quite complex making it
hard to easy discern the remnant's emission features. This led us to initially refer to
this remnant as the ``hidden SNR''. However,
when imaged in \O3, the remnant is far from
hidden, with extensive sharp \O3 filaments
across much of the remnant's structure.

Interesting, the \O3 image shows two `breakout'
features, a small one along the remnant's
northeastern tip, and a much larger but 
less well defined along its western and southwestern limbs.
These breakout features are better seen
in the enlargement presented
in Figure~\ref{G321_blowups}.
We note that neither breakout features are seen or hinted at in the radio data.
While both breakouts are seen to some extent in both H$\alpha$ and \O3, they are best
seen in \O3 emission.

The remnant's optical emission, especially its \O3 emission, lines up well with
its radio emission. This is shown in the top panels of Figure~\ref{G321_color}
where we show its \O3 structure side by side with its 843 MHz radio emission as
seen in the Sydney University Molonglo Sky Survey (SUMSS) and the 2nd epoch
Molonglo Galactic Plane Survey (MGPS-2) \citep{Green2014}. We find good
optical-radio correlations along the remnant's southern limb but poor along the
northern regions.

Figure~\ref{G321_color} also shows a color composite image made from our \O3, H$\alpha$,
[\ion{S}{2}], and RGB images (bottom panel). For a SNR with no known optical emission, G321.3-3.9 exhibits an impressive amount
of optical filaments and emission. Much of the remnant's H$\alpha$ bright filaments are seen in its northern half, with the southern regions dominated by strong \O3 emissions. Although almost half the angular dimensions of the Cygnus Loop,  this remnant has nearly as much optical emission filaments
arranged in a similar elliptical fashion, 
with a southern emission `gap' similar to that seen in the Cygnus Loop. Overall, we find G321.3-3.9 to be a spectacular optical remnant in the southern sky.

\bigskip

\subsection{New Galactic SNRs}

As mentioned in Section 1.4, during the course of our Galactic SNR imaging
program, we discovered three new SNRs,  namely G107.5-5.1, G209.9-8.2,
and G210.5+1.3. Below we briefly discuss their optical emission structures and
properties and why we believe they are SNRs despite a lack of supporting radio
data.

\begin{figure*}[ht]
\hspace{1.0cm}
\begin{center}
\includegraphics[angle=0,width=12.5cm]{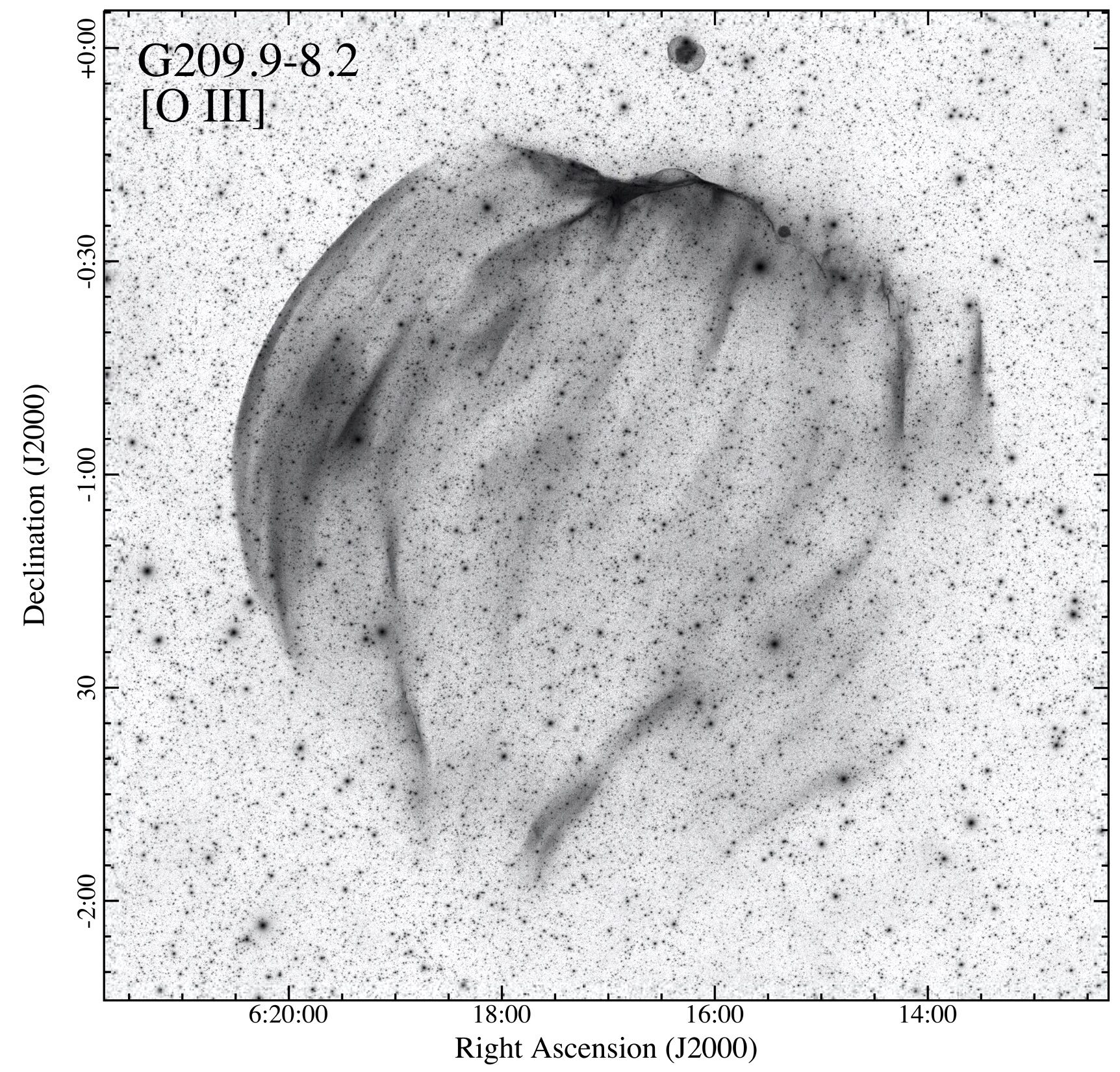} \\
\includegraphics[angle=0,width=12.5cm]{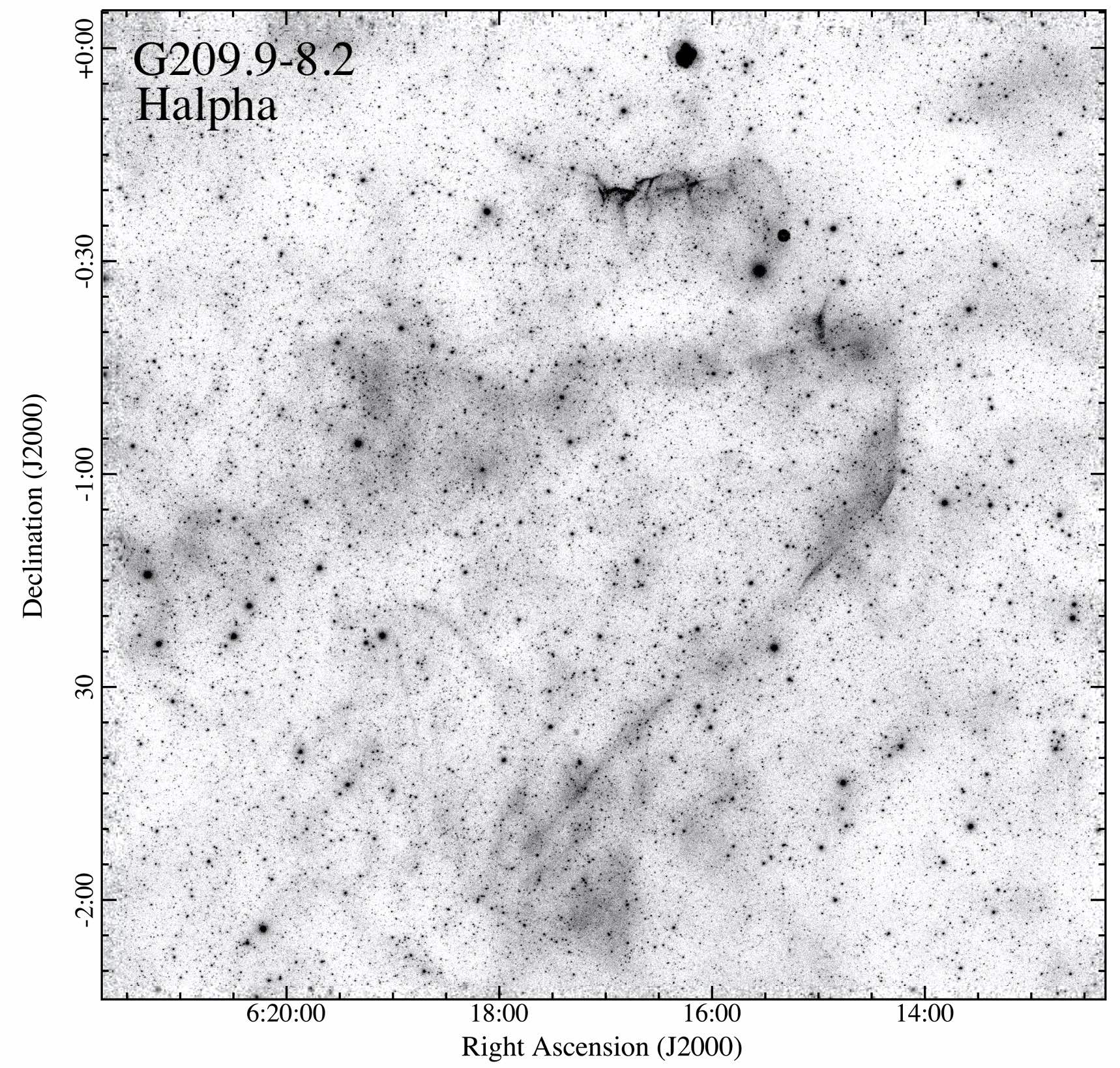}  
\caption{\O3 and H$\alpha$ images of SNR G209.9-8.2.
\label{G209_plots} 
} 
\end{center}
\end{figure*}

\begin{figure*}[ht]
\hspace{1.0cm}
\begin{center}
\vspace{0.30cm}
\centerline{\includegraphics[angle=0,width=17.0cm]{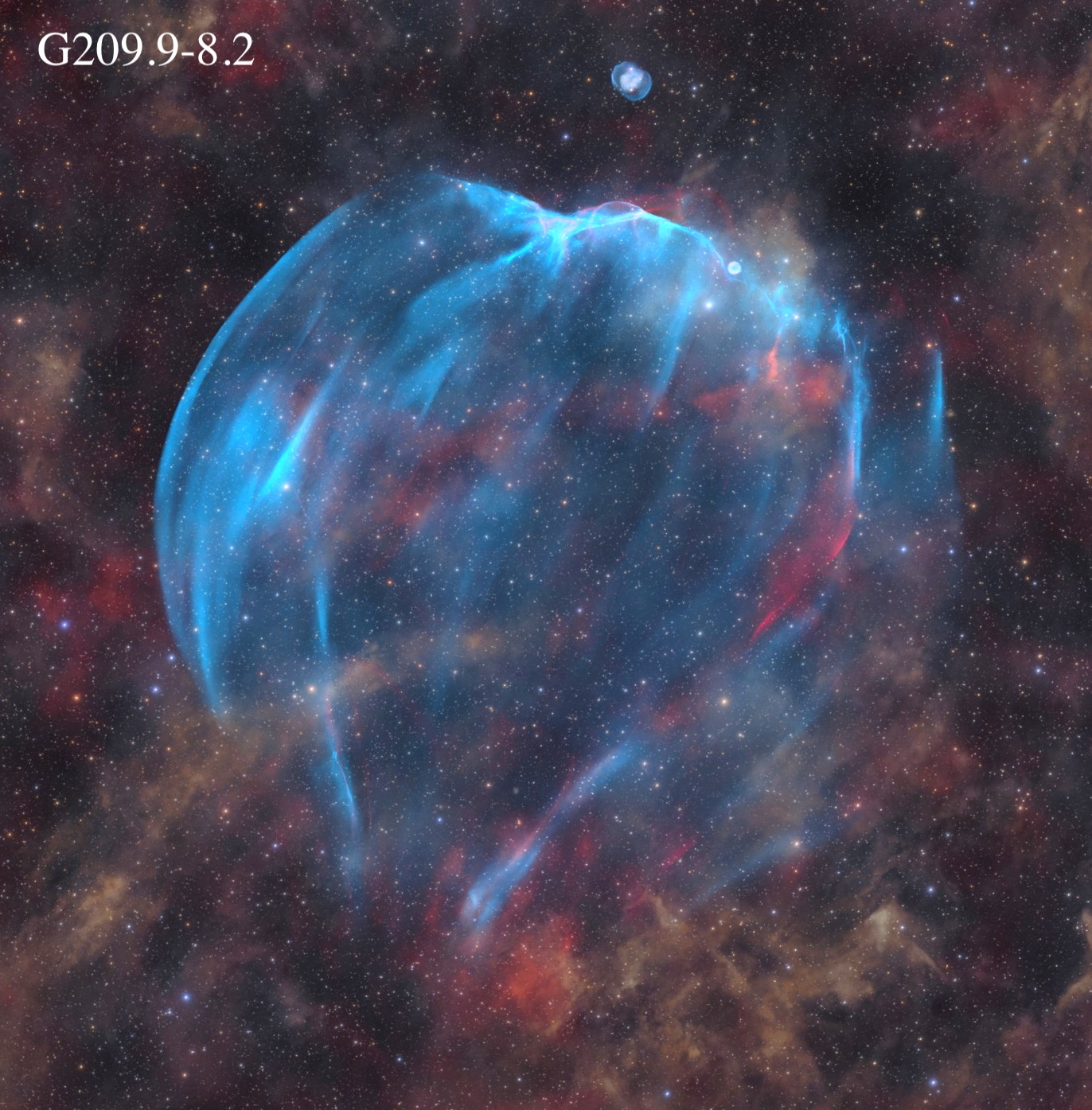}}
\caption{A color composite image of SNR G209.9-8.2. North is up, east to the left.
\label{G209_color} 
} 
\end{center}
\end{figure*}

\begin{figure*}
\begin{center}
\includegraphics[angle=0,width=17.5cm]{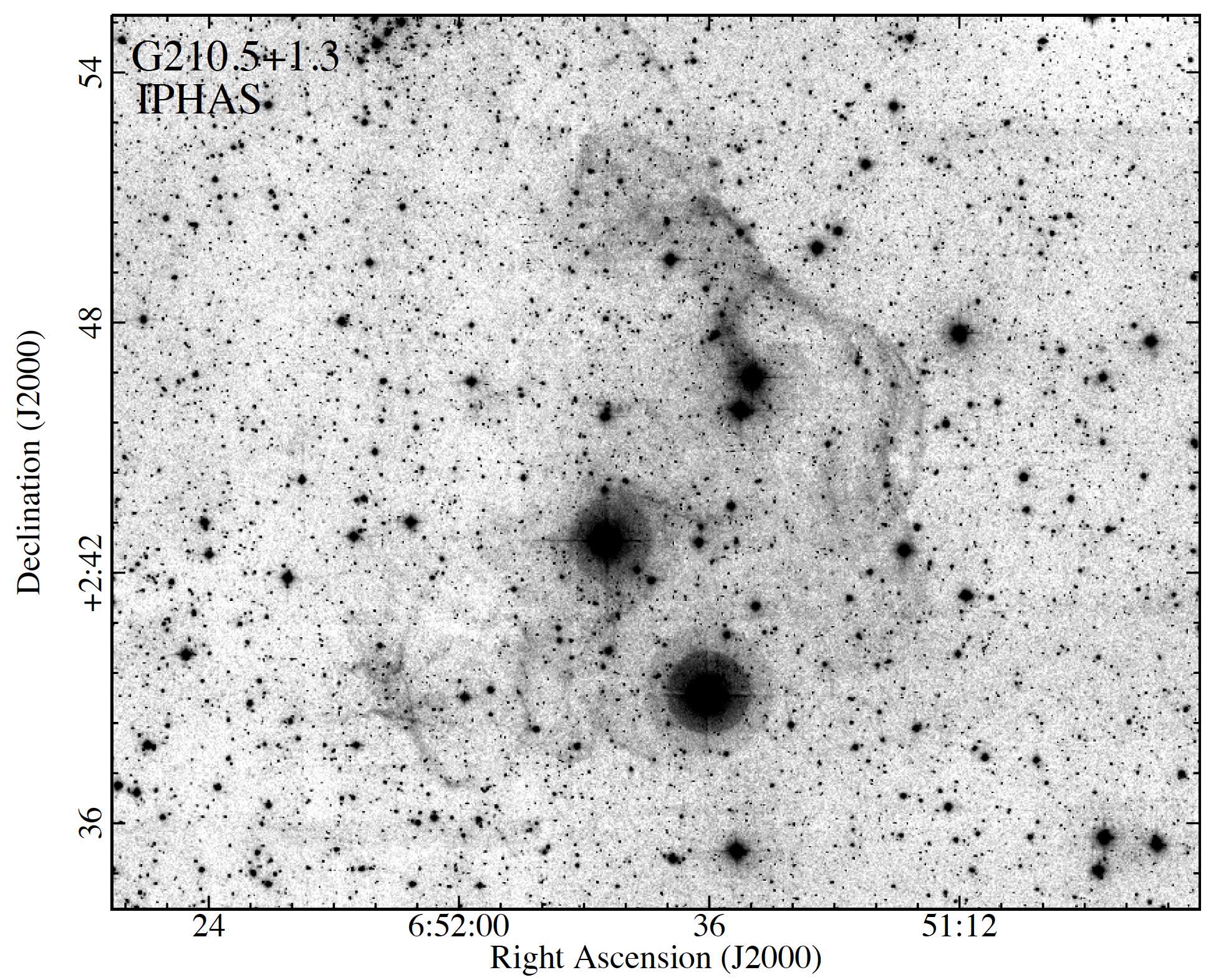} \\
\includegraphics[angle=0,width=8.0cm]{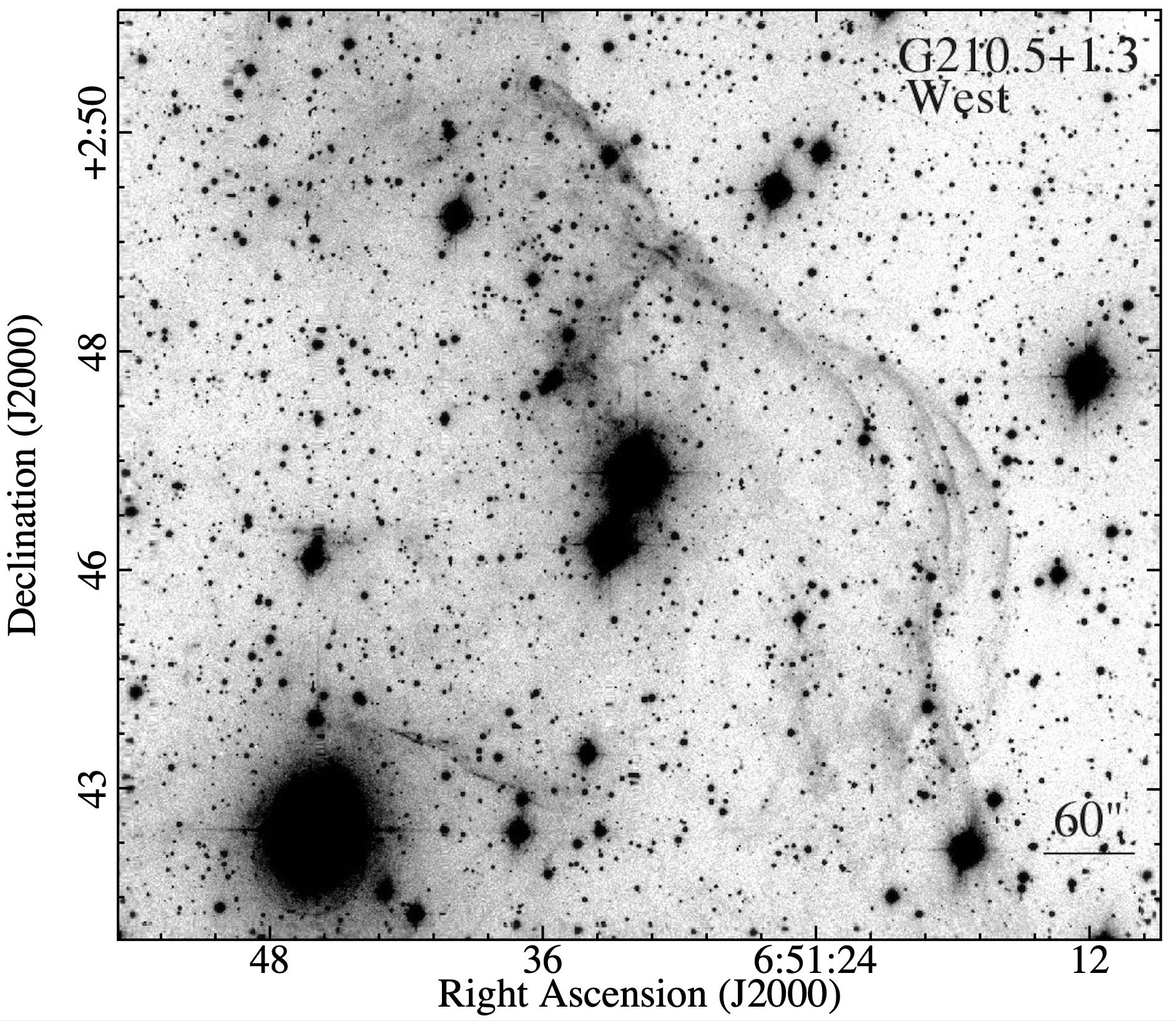}
\includegraphics[angle=0,width=9.35cm]{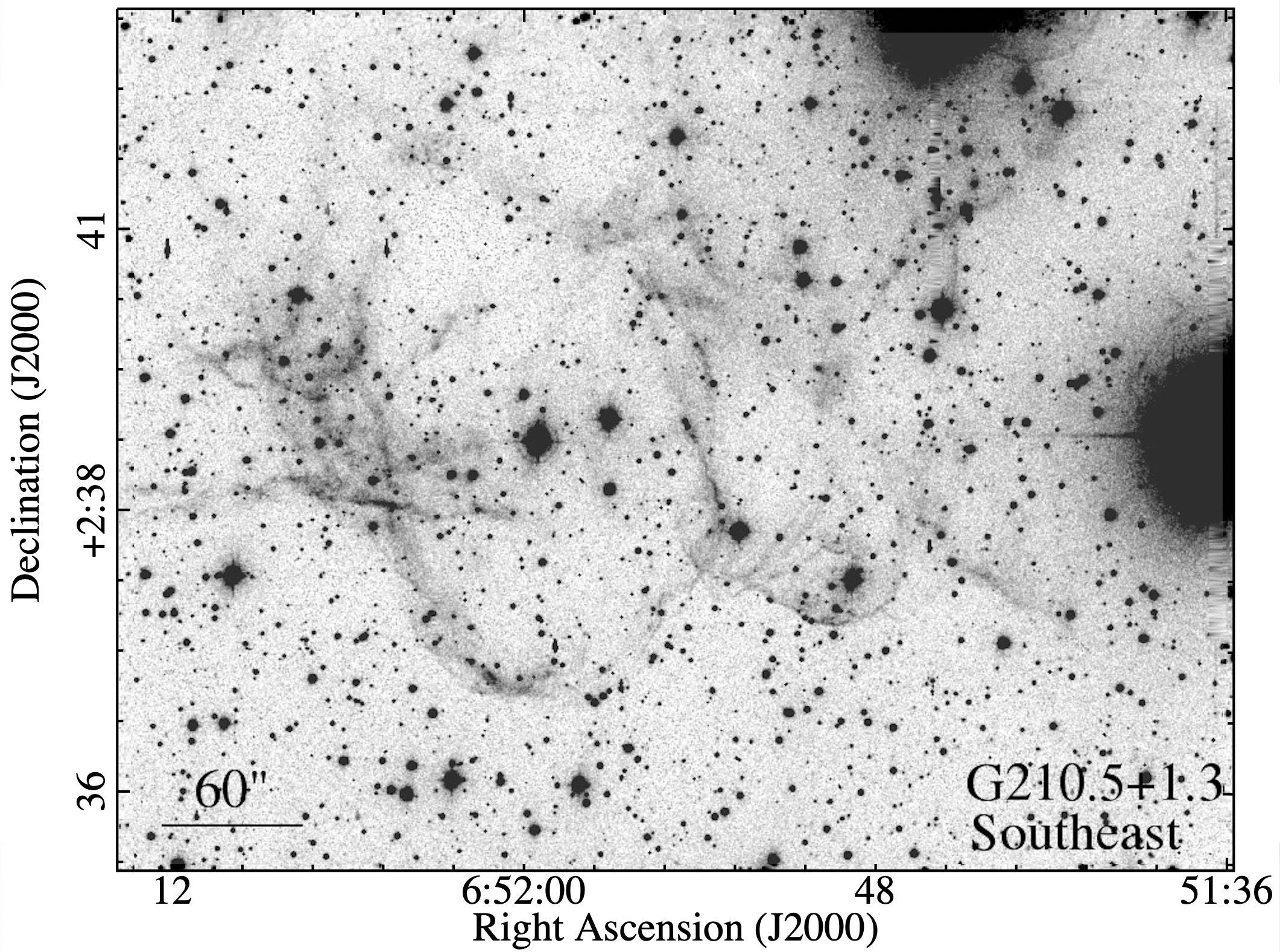}
\caption{Top: IPHAS H$\alpha$ image of G210.5+1.3. Bottom: MDM higher resolution H$\alpha$ images of west and southeast regions.
\label{G210_Ha} 
} 
\end{center}
\end{figure*}

\begin{figure*}
\begin{center}
\includegraphics[angle=0,width=18.5cm]{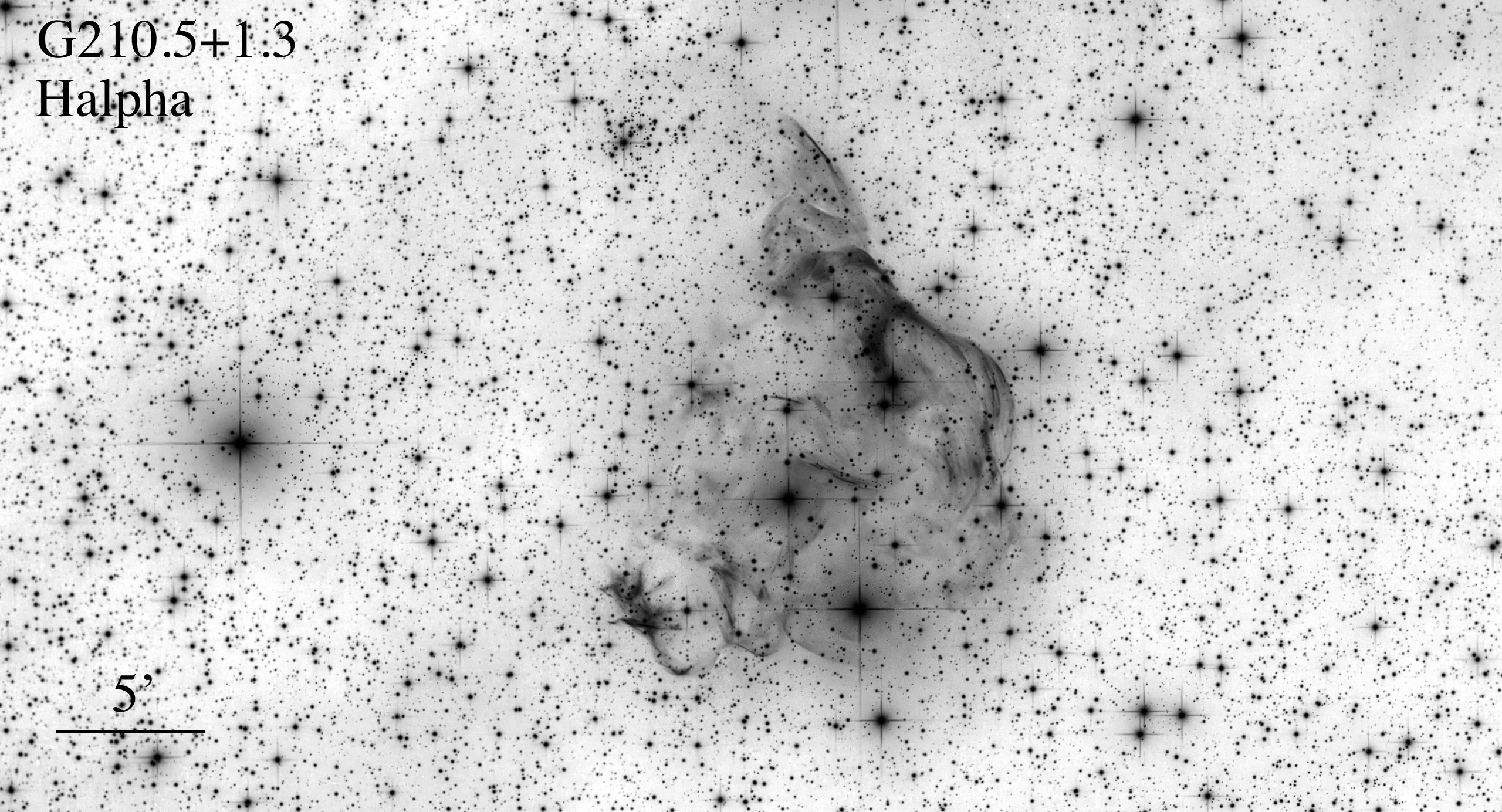} \\
\includegraphics[angle=0,width=18.5cm]{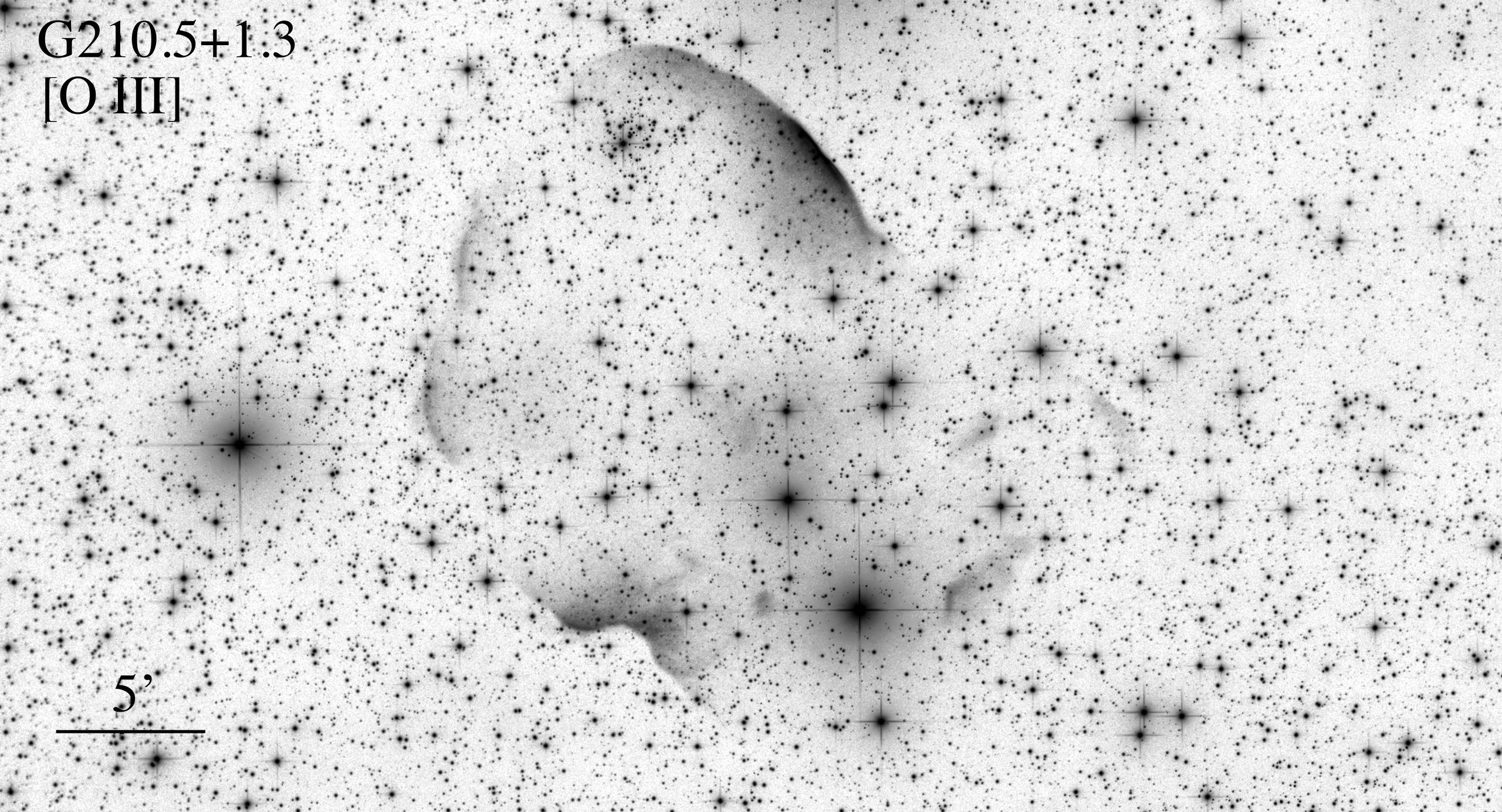} \\
\caption{H$\alpha$ and \O3 images  of G210.5+1.3. North is up, east to the left.
\label{G210_Ha_O3} 
} 
\end{center}
\end{figure*}

\begin{figure*}
\begin{center}
\includegraphics[angle=0,width=12.5cm]{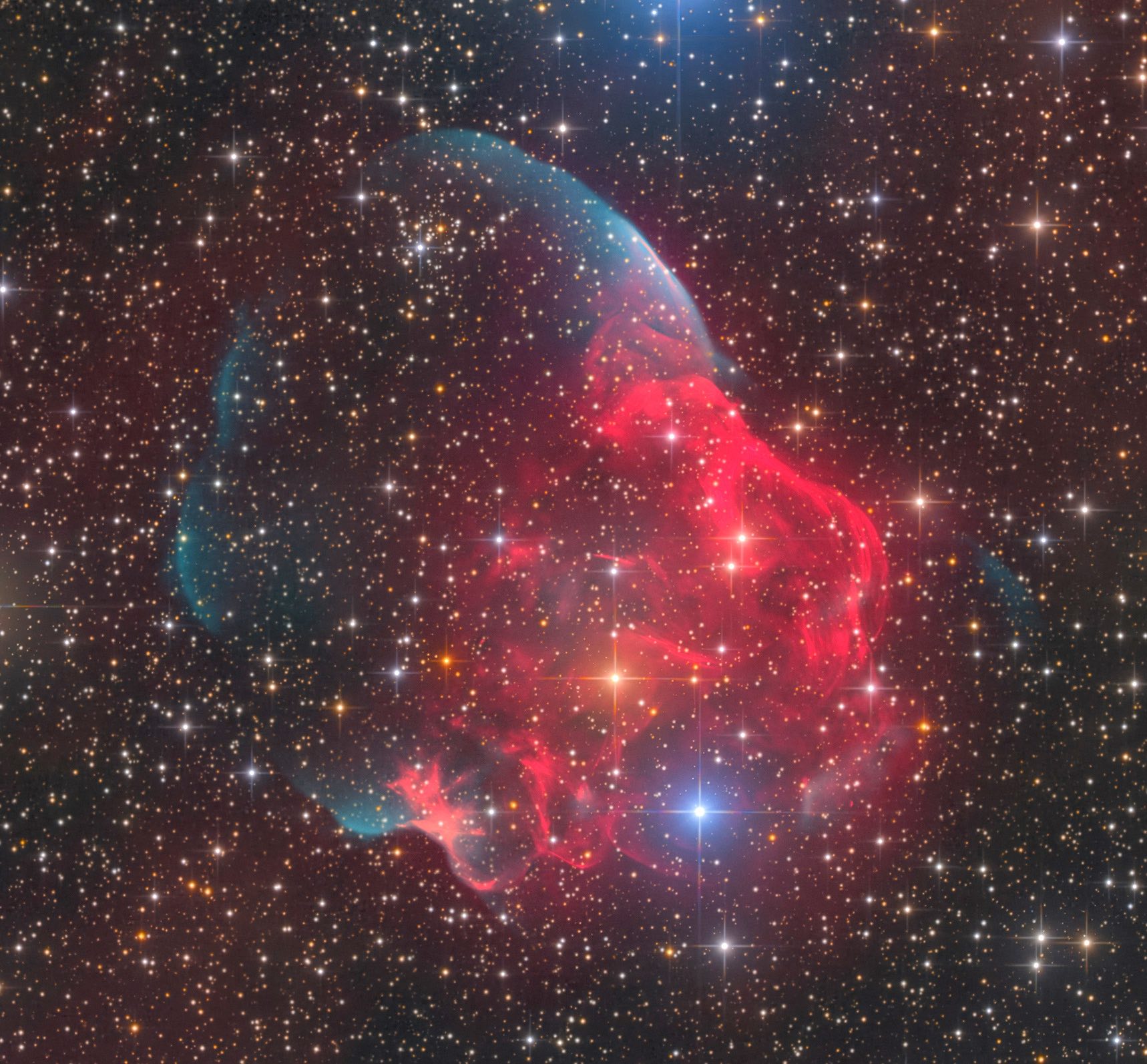} \\
\includegraphics[angle=0,width=12.5cm]{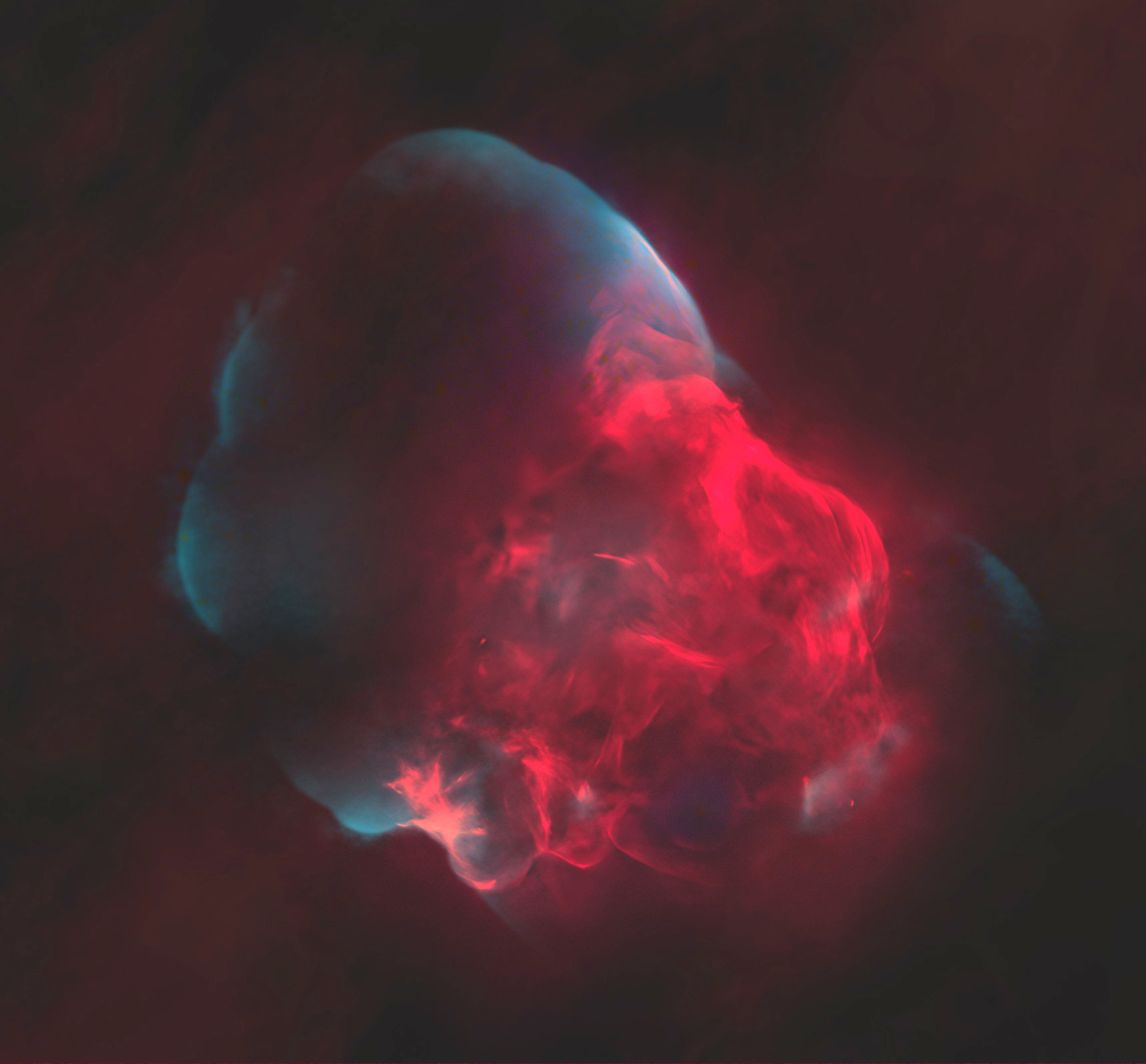}
\caption{Color H$\alpha$ and \O3 composite images of G210.5+1.3 with and without stars. North is up, east to the left.
\label{G210_color} 
} 
\end{center}
\end{figure*}

\subsubsection{G107.7-5.1: The Nereides Nebula\footnote{Due to its location in the Cassiopeia constellation 
and its many beautiful \O3 filaments, the SNR was named by its discoverers B.\ Falls and M.\ Drechsler after the Greek mythology daughters of the sea god Nereus known for their beauty
who were angered by Cassiopeia who boasted that her daughter, Andromeda, was more beautiful than any of the many Nereides.}}

During a wide FOV imaging program undertaken during the summer of 2022 to generate an emission-line mosaic of the Milky Way in the Cassiopeia and Cepheus 
region\footnote{https://www.astrobin.com/full/okz4di/0/?nc=\&nce=} 
a large, 2$\degr$+ wide faint nebula brightest in \O3 but also weakly seen in deep H$\alpha$ images was detected along the edge of the imaged region. 
This was followed-up by an intense
imaging campaign beginning in late 2023 July and ending
in mid-November 2023 with a total of over 250 hr  spent obtaining H$\alpha$, \O3, and RGB of images. 

Co-addition plus software processing of these data revealed a large and remarkable emission shell composed of dozens of fine and delicate
\O3 filaments roughly $2.4\degr \times 2.7\deg$ in size centered at
$l = 107.7\degr b = -5.1\degr$ (J2000: RA = 23:03:48; Dec.\ = +54:33).   A check of the \citet{Green2019} and \citet{Safi2012} catalogs showed no known SNR in this region.
Examination of several radio surveys also did not turn up
evidence for an obvious coincident radio source.

The fact that this SNR candidate lies some five degrees off the plane and is relatively near the bright radio remnant Cassiopeia A might help explain why it has not been noticed in radio surveys. However, 
this remnant is extremely faint optically and maybe equally faint in the radio and X-rays.  
As an aside, we note that the nearby suspected SNR G107.5-1.5 
(\citealt{Kothes2003}; d $\simeq 0.6\degr$) was recently discovered to show considerable optical emission suggestive of shock emission \citep{Bakis2023}.

Figure~\ref{G107_plots} shows wide FOV H$\alpha$ and \O3 images
of G107.7-5.1. Whereas only a broken shell of H$\alpha$ emission
is seen with mostly diffuse emission and a few sharp filaments, the remnant displays an extensive and highly complex emission structure consisting of numerous \O3 filaments forming a nearly complete emission shell. Little emission is seen  near the center of the shell in either emission line. 

Most of the shell's H$\alpha$ emission is diffuse with only a handful of short, sharp filaments. This can be seen in the upper panel of Fig.~\ref{G107_plots}. Whereas the broad and nearly one degree long NW H$\alpha$ emission exhibits no sharp filamentary features, we do find a few in the
shell's bright northeast and southwest nebulosities. 
But even there the emission is mainly diffuse.

In contrast, the shell's morphology as viewed in \O3 
(Fig.~\ref{G107_plots}) is quite filamentary giving the  impression of multiple interstellar shocks. This impression is made even clearer in the enlargements of east and west limbs of the shell's \O3 emission shown in Figures~\ref{G107_east} and \ref{G107_west}. The remnant's \O3 filamentary structure is both exceptional and extensive with few other SNRs exhibiting such a bright \O3 structure with the possible exceptions of the Cygnus Loop and CTA~1. Nearly all of the shell's \O3 emissions are in the form of sharp, thin filaments. And even when there is diffuse \O3 emission it is associated with sharp filaments. 
A color composite image of the remnant's 
H$\alpha$, \O3, and RGB images is shown in Figure~\ref{G107_color}.

Although the object's \O3 filaments appear in these images sharp, we note that many filaments are actually not so thin and sharp at higher spatial resolution. This is especially true for the bright
complex of filaments along the remnant's northeastern limb. \O3 images taken with the MDM 2.4m telescope show most of these
filaments to be broad on scales of 2-5 arcsec rather than thin emission filaments on the scale of 1 to 2 arcsec. On the other hand, there are indeed cases where the remnant's \O3 emission is in the form of  thin and sharp filaments. These instances include filaments along the southeast and southwest regions.

Given the highly filamentary structure of this large, nearly 3 degree emission
shell, the lack of neighboring H~II regions and obvious local sources of photoionization five degrees off the Galactic plane, a SNR nature appears highly likely. In fact, its morphology is not unlike that seen for other Galactic SNRs (see Figs.\ \ref{CTA1_O3_n_Ha} and  \ref{CTA1_pos}) and its obvious empty shell structure with no OB association inside
virtually eliminates an H~II region explanation.

On the other hand, the lack of known nonthermal radio emission is a problem for its firm SNR identification. However, despite
its bright appearance in our images, the remnant's emission
is actually very faint, helping to explain it being missed
until now. Twenty minute \O3 exposures taken with the MDM 2.4m telescope of the remnant's bright NE filaments detected only the
very brightest features shown in Fig.~\ref{G107_east}, and
nothing of the extensive surrounding but fainter emissions. Consequently,
it is possible that the nebula's radio emission is similarly
very faint and below detection levels of general galactic radio
surveys especially  given the object's $-5.1\degr$ Galactic latitude.

Optical spectra of some of the nebula's brightest \O3 and/or
H$\alpha$ would help confirm its SNR nature.
One short exposure, low S/N red spectrum taken with the MDM 2.4m telescope of a portion of the nebula's
H$\alpha$ emission along its northwestern limb 
(see top panel of Fig.~\ref{G107_plots}) suggests
a I([\ion{S}{2}]/I(H$\alpha$) ratio value above 0.4 which would
indicate the presence of shocks. However this measurement
needs to be investigated with much better data.

\subsubsection{G209.9-8.2}

Following the identification by M.\ Drechsler of a large emission shell some
eight degrees off the Galactic plane in the 1.4 GHz NRAO VLA Sky Survey (NVSS;
\citealt{Condon1998}), exploratory images by B.\ Falls led to the discovery 
of faint optical emission matching the position of the faint radio
shell.  The remnant, roughly 1.8 degrees in diameter and centered at $l =
209.9\degr, b = -8.2\degr$, was subsequently found to be mainly visible
optically via \O3 emission.  Follow-up H$\alpha$ and \O3 images revealed a
nearly complete emission shell brightest in the north and east.  Nicknamed the
`Atlas' remnant due to the Atlas mountains of Morocco from where it was most often
imaged, Figure~\ref{G209_plots} shows our \O3 (top) and H$\alpha$ (bottom)
images.  In \O3, the  remnant exhibits a bright, flat northern rim, with its
eastern edge unusually sharp and well defined. This sharp eastern edge of \O3 emission matches exactly 
bright eastern emission arc seen in the 1.4 GHz NVSS map, whereas the
much fainter western 1.4 GHz emission arc lies some $10'$ farther west than the western edge of the detected \O3 emission. Although parts of the western limb is faint
and/or missing in \O3, there is an unusual amount of internal diffuse emission near and north of the remnant center.

As we have seen in other remnants, G209.9-8.2's H$\alpha$ emission, shown in the bottom panel of 
Figure~\ref{G209_plots}, is much less
extensive in contrast to that seen in \O3. While we find a few filaments along the remnant's northern and western edges, most of the
H$\alpha$ emission is diffuse. Consequently, except for a few sharp H$\alpha$ filaments, we are not sure if all the diffuse H$\alpha$ seen here is actually related to remnant. 

Figure~\ref{G209_color} shows a color composite image where \O3 emission
is blue, H$\alpha$ emission is red, and RGB images form a broadband stellar background.
Here one can see an H$\alpha$ filament along the western limb coincides with a long \O3 filament, and that a large H$\alpha$ diffuse emission patch lies at the remnant's southern tip.
There is also a small curved filament at the remnant's north-central edge along with very faint H$\alpha$ emission outside (north) of the 
bright E-W line of \O3 filamentary emission.

Although we do not have optical spectra or radio data that confirms this
nearly 2 degree diameter nebula is a true SNR, we feel there is little doubt in its SNR nature. 
Given its large angular size, location eight degrees off the Galactic plane, 
and its highly filamentary nebulosity, especially in \O3, it has many of the expected characteristics of a SNR. 
There is no OB association inside or nearby
the shell which would raise the notion of a H~II region. 
If, on the other hand, it were a planetary nebula (PN), it displays an unexpectedly strong \O3 vs.\ H$\alpha$ emission ratio, and an inverse morphology where \O3 is strongest at larger radii with little H$\alpha$. Also, with a diameter of 1.8$\degr$ it would be the largest PN known, edging out the current
record setter SH2-216 at a diameter of 1.6$\degr$.

To give some perspective on 
the size of G209.9-8.2 if viewed as a possible PN, the small round
nebula visible $\sim17'$ north of G209.9-8.2 is the planetary nebula
PN~G208.9-8.0 \citep{TaWe1995,Ali1999}. Its bright inner region is
$145''$ in diameter but has fainter outer emission to the NE and SW giving
it a full diameter of $330''$, nearly as large as the Dumbbell Nebula
(M27, NGC 6853). Yet, as Figure~\ref{G209_color} shows it is completely dwarfed
by G209.9-8.2. 

Consequently, we believe G209.9-8.2 is a new Galactic SNR. Direct confirmation
through nonthermal radio data or optical spectra may be difficult to obtain.
The NVSS radio map showed little in the way of 1.4 GHz emission except for faint emission coincident with the remnant's
sharp eastern edge of \O3 emission. Optically, the remnant is also quite faint
even in \O3, and we estimate its brightest features to be around 5 Rayleighs. 
The fact that we took 200+ hours to image this nebula is a sign of just how faint this object is optically. Thus, obtaining an accurate measurement of H$\alpha$ and [\ion{S}{2}] line emissions might be a non-trivial task.

\subsubsection{G210.5+1.3}

A search of filamentary nebulae
visible on IPHAS images by M.\ Drechsler 
and X. Strottner turned up a possible SNR-like nebulosity
centered
at RA = 6h 36m, Dec. = $+2\degr$ $46'$
($l = 210.5, b = +1.3$) 
(see top panel of Fig.~\ref{G210_Ha}).
Subsequent higher resolution MDM images confirmed
the filamentary nature of the emission. 

No Galactic SNR appears at this nebula's position in SNR list of \citet{Green2019} or in lists of SNR candidates.
However, some NVSS 1.4 GHz emission is seen coincident with
the nebula's H$\alpha$ emission.

Four exploratory long slit spectra taken along the nebula's western 
limb using the MDM 2.4 telescope with the OSMOS spectrograph/camera
found I({\ion{S}2}]/I(H$\alpha$)
ratio between 0.44 - 0.62 ($\pm 0.06$) confirming a shock nature. The [\ion{S}{2}]
$\lambda$6716/$\lambda$6731 ratio
was also found to be between $\simeq$ 1.30 and 1.45
suggesting electron densities
at or near the low density limit,
i.e., n$_{\rm e}$ = $10 - 250$ cm$^{-3}$.

Figure~\ref{G210_Ha} shows higher resolution MDM images of the remnant's west and southeastern limb.
These images show more clearly the
largely filamentary structure of the nebula. Deep MDM \O3 images showed little
emission with the exception
of faint emission in the southeast
and northwest. 

Follow-up wide FOV exposure \O3 images suggested additional 
emission farther to the north and northeast, 
but little in the way of coincident H$\alpha$ emission. 
This lead to a campaign to obtain deeper \O3 images 
as well as in H$\alpha$ in order to map out this new remnant's
full optical emission structure.

The results of these follow-up images
is shown in Figure~\ref{G210_Ha_O3}.
Matching deep H$\alpha$ and \O3 images
reveal a far more interesting and extended nebulosity than at first suggested by the IPHAS images.
The extended \O3 emission to the north and east with little in the way of associated H$\alpha$ is highly unusual in SNRs and suggests
a bubble-like expansion due to an adjacent low density ISM 
region.

The remnant's unusual strong H$\alpha$ but weak \O3 southwest region and the strong \O3 but weak H$\alpha$ NE structure is highlighted in the color composite images shown in Figure~\ref{G210_color}. The upper panel shows the 
remnant with stars while the lower panel shows it with
the stars removed via software. The bipolar emission
structure is striking. 

Lastly, we note that the blue star seen
in the lower western edge of the remnant lies in an apparent ``hole" in
the remnant's H$\alpha$ emission giving the impression that
the star and the nebula are physically connected.
However, this is unlikely. The star, HD~50039, a F5~V star (B=7.75, V=7.28), has a Gaia DR3 parallax of
6.629 mas indicating a distance of around 150 pc. If the SNR
were at this distance, it would be by far the nearest remnant known yet only less than a half a degree in angular size.


\section{Discussion and Conclusions}

While the majority of Galactic SNRs are discovered through radio surveys, a remnant's
optical emission typically offers  higher resolution of its structure as well
as information concerning its shock velocity through both optical spectra and
the location of higher ionization lines such as strong \O3 filaments.  Optical
images taken in H$\alpha$ and other emission lines, especially \O3 and
[\ion{S}{2}] are also sensitive to low density remnant regions where optical
emission may be enhanced relative to radio emission levels.  In addition,
optical imaging surveys sometimes detect Galactic SNR emission that are either
too radio faint or simply are located at especially high Galactic latitudes
which fall outside many radio surveys. Optical images can also detect remnants,
like G210.5+1.3, which may have missed due to a lack of any obvious radio shell
morphology.

Here we have presented deep, wide-field optical emission-line images of a small 
sample of large Galactic SNRs (dia. $\geq 1\degr$).
We estimate our images detect emission-line levels of around 3-5 Rayleighs
for remnants that were imaged for 10-20 hours (e.g., G13.3-1.3) and 
down to $\sim1$ Rayleighs for objects like G107-5.1 and G209.9-8.2
where 200+ hours of exposures were taken.

\begin{figure}
\hspace{1.0cm}
\begin{center}
\includegraphics[angle=0,width=9.0cm]{G209_Ha_overlay_v2.jpg}
\includegraphics[angle=0,width=9.1cm]{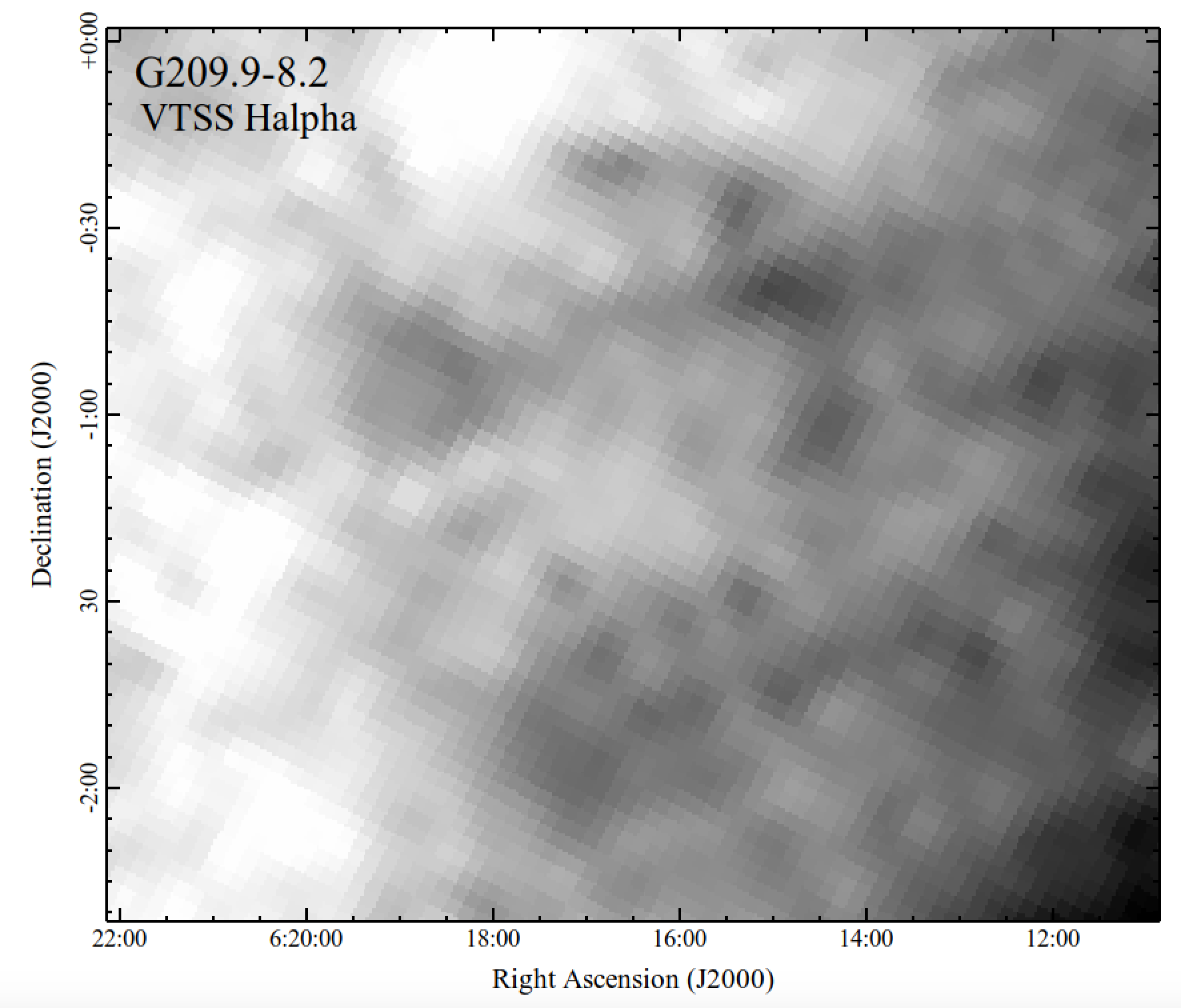} \\
\caption{Comparison of our H$\alpha$ image of G209.9-8.2 versus the VTSS H$\alpha$ image.
\label{G209_VTSS} 
} 
\end{center}
\end{figure} 

As an example of the depth of our images, we show in Figure~\ref{G209_VTSS} a
comparison of our H$\alpha$ image with that of the VTSS on the G209.9-8.2
remnant.  VTSS claims an H$\alpha$ detection limit down to 2 Rayleighs in a 2
hr exposure using a 17 \AA \ wide filter.  Our H$\alpha$ image is the product
of $\sim$120 hr using 2 to 3 times wider filters with a resolution of 2.4$''$,
some 40 times better than the 98$''$ listed for VTSS.  While both images show
similar emission features, our image with its substantially better angular
resolution has a detection limit at least as deep as the VTSS.  Moreover,
remnant features can be more readily distinguished in our image from background
emissions such as the emission along much of the western portion of the VTSS
image arising from a bright and unrelated nearby H~II region.

One aim of our work was to obtain high-quality, wide FOV optical emission-line images 
of several large but poorly studied Galactic SNRs in order to characterize their overall morphology and fine-scale structure. Indeed, 
most of the objects we examined are neither famous or known to have interesting optical emission structures.
Of the nine known Galactic SNRs we imaged, only three
could be considered well studied, namely G82.2+5.3 (W63),
G89.0+4.7 (HB21), and G119.5+10.2 (CTA~1).
Four of the other six remnants have fewer than six published
papers on them, while the two remaining remnants
do not appear in the on-line 2022 updated \citet{Green2019} catalog.

Nonetheless, our images of these nine SNRs revealed  many new features and sometimes quite unexpected and beautiful emission structures. New discoveries include:
\begin{itemize}
    \item G13.3-1.3: Wide-field \O3 images show a remarkable group of long, thin filaments coincident with the remnant's radio emission
    which terminates in broad, extended diffuse \O3 emission. Our images also detected more extensive emission along its southern and eastern boundary than previous known. 
    \item G70.0-21.5: Deep H$\alpha$ and \O3 imaging of nearly all of this large $4.0\degr \times 5.5\degr$ remnant shows large angular separations between bright \O3 and H$\alpha$ filaments in some remnant regions.
    \item G82.2+5.3 (W63): Our images allow clear identification of this remnant's emission structure from the very complex Galactic H$\alpha$ emission seen in this direction. Our \O3 images also show optical shock emission coincident with radio emission features extending outside the remnant's main optical/radio emission shell both to the north and west.
    \item G89.0+4.7 (HB21): Like that seen of G82.2+5.3, our images, especially \O3 images, better define the remnant's
    extent and structure. Our deep \O3 image (Fig.~\ref{G89_radio}) offers considerable new detail and a new look to this remnant's structure along its northern and western limbs.
    \item G119.5+10.5 (CTA~1): This large and high latitude SNR was previously known to be brightest in \O3 emission. However, our wide-field and high-resolution H$\alpha$ and \O3 images
    reveals a far more and highly complex structure with its H$\alpha$ and \O3 filaments closely aligned. A color composite image highlights this complexity, showing a web of tangled filaments alternatively bright in \O3 and H$\alpha$ emissions.
    \item G150.3+4.5: Our images of this poorly studied remnant represents the first optical investigation. As we have seen in other SNRs in our survey, deep \O3 images proved invaluable in determining the remnant's structure at a much high resolution than that of radio data.  Our \O3 images also unexpectedly
    uncovered a complete emission shell surrounding the 
    variable X-ray binary source CI Cam which appears to have not been detected before.
    \item G181.1+9.5: Discovered only recently as one of the faintest radio SNRs known, this high latitude SNR exhibits considerable optical emission. Our color composite image of H$\alpha$ + \O3 emissions shows a complete shell along with substantial emission to the south. A wide-field H$\alpha$ mosaic of the region around this remnant revealed the presence of a
    large (dia.\ $\sim9\degr$) partial H$\alpha$ emission shell,
    which, if a SNR shell, would rank among the largest Galactic remnants known.
    \item G288.8-6.3: Although not listed among the confirmed Galactic SNRs \citep{Green2019} and only previously known to show one small H$\alpha$ emission patch plus a short, faint H$\alpha$ filament, our \O3 images instead show an extensive optical emission structure which correlates with the recent radio map of \citet{Filipovic2023}.
    \item G321.3-3.9: This remnant displays an amazing amount of optical emission, and like that seen for some other remnants, it is particularly bright in \O3. It also displays two shell breakout features. While the remnant's 
    optical emission is in excellent agreement with the remnant's brightest radio emission, it exhibits an enormous wealth of additional details unseen in the radio.
\end{itemize}

We also announce the discovery and present here deep H$\alpha$ and \O3 images
of three new apparent Galactic remnants. Each of these new SNRs have extensive
optical structures comparable in size and detail to that seen for the best
studied and imaged known SNRs.  For example, the extensive and beautiful \O3
filaments of the large SNR G107.7-5.1 have few rivals in the other 300+
Galactic SNRs.  The exceedingly faint but beautiful G209.9-8.2 remnant is one
more example where \O3 imaging reveals a SNR far better than deep
H$\alpha$ images.  And the remarkable \O3 `bubble' in the G210.5+1.3 distinct from
the its H$\alpha$ emission filaments, best seen in the starless version
presented here, gives a new prospective regarding the size of expansion
breakouts.

Although our optical imaging survey only investigated a dozen SNRs, there are several key findings from this work. These include:

1) While H$\alpha$ emission has long been the main criteria for defining whether a remnant exhibits optical emission, deep \O3 imaging is actually just as important, and in some cases, much more so. Due to few  bright telluric emission lines around  the  [\ion{O}{3}] $\lambda$5007 emission line, \O3 images confront a much fainter background compared to H$\alpha$ and thus in some cases can map a remnant's optical emission structure better than even deep H$\alpha$ images. Without \O3 images, several SNRs in our survey would have displayed far less -- and less interesting -- emission structures.

2) The great complexity of some remnants in terms of their H$\alpha$ and \O3 line emissions, seen in G70.0-21.5, G107.7-5.1, G119.5+10.5 (CTA~1), and G321.3-3.9, raises serious questions about the usefulness
of spectral analysis results from just a handful of slit spectra taken on a few filaments. The \O3 bubble of 
G210.5+1.3 dramatically shows that without deep emission line images besides H$\alpha$ one may miss large portions of a remnant whose spectral properties differ greatly from other regions.

3) Remnant ``breakout'' features do not seem to be especially rare. In our survey of nine known and three new remnants, about 30\% exhibited significant extended features outside their main shell; these include
G70.0-21.5, G82.2+5.3, G210.5+1.3, and G321.3-3.9.
breakout features also appear common in radio data. A recent study
of 36 SNRs using MeerKAT 1.3 GHz observations found nearly 50\% of the remnants showed breakout or ``ears'' \citep{Cotton2024}. However, we caution that our small survey
focused on particularly large remnants where their large size might be more sensitive to local interstellar density variations which subsequently show up as shock expansion like bubbles. 

4) Lastly, we note that the recent revolution in emission-line imaging by both professional and amateurs brought on by new sensitive detectors and high throughput narrow
bandpass filter technology will likely increase significantly  the depth of future optical emission-line 
surveys, both along and far away from the Galactic plane. The recent discovery of a very large and faint \O3 emission nebula near M31 
\citep{Drechsler2023,Fesen2023} is just one sign of the discovery power of deep imaging of an otherwise  very well studied object through the combination of hundreds of exposures taken using small and modest aperture telescopes and processed by powerful software. \\

We wish to thank the many staff and support personnel at the various observatories and observing sites across several continents that made these deep observations possible. 
We also thank Eric Galayda and the entire MDM staff for supporting the MDM observations. This work made use of the Simbad database, NASA's Skyview online data archives, the on-line WCS astrometry website nova.astrometry.net, and the 
Survey Sampler at the Max-Planck-Institut f\"ur Radioastronomie.
This work is part of R.A.F's Archangel III Research Program at Dartmouth. Our collection of deep optical SNR images was inspired by the 1973 van den Bergh et al.\ paper ``An Optical Atlas of Galactic Supernova Remnants'' published fifty years ago which stood for many years as an invaluable reference for SNR researchers across many wavelength regimes.

\facilities{Hakos Astrofarm, Oukaimeden Observatories, MDM Observatory at Kitt Peak, New Mexico Skies Observatory, Ross Creek Observatory, Sierra Remote Observatory, and Ursa Major Observatory at ROSA}.

\software{PYRAF \citep{pyrafcite}, Astropy v4.0 \citep{AstropyCiteA,AstropyCiteB}, ds9 \citep{ds9cite}, L.A.\ Cosmic \citep{vanDokkum2001}, OSMOS Pipeline (thorosmos: \url{https://github.com/jrthorstensen/thorosmos}), Photoshop, PixInsight}

\bibliography{ref4}

\begin{thebibliography}{}
\expandafter\ifx\csname natexlab\endcsname\relax\def\natexlab#1{#1}\fi
\providecommand{\url}[1]{\href{#1}{#1}}
\providecommand{\dodoi}[1]{doi:~\href{http://doi.org/#1}{\nolinkurl{#1}}}
\providecommand{\doeprint}[1]{\href{http://ascl.net/#1}{\nolinkurl{http://ascl.net/#1}}}
\providecommand{\doarXiv}[1]{\href{https://arxiv.org/abs/#1}{\nolinkurl{https://arxiv.org/abs/#1}}}

\bibitem[{{Abdo} {et~al.}(2008){Abdo}, {Ackermann}, {Atwood}, {Baldini}, {Ballet}, {Barbiellini}, {Baring}, {Bastieri}, {Baughman}, {Bechtol}, {Bellazzini}, {Berenji}, {Blandford}, {Bloom}, {Bogaert}, {Bonamente}, {Borgland}, {Bregeon}, {Brez}, {Brigida}, {Bruel}, {Burnett}, {Caliandro}, {Cameron}, {Caraveo}, {Carlson}, {Casandjian}, {Cecchi}, {Charles}, {Chekhtman}, {Cheung}, {Chiang}, {Ciprini}, {Claus}, {Cohen-Tanugi}, {Cominsky}, {Conrad}, {Cutini}, {Davis}, {Dermer}, {de Angelis}, {de Palma}, {Digel}, {Dormody}, {do Couto e Silva}, {Drell}, {Dubois}, {Dumora}, {Edmonds}, {Farnier}, {Focke}, {Fukazawa}, {Funk}, {Fusco}, {Gargano}, {Gasparrini}, {Gehrels}, {Germani}, {Giebels}, {Giglietto}, {Giordano}, {Glanzman}, {Godfrey}, {Grenier}, {Grondin}, {Grove}, {Guillemot}, {Guiriec}, {Harding}, {Hartman}, {Hays}, {Hughes}, {J{\'o}hannesson}, {Johnson}, {Johnson}, {Johnson}, {Johnson}, {Kamae}, {Kanai}, {Kanbach}, {Katagiri}, {Kawai}, {Kerr}, {Kishishita}, {Kiziltan}, {Kn{\"o}dlseder}, {Kocian}, {Komin},
  {Kuehn}, {Kuss}, {Latronico}, {Lemoine-Goumard}, {Longo}, {Lonjou}, {Loparco}, {Lott}, {Lovellette}, {Lubrano}, {Makeev}, {Marelli}, {Mazziotta}, {McEnery}, {McGlynn}, {Meurer}, {Michelson}, {Mineo}, {Mitthumsiri}, {Mizuno}, {Moiseev}, {Monte}, {Monzani}, {Morselli}, {Moskalenko}, {Murgia}, {Nakamori}, {Nolan}, {Nuss}, {Ohno}, {Ohsugi}, {Okumura}, {Omodei}, {Orlando}, {Ormes}, {Ozaki}, {Paneque}, {Panetta}, {Parent}, {Pelassa}, {Pepe}, {Pesce-Rollins}, {Piano}, {Pieri}, {Piron}, {Porter}, {Rain{\`o}}, {Rando}, {Ray}, {Razzano}, {Reimer}, {Reimer}, {Reposeur}, {Ritz}, {Rochester}, {Rodriguez}, {Romani}, {Roth}, {Ryde}, {Sadrozinski}, {Sanchez}, {Sander}, {Parkinson}, {Schalk}, {Sellerholm}, {Sgr{\`o}}, {Siskind}, {Smith}, {Smith}, {Spandre}, {Spinelli}, {Starck}, {Strickman}, {Suson}, {Tajima}, {Takahashi}, {Takahashi}, {Tanaka}, {Thayer}, {Thayer}, {Thompson}, {Thorsett}, {Tibaldo}, {Torres}, {Tosti}, {Tramacere}, {Usher}, {Van Etten}, {Vilchez}, {Vitale}, {Wang}, {Watters}, {Winer}, {Wood}, {Yasuda},
  {Ylinen}, \& {Ziegler}}]{Abdo2008}
{Abdo}, A.~A., {Ackermann}, M., {Atwood}, W.~B., {et~al.} 2008, Science, 322, 1218, \dodoi{10.1126/science.1165572}

\bibitem[{{Abraham} {et~al.}(2017){Abraham}, {van Dokkum}, {Conroy}, {Merritt}, {Zhang}, {Lokhorst}, {Danieli}, \& {Mowla}}]{Abraham2017}
{Abraham}, R., {van Dokkum}, P., {Conroy}, C., {et~al.} 2017, in Astrophysics and Space Science Library, Vol. 434, Outskirts of Galaxies, ed. J.~H. {Knapen}, J.~C. {Lee}, \& A.~{Gil de Paz}, 333, \dodoi{10.1007/978-3-319-56570-5_10}

\bibitem[{{Abraham} \& {van Dokkum}(2014)}]{Abraham2014}
{Abraham}, R.~G., \& {van Dokkum}, P.~G. 2014, \pasp, 126, 55, \dodoi{10.1086/674875}

\bibitem[{{Ali}(1999)}]{Ali1999}
{Ali}, A. 1999, \na, 4, 95, \dodoi{10.1016/S1384-1076(99)00003-2}

\bibitem[{{Aret} {et~al.}(2020){Aret}, {Arias}, {Torres}, {Cidale}, \& {Eenm{\"a}e}}]{Aret2020}
{Aret}, A., {Arias}, M.~L., {Torres}, A., {Cidale}, L.~S., \& {Eenm{\"a}e}, T. 2020, Boletin de la Asociacion Argentina de Astronomia La Plata Argentina, 61B, 99

\bibitem[{{Astropy Collaboration} {et~al.}(2013){Astropy Collaboration}, {Robitaille}, {Tollerud}, {Greenfield}, {Droettboom}, {Bray}, {Aldcroft}, {Davis}, {Ginsburg}, \& {Price-Whelan}}]{AstropyCiteA}
{Astropy Collaboration}, {Robitaille}, T.~P., {Tollerud}, E.~J., {et~al.} 2013, \aap, 558, A33, \dodoi{10.1051/0004-6361/201322068}

\bibitem[{{Astropy Collaboration} {et~al.}(2018){Astropy Collaboration}, {Price-Whelan}, {Sip{\H{o}}cz}, {G{\"u}nther}, {Lim}, {Crawford}, {Conseil}, {Shupe}, {Craig}, \& {Dencheva}}]{AstropyCiteB}
{Astropy Collaboration}, {Price-Whelan}, A.~M., {Sip{\H{o}}cz}, B.~M., {et~al.} 2018, \aj, 156, 123, \dodoi{10.3847/1538-3881/aabc4f}

\bibitem[{{Bak{\i}{\c{s}}} {et~al.}(2023){Bak{\i}{\c{s}}}, {Bulut}, {Bak{\i}{\c{s}}}, {Sano}, \& {Sezer}}]{Bakis2023}
{Bak{\i}{\c{s}}}, H., {Bulut}, G., {Bak{\i}{\c{s}}}, V., {Sano}, H., \& {Sezer}, A. 2023, \mnras, 521, 1099, \dodoi{10.1093/mnras/stad576}

\bibitem[{{Barsukova} {et~al.}(2023){Barsukova}, {Burenkov}, {Goranskij}, {Zharikov}, {Iliev}, {Manset}, {Metlova}, {Miroshnichenko}, {Moiseeva}, {Nedialkov}, {Semenko}, {Stoyanov}, \& {Yakunin}}]{Barsukova2023}
{Barsukova}, E.~A., {Burenkov}, A.~N., {Goranskij}, V.~P., {et~al.} 2023, Astrophysical Bulletin, 78, 1, \dodoi{10.1134/S1990341323010029}

\bibitem[{{Bartlett} {et~al.}(2019){Bartlett}, {Clark}, \& {Negueruela}}]{Bartlett2019}
{Bartlett}, E.~S., {Clark}, J.~S., \& {Negueruela}, I. 2019, \aap, 622, A93, \dodoi{10.1051/0004-6361/201834315}

\bibitem[{{Ben-Ami} {et~al.}(2023){Ben-Ami}, {Ofek}, {Polishook}, {Franckowiak}, {Hallakoun}, {Segre}, {Shvartzvald}, {Strotjohann}, {Yaron}, {Aharonson}, {Arcavi}, {Berge}, {Ramazani}, {Gal-Yam}, {Garrappa}, {Hershko}, {Nir}, {Ohm}, {Rybicki}, {Sadeh}, {Segev}, {Shani}, {Sofer-Rimalt}, \& {Weimann}}]{Ben-Ami023}
{Ben-Ami}, S., {Ofek}, E.~O., {Polishook}, D., {et~al.} 2023, \pasp, 135, 085002, \dodoi{10.1088/1538-3873/aceb30}

\bibitem[{{Blair} {et~al.}(1981){Blair}, {Kirshner}, \& {Chevalier}}]{Blair1981}
{Blair}, W.~P., {Kirshner}, R.~P., \& {Chevalier}, R.~A. 1981, \apj, 247, 879, \dodoi{10.1086/159098}

\bibitem[{{Boumis} {et~al.}(2002){Boumis}, {Mavromatakis}, {Paleologou}, \& {Becker}}]{Boumis2002}
{Boumis}, P., {Mavromatakis}, F., {Paleologou}, E.~V., \& {Becker}, W. 2002, \aap, 396, 225, \dodoi{10.1051/0004-6361:20021365}

\bibitem[{{Boumis} {et~al.}(2009){Boumis}, {Xilouris}, {Alikakos}, {Christopoulou}, {Mavromatakis}, {Katsiyannis}, \& {Goudis}}]{Boumis2009}
{Boumis}, P., {Xilouris}, E.~M., {Alikakos}, J., {et~al.} 2009, \aap, 499, 789, \dodoi{10.1051/0004-6361/200811474}

\bibitem[{{Burger-Scheidlin} {et~al.}(2023){Burger-Scheidlin}, {Brose}, {Mackey}, {Filipovi{\'c}}, {Goswami}, {Mestre Guillen}, {de O{\~n}a Wilhelmi}, \& {Sushch}}]{BS2023}
{Burger-Scheidlin}, C., {Brose}, R., {Mackey}, J., {et~al.} 2023, arXiv e-prints, arXiv:2310.14431, \dodoi{10.48550/arXiv.2310.14431}

\bibitem[{{Condon} {et~al.}(1998){Condon}, {Cotton}, {Greisen}, {Yin}, {Perley}, {Taylor}, \& {Broderick}}]{Condon1998}
{Condon}, J.~J., {Cotton}, W.~D., {Greisen}, E.~W., {et~al.} 1998, \aj, 115, 1693, \dodoi{10.1086/300337}

\bibitem[{{Cotton} {et~al.}(2024){Cotton}, {Kothes}, {Camilo}, {Chandra}, {Buchner}, \& {Nyamai}}]{Cotton2024}
{Cotton}, W.~D., {Kothes}, R., {Camilo}, F., {et~al.} 2024, \apjs, 270, 21, \dodoi{10.3847/1538-4365/ad0ecb}

\bibitem[{{Dennison} {et~al.}(1998){Dennison}, {Simonetti}, \& {Topasna}}]{Dennison1998}
{Dennison}, B., {Simonetti}, J.~H., \& {Topasna}, G.~A. 1998, \pasa, 15, 147, \dodoi{10.1071/AS98147}

\bibitem[{{Devin} {et~al.}(2020){Devin}, {Lemoine-Goumard}, {Grondin}, {Castro}, {Ballet}, {Cohen}, \& {Hewitt}}]{Devin2020}
{Devin}, J., {Lemoine-Goumard}, M., {Grondin}, M.~H., {et~al.} 2020, \aap, 643, A28, \dodoi{10.1051/0004-6361/202038503}

\bibitem[{{Dickel} {et~al.}(1969){Dickel}, {Wendker}, \& {Bieritz}}]{Dickel1969}
{Dickel}, H.~R., {Wendker}, H., \& {Bieritz}, J.~H. 1969, \aap, 1, 270

\bibitem[{{Dopita} {et~al.}(1984){Dopita}, {Binette}, {Dodorico}, \& {Benvenuti}}]{Dopita1984}
{Dopita}, M.~A., {Binette}, L., {Dodorico}, S., \& {Benvenuti}, P. 1984, \apj, 276, 653, \dodoi{10.1086/161653}

\bibitem[{{Drechsler} {et~al.}(2023){Drechsler}, {Strottner}, {Sainty}, {Fesen}, {Kimeswenger}, {Shull}, {Falls}, {Vergnes}, {Martino}, \& {Walker}}]{Drechsler2023}
{Drechsler}, M., {Strottner}, X., {Sainty}, Y., {et~al.} 2023, Research Notes of the American Astronomical Society, 7, 1, \dodoi{10.3847/2515-5172/acaf7e}

\bibitem[{{Drew} {et~al.}(2005){Drew}, {Greimel}, {Irwin}, {Aungwerojwit}, {Barlow}, {Corradi}, {Drake}, {G{\"a}nsicke}, {Groot}, {Hales}, {Hopewell}, {Irwin}, {Knigge}, {Leisy}, {Lennon}, {Mampaso}, {Masheder}, {Matsuura}, {Morales-Rueda}, {Morris}, {Parker}, {Phillipps}, {Rodriguez-Gil}, {Roelofs}, {Skillen}, {Sokoloski}, {Steeghs}, {Unruh}, {Viironen}, {Vink}, {Walton}, {Witham}, {Wright}, {Zijlstra}, \& {Zurita}}]{Drew2005}
{Drew}, J.~E., {Greimel}, R., {Irwin}, M.~J., {et~al.} 2005, \mnras, 362, 753, \dodoi{10.1111/j.1365-2966.2005.09330.x}

\bibitem[{{Dubner} \& {Giacani}(2015)}]{Dubner2015}
{Dubner}, G., \& {Giacani}, E. 2015, \aapr, 23, 3, \dodoi{10.1007/s00159-015-0083-5}

\bibitem[{{Duin} \& {van der Laan}(1975)}]{Duin1975}
{Duin}, R.~M., \& {van der Laan}, H. 1975, \aap, 40, 111

\bibitem[{{Duncan} {et~al.}(1997){Duncan}, {Stewart}, {Haynes}, \& {Jones}}]{Duncan1997}
{Duncan}, A.~R., {Stewart}, R.~T., {Haynes}, R.~F., \& {Jones}, K.~L. 1997, \mnras, 287, 722, \dodoi{10.1093/mnras/287.4.722}

\bibitem[{{Feng} {et~al.}(2024){Feng}, {Chen}, {Su}, {Sun}, {Zhang}, {Zhou}, \& {Guo}}]{Feng2024}
{Feng}, J.-C., {Chen}, X., {Su}, Y., {et~al.} 2024, arXiv e-prints, arXiv:2403.19788, \dodoi{10.48550/arXiv.2403.19788}

\bibitem[{{Ferrand} \& {Safi-Harb}(2012)}]{Safi2012}
{Ferrand}, G., \& {Safi-Harb}, S. 2012, Advances in Space Research, 49, 1313, \dodoi{10.1016/j.asr.2012.02.004}

\bibitem[{{Fesen} {et~al.}(1985){Fesen}, {Blair}, \& {Kirshner}}]{Fesen1985}
{Fesen}, R.~A., {Blair}, W.~P., \& {Kirshner}, R.~P. 1985, \apj, 292, 29, \dodoi{10.1086/163130}

\bibitem[{{Fesen} {et~al.}(1983){Fesen}, {Gull}, \& {Ketelsen}}]{Fesen1983}
{Fesen}, R.~A., {Gull}, T.~R., \& {Ketelsen}, D.~A. 1983, \apjs, 51, 337, \dodoi{10.1086/190853}

\bibitem[{{Fesen} {et~al.}(2015){Fesen}, {Neustadt}, {Black}, \& {Koeppel}}]{Fesen2015}
{Fesen}, R.~A., {Neustadt}, J. M.~M., {Black}, C.~S., \& {Koeppel}, A. H.~D. 2015, \apj, 812, 37, \dodoi{10.1088/0004-637X/812/1/37}

\bibitem[{{Fesen} {et~al.}(2019){Fesen}, {Neustadt}, {How}, \& {Black}}]{Fesen2019}
{Fesen}, R.~A., {Neustadt}, J. M.~M., {How}, T.~G., \& {Black}, C.~S. 2019, \mnras, 486, 4701, \dodoi{10.1093/mnras/stz1140}

\bibitem[{{Fesen} {et~al.}(2020){Fesen}, {Weil}, {Raymond}, {Huet}, {Rusterholz}, {di Cicco}, {Mittelman}, {Walker}, {Drechsler}, \& {Faworski}}]{Fesen2020}
{Fesen}, R.~A., {Weil}, K.~E., {Raymond}, J.~C., {et~al.} 2020, \mnras, 498, 5194, \dodoi{10.1093/mnras/staa2765}

\bibitem[{{Fesen} {et~al.}(2021){Fesen}, {Drechsler}, {Weil}, {Strottner}, {Raymond}, {Rupert}, {Milisavljevic}, {Subrayan}, {di Cicco}, {Walker}, {Mittelman}, \& {Ludgate}}]{Fesen2021}
{Fesen}, R.~A., {Drechsler}, M., {Weil}, K.~E., {et~al.} 2021, \apj, 920, 90, \dodoi{10.3847/1538-4357/ac0ada}

\bibitem[{{Fesen} {et~al.}(2023){Fesen}, {Kimeswenger}, {Shull}, {Drechsler}, {Strottner}, {Sainty}, {Falls}, {Vergnes}, {Martino}, {Walker}, \& {Rupert}}]{Fesen2023}
{Fesen}, R.~A., {Kimeswenger}, S., {Shull}, J.~M., {et~al.} 2023, \apj, 957, 82, \dodoi{10.3847/1538-4357/acfe0d}

\bibitem[{{Filipovi{\'c}} {et~al.}(2023){Filipovi{\'c}}, {Dai}, {Arbutina}, {Hurley-Walker}, {Brose}, {Becker}, {Sano}, {Uro{\v{s}}evi{\'c}}, {Jarrett}, {Hopkins}, {Alsaberi}, {Alsulami}, {Bordiu}, {Ball}, {Bufano}, {Burger-Scheidlin}, {Crawford}, {English}, {Haberl}, {Ingallinera}, {Kapinska}, {Kavanagh}, {Koribalski}, {Kothes}, {Lazarevi{\'c}}, {Mackey}, {Rowell}, {Leahy}, {Loru}, {Macgregor}, {Nicastro}, {Norris}, {Riggi}, {Sasaki}, {Stupar}, {Trigilio}, {Umana}, {Vernstrom}, \& {Vukoti{\'c}}}]{Filipovic2023}
{Filipovi{\'c}}, M.~D., {Dai}, S., {Arbutina}, B., {et~al.} 2023, \aj, 166, 149, \dodoi{10.3847/1538-3881/acf19c}

\bibitem[{{Gaetz} {et~al.}(1988){Gaetz}, {Edgar}, \& {Chevalier}}]{Gaetz1988}
{Gaetz}, T.~J., {Edgar}, R.~J., \& {Chevalier}, R.~A. 1988, \apj, 329, 927, \dodoi{10.1086/166437}

\bibitem[{{Gao} \& {Han}(2014)}]{Gao2014}
{Gao}, X.~Y., \& {Han}, J.~L. 2014, \aap, 567, A59, \dodoi{10.1051/0004-6361/201424128}

\bibitem[{{Gaustad} {et~al.}(2001){Gaustad}, {McCullough}, {Rosing}, \& {Van Buren}}]{Gaustad2001}
{Gaustad}, J.~E., {McCullough}, P.~R., {Rosing}, W., \& {Van Buren}, D. 2001, \pasp, 113, 1326, \dodoi{10.1086/323969}

\bibitem[{{Georgelin} {et~al.}(1994){Georgelin}, {Amram}, {Georgelin}, {Le Coarer}, \& {Marcelin}}]{Georgelin1994}
{Georgelin}, Y.~M., {Amram}, P., {Georgelin}, Y.~P., {Le Coarer}, E., \& {Marcelin}, M. 1994, \aaps, 108, 513

\bibitem[{{Georgelin} {et~al.}(2000){Georgelin}, {Russeil}, {Amram}, {Georgelin}, {Marcelin}, {Parker}, \& {Viale}}]{Georgelin2000A}
{Georgelin}, Y.~M., {Russeil}, D., {Amram}, P., {et~al.} 2000, \aap, 357, 308

\bibitem[{{Georgelin} {et~al.}(1996){Georgelin}, {Russeil}, {Marcelin}, {Amram}, {Georgelin}, {Goldes}, {Le Coarer}, \& {Morandini}}]{Georgelin1996}
{Georgelin}, Y.~M., {Russeil}, D., {Marcelin}, M., {et~al.} 1996, \aaps, 120, 41

\bibitem[{{Gerbrandt} {et~al.}(2014){Gerbrandt}, {Foster}, {Kothes}, {Geisb{\"u}sch}, \& {Tung}}]{Gerbrandt2014}
{Gerbrandt}, S., {Foster}, T.~J., {Kothes}, R., {Geisb{\"u}sch}, J., \& {Tung}, A. 2014, \aap, 566, A76, \dodoi{10.1051/0004-6361/201423679}

\bibitem[{{Gilhuly} {et~al.}(2022){Gilhuly}, {Merritt}, {Abraham}, {Danieli}, {Lokhorst}, {Liu}, {van Dokkum}, {Conroy}, \& {Greco}}]{Gilhuly2022}
{Gilhuly}, C., {Merritt}, A., {Abraham}, R., {et~al.} 2022, \apj, 932, 44, \dodoi{10.3847/1538-4357/ac6750}

\bibitem[{{Green} {et~al.}(2014){Green}, {Reeves}, \& {Murphy}}]{Green2014}
{Green}, A.~J., {Reeves}, S.~N., \& {Murphy}, T. 2014, \pasa, 31, e042, \dodoi{10.1017/pasa.2014.37}

\bibitem[{{Green}(2004)}]{Green2004}
{Green}, D.~A. 2004, Bulletin of the Astronomical Society of India, 32, 335.
\newblock \doarXiv{astro-ph/0411083}

\bibitem[{{Green}(2019)}]{Green2019}
---. 2019, Journal of Astrophysics and Astronomy, 40, 36, \dodoi{10.1007/s12036-019-9601-6}

\bibitem[{{Green}(2012)}]{pyrafcite}
{Green}, W. 2012, Society for Astronomical Sciences Annual Symposium, 31, 159

\bibitem[{{Haffner} {et~al.}(2003){Haffner}, {Reynolds}, {Tufte}, {Madsen}, {Jaehnig}, \& {Percival}}]{Haffner2003}
{Haffner}, L.~M., {Reynolds}, R.~J., {Tufte}, S.~L., {et~al.} 2003, \apjs, 149, 405, \dodoi{10.1086/378850}

\bibitem[{{Hanbury Brown} \& {Hazard}(1953)}]{HB1953}
{Hanbury Brown}, R., \& {Hazard}, C. 1953, \mnras, 113, 123, \dodoi{10.1093/mnras/113.2.123}

\bibitem[{{Harris}(1962)}]{Harris1962}
{Harris}, D.~E. 1962, \apj, 135, 661, \dodoi{10.1086/147310}

\bibitem[{{Harris} \& {Roberts}(1960)}]{Harris1960}
{Harris}, D.~E., \& {Roberts}, J.~A. 1960, \pasp, 72, 237, \dodoi{10.1086/127521}

\bibitem[{{How} {et~al.}(2018){How}, {Fesen}, {Neustadt}, {Black}, \& {Outters}}]{How2018}
{How}, T.~G., {Fesen}, R.~A., {Neustadt}, J. M.~M., {Black}, C.~S., \& {Outters}, N. 2018, \mnras, 478, 1987, \dodoi{10.1093/mnras/sty1007}

\bibitem[{{Javanmardi} {et~al.}(2016){Javanmardi}, {Martinez-Delgado}, {Kroupa}, {Henkel}, {Crawford}, {Teuwen}, {Gabany}, {Hanson}, {Chonis}, \& {Neyer}}]{Java2016}
{Javanmardi}, B., {Martinez-Delgado}, D., {Kroupa}, P., {et~al.} 2016, \aap, 588, A89, \dodoi{10.1051/0004-6361/201527745}

\bibitem[{{Kopsacheili} {et~al.}(2020){Kopsacheili}, {Zezas}, \& {Leonidaki}}]{Kop2020}
{Kopsacheili}, M., {Zezas}, A., \& {Leonidaki}, I. 2020, \mnras, 491, 889, \dodoi{10.1093/mnras/stz2594}

\bibitem[{{Kothes}(2003)}]{Kothes2003}
{Kothes}, R. 2003, \aap, 408, 187, \dodoi{10.1051/0004-6361:20030853}

\bibitem[{{Kothes} {et~al.}(2006){Kothes}, {Fedotov}, {Foster}, \& {Uyan{\i}ker}}]{Kothes2006}
{Kothes}, R., {Fedotov}, K., {Foster}, T.~J., \& {Uyan{\i}ker}, B. 2006, \aap, 457, 1081, \dodoi{10.1051/0004-6361:20065062}

\bibitem[{{Kothes} {et~al.}(2017){Kothes}, {Reich}, {Foster}, \& {Reich}}]{Kothes2017}
{Kothes}, R., {Reich}, P., {Foster}, T.~J., \& {Reich}, W. 2017, \aap, 597, A116, \dodoi{10.1051/0004-6361/201629848}

\bibitem[{{Kuijken} {et~al.}(2019){Kuijken}, {Heymans}, {Dvornik}, {Hildebrandt}, {de Jong}, {Wright}, {Erben}, {Bilicki}, {Giblin}, {Shan}, {Getman}, {Grado}, {Hoekstra}, {Miller}, {Napolitano}, {Paolilo}, {Radovich}, {Schneider}, {Sutherland}, {Tewes}, {Tortora}, {Valentijn}, \& {Verdoes Kleijn}}]{Kuijken2019}
{Kuijken}, K., {Heymans}, C., {Dvornik}, A., {et~al.} 2019, \aap, 625, A2, \dodoi{10.1051/0004-6361/201834918}

\bibitem[{{Lanzetta} {et~al.}(2023){Lanzetta}, {Gromoll}, {Shara}, {Berg}, {Valls-Gabaud}, {Walter}, \& {Webb}}]{Lanzetta2023}
{Lanzetta}, K.~M., {Gromoll}, S., {Shara}, M.~M., {et~al.} 2023, \pasp, 135, 015002, \dodoi{10.1088/1538-3873/acaee6}

\bibitem[{{Le Coarer} {et~al.}(1992){Le Coarer}, {Amram}, {Boulesteix}, {Georgelin}, {Georgelin}, {Marcelin}, {Joulie}, \& {Urios}}]{LeCoarer1992}
{Le Coarer}, E., {Amram}, P., {Boulesteix}, J., {et~al.} 1992, \aap, 257, 389

\bibitem[{{Leonidaki} {et~al.}(2013){Leonidaki}, {Boumis}, \& {Zezas}}]{Leonidaki2013}
{Leonidaki}, I., {Boumis}, P., \& {Zezas}, A. 2013, \mnras, 429, 189, \dodoi{10.1093/mnras/sts324}

\bibitem[{{Li} {et~al.}(2016){Li}, {Torres}, {de O{\~n}a Wilhelmi}, {Rea}, \& {Martin}}]{Li2016}
{Li}, J., {Torres}, D.~F., {de O{\~n}a Wilhelmi}, E., {Rea}, N., \& {Martin}, J. 2016, \apj, 831, 19, \dodoi{10.3847/0004-637X/831/1/19}

\bibitem[{{Long}(2017)}]{Long2017}
{Long}, K.~S. 2017, {Galactic and Extragalactic Samples of Supernova Remnants: How They Are Identified and What They Tell Us}, ed. A.~W. {Alsabti} \& P.~{Murdin}, 2005, \dodoi{10.1007/978-3-319-21846-5_90}

\bibitem[{{Lozinskaia}(1980)}]{Lozinskaya1980}
{Lozinskaia}, T.~A. 1980, \aap, 84, 26

\bibitem[{{Lozinskaya}(1972)}]{Lozinskaya1972}
{Lozinskaya}, T.~A. 1972, \sovast, 16, 219

\bibitem[{{Mantovanini} {et~al.}(2024){Mantovanini}, {Becker}, {Khokhriakova}, {Hurley-Walker}, {Anderson}, \& {Nicastro}}]{Mant2024}
{Mantovanini}, S., {Becker}, W., {Khokhriakova}, A., {et~al.} 2024, arXiv e-prints, arXiv:2401.17294, \dodoi{10.48550/arXiv.2401.17294}

\bibitem[{{Martinez-Delgado}(2020)}]{Martinez2020}
{Martinez-Delgado}, D. 2020, arXiv e-prints, arXiv:2001.05746, \dodoi{10.48550/arXiv.2001.05746}

\bibitem[{{Mart{\'\i}nez-Delgado} {et~al.}(2015){Mart{\'\i}nez-Delgado}, {D'Onghia}, {Chonis}, {Beaton}, {Teuwen}, {GaBany}, {Grebel}, \& {Morales}}]{Martinez2015}
{Mart{\'\i}nez-Delgado}, D., {D'Onghia}, E., {Chonis}, T.~S., {et~al.} 2015, \aj, 150, 116, \dodoi{10.1088/0004-6256/150/4/116}

\bibitem[{{Mart{\'\i}nez-Delgado} {et~al.}(2021){Mart{\'\i}nez-Delgado}, {Rom{\'a}n}, {Erkal}, {Schirmer}, {Roca-F{\`a}brega}, {Laine}, {Donatiello}, {Jimenez}, {Malin}, \& {Carballo-Bello}}]{Martinez2021}
{Mart{\'\i}nez-Delgado}, D., {Rom{\'a}n}, J., {Erkal}, D., {et~al.} 2021, \mnras, 506, 5030, \dodoi{10.1093/mnras/stab1874}

\bibitem[{{Martini} {et~al.}(2011){Martini}, {Stoll}, {Derwent}, {Zhelem}, {Atwood}, {Gonzalez}, {Mason}, {O'Brien}, {Pappalardo}, {Pogge}, {Ward}, \& {Wong}}]{Martini2011}
{Martini}, P., {Stoll}, R., {Derwent}, M.~A., {et~al.} 2011, \pasp, 123, 187, \dodoi{10.1086/658357}

\bibitem[{{Massey} \& {Gronwall}(1990)}]{Massey1990}
{Massey}, P., \& {Gronwall}, C. 1990, \apj, 358, 344, \dodoi{10.1086/168991}

\bibitem[{{Mavromatakis} {et~al.}(2004){Mavromatakis}, {Aschenbach}, {Boumis}, \& {Papamastorakis}}]{Mav2004}
{Mavromatakis}, F., {Aschenbach}, B., {Boumis}, P., \& {Papamastorakis}, J. 2004, \aap, 415, 1051, \dodoi{10.1051/0004-6361:20031694}

\bibitem[{{Mavromatakis} {et~al.}(2009){Mavromatakis}, {Boumis}, {Meaburn}, \& {Caulet}}]{Mav2009}
{Mavromatakis}, F., {Boumis}, P., {Meaburn}, J., \& {Caulet}, A. 2009, \aap, 503, 129, \dodoi{10.1051/0004-6361/200912211}

\bibitem[{{Mavromatakis} {et~al.}(2005){Mavromatakis}, {Boumis}, {Xilouris}, {Papamastorakis}, \& {Alikakos}}]{Mav2005}
{Mavromatakis}, F., {Boumis}, P., {Xilouris}, E., {Papamastorakis}, J., \& {Alikakos}, J. 2005, \aap, 435, 141, \dodoi{10.1051/0004-6361:20042187}

\bibitem[{{Mavromatakis} {et~al.}(2000){Mavromatakis}, {Papamastorakis}, {Paleologou}, \& {Ventura}}]{Mav2000}
{Mavromatakis}, F., {Papamastorakis}, J., {Paleologou}, E.~V., \& {Ventura}, J. 2000, \aap, 353, 371

\bibitem[{{Mavromatakis} {et~al.}(2001){Mavromatakis}, {Papamastorakis}, {Ventura}, {Becker}, {Paleologou}, \& {Schaudel}}]{Mav2001}
{Mavromatakis}, F., {Papamastorakis}, J., {Ventura}, J., {et~al.} 2001, \aap, 370, 265, \dodoi{10.1051/0004-6361:20010137}

\bibitem[{{Mavromatakis} {et~al.}(2007){Mavromatakis}, {Xilouris}, \& {Boumis}}]{Mav2007}
{Mavromatakis}, F., {Xilouris}, E.~M., \& {Boumis}, P. 2007, \aap, 461, 991, \dodoi{10.1051/0004-6361:20054786}

\bibitem[{{Meaburn}(1978)}]{Meaburn1978}
{Meaburn}, J. 1978, \ao, 17, 1271, \dodoi{10.1364/AO.17.001271}

\bibitem[{{Minkowski}(1958)}]{Minkowski1958}
{Minkowski}, R. 1958, Reviews of Modern Physics, 30, 1048, \dodoi{10.1103/RevModPhys.30.1048}

\bibitem[{{Oke}(1974)}]{Oke1974}
{Oke}, J.~B. 1974, \apjs, 27, 21, \dodoi{10.1086/190287}

\bibitem[{{Parker} {et~al.}(2005){Parker}, {Phillipps}, {Pierce}, {Hartley}, {Hambly}, {Read}, {MacGillivray}, {Tritton}, {Cass}, {Cannon}, {Cohen}, {Drew}, {Frew}, {Hopewell}, {Mader}, {Malin}, {Masheder}, {Morgan}, {Morris}, {Russeil}, {Russell}, \& {Walker}}]{Parker2005}
{Parker}, Q.~A., {Phillipps}, S., {Pierce}, M.~J., {et~al.} 2005, \mnras, 362, 689, \dodoi{10.1111/j.1365-2966.2005.09350.x}

\bibitem[{{Parker} {et~al.}(1979){Parker}, {Gull}, \& {Kirshner}}]{Parker1979}
{Parker}, R.~A.~R., {Gull}, T.~R., \& {Kirshner}, R.~P. 1979, {An emission-line survey of the Milky Way}, Vol. 434

\bibitem[{{Pineault} {et~al.}(1997){Pineault}, {Landecker}, {Swerdlyk}, \& {Reich}}]{Pineault1997}
{Pineault}, S., {Landecker}, T.~L., {Swerdlyk}, C.~M., \& {Reich}, W. 1997, \aap, 324, 1152

\bibitem[{{Ranasinghe} \& {Leahy}(2022)}]{Leahy2022}
{Ranasinghe}, S., \& {Leahy}, D. 2022, \apj, 940, 63, \dodoi{10.3847/1538-4357/ac940a}

\bibitem[{{Raymond} {et~al.}(1983){Raymond}, {Blair}, {Fesen}, \& {Gull}}]{Raymond1983}
{Raymond}, J.~C., {Blair}, W.~P., {Fesen}, R.~A., \& {Gull}, T.~R. 1983, \apj, 275, 636, \dodoi{10.1086/161561}

\bibitem[{{Raymond} {et~al.}(2020){Raymond}, {Caldwell}, {Fesen}, {Weil}, {Boumis}, {di Cicco}, {Mittelman}, \& {Walker}}]{Raymond2020}
{Raymond}, J.~C., {Caldwell}, N., {Fesen}, R.~A., {et~al.} 2020, \apj, 888, 90, \dodoi{10.3847/1538-4357/ab5e84}

\bibitem[{{Reich} {et~al.}(1990{\natexlab{a}}){Reich}, {Fuerst}, {Reich}, \& {Reif}}]{Reich90a}
{Reich}, W., {Fuerst}, E., {Reich}, P., \& {Reif}, K. 1990{\natexlab{a}}, \aaps, 85, 633

\bibitem[{{Reich} {et~al.}(1990{\natexlab{b}}){Reich}, {Reich}, \& {Fuerst}}]{Reich90b}
{Reich}, W., {Reich}, P., \& {Fuerst}, E. 1990{\natexlab{b}}, \aaps, 83, 539

\bibitem[{{Rengelink} {et~al.}(1997){Rengelink}, {Tang}, {de Bruyn}, {Miley}, {Bremer}, {Roettgering}, \& {Bremer}}]{Rengelink1997}
{Rengelink}, R.~B., {Tang}, Y., {de Bruyn}, A.~G., {et~al.} 1997, \aaps, 124, 259, \dodoi{10.1051/aas:1997358}

\bibitem[{{Reynolds}(2011)}]{Reynolds2011}
{Reynolds}, S.~P. 2011, \apss, 336, 257, \dodoi{10.1007/s10509-010-0559-8}

\bibitem[{{Rosado} \& {Gonzalez}(1981)}]{Rosado1981}
{Rosado}, M., \& {Gonzalez}, J. 1981, \rmxaa, 5, 93

\bibitem[{{Russeil} {et~al.}(1998){Russeil}, {Georgelin}, {Amram}, {Gach}, {Georgelin}, \& {Marcelin}}]{Russeil1998}
{Russeil}, D., {Georgelin}, Y.~M., {Amram}, P., {et~al.} 1998, \aaps, 130, 119, \dodoi{10.1051/aas:1998406}

\bibitem[{{Sabbadin}(1976)}]{Sabbadin1976}
{Sabbadin}, F. 1976, \aap, 51, 159

\bibitem[{{Sabin} {et~al.}(2013){Sabin}, {Parker}, {Contreras}, {Olgu{\'\i}n}, {Frew}, {Stupar}, {V{\'a}zquez}, {Wright}, {Corradi}, \& {Morris}}]{Sabin2013}
{Sabin}, L., {Parker}, Q.~A., {Contreras}, M.~E., {et~al.} 2013, \mnras, 431, 279, \dodoi{10.1093/mnras/stt160}

\bibitem[{{Seward} {et~al.}(1995){Seward}, {Dame}, {Fesen}, \& {Aschenbach}}]{Seward1995}
{Seward}, F.~D., {Dame}, T.~M., {Fesen}, R.~A., \& {Aschenbach}, B. 1995, \apj, 449, 681, \dodoi{10.1086/176089}

\bibitem[{{Sezer} {et~al.}(2012){Sezer}, {G{\"o}k}, \& {Aktekin}}]{Sezer2012}
{Sezer}, A., {G{\"o}k}, F., \& {Aktekin}, E. 2012, \mnras, 427, 1168, \dodoi{10.1111/j.1365-2966.2012.22015.x}

\bibitem[{{Shan} {et~al.}(2018){Shan}, {Zhu}, {Tian}, {Zhang}, {Zhang}, {Wu}, \& {Yang}}]{Shan2018}
{Shan}, S.~S., {Zhu}, H., {Tian}, W.~W., {et~al.} 2018, \apjs, 238, 35, \dodoi{10.3847/1538-4365/aae07a}

\bibitem[{{Sieber} {et~al.}(1979){Sieber}, {Haslam}, \& {Salter}}]{Sieber1979}
{Sieber}, W., {Haslam}, C.~G.~T., \& {Salter}, C.~J. 1979, \aap, 74, 361

\bibitem[{{Slane} {et~al.}(1997){Slane}, {Seward}, {Bandiera}, {Torii}, \& {Tsunemi}}]{Slane1997}
{Slane}, P., {Seward}, F.~D., {Bandiera}, R., {Torii}, K., \& {Tsunemi}, H. 1997, \apj, 485, 221, \dodoi{10.1086/304416}

\bibitem[{{Smithsonian Astrophysical Observatory}(2000)}]{ds9cite}
{Smithsonian Astrophysical Observatory}. 2000, {SAOImage DS9: A utility for displaying astronomical images in the X11 window environment}.
\newblock \doeprint{0003.002}

\bibitem[{{Sofue} \& {Reich}(1979)}]{Sofue1979}
{Sofue}, Y., \& {Reich}, W. 1979, \aaps, 38, 251

\bibitem[{{Spitler} {et~al.}(2019){Spitler}, {Longbottom}, {Alvarado-Montes}, {Bazkiaei}, {Caddy}, {Gee}, {Horton}, {Lee}, \& {Prole}}]{Spitler2019}
{Spitler}, L.~R., {Longbottom}, F.~D., {Alvarado-Montes}, J.~A., {et~al.} 2019, arXiv e-prints, arXiv:1911.11579, \dodoi{10.48550/arXiv.1911.11579}

\bibitem[{{Stupar} \& {Parker}(2011)}]{Stupar2011}
{Stupar}, M., \& {Parker}, Q.~A. 2011, \mnras, 414, 2282, \dodoi{10.1111/j.1365-2966.2011.18547.x}

\bibitem[{{Stupar} \& {Parker}(2012)}]{Stupar2012}
---. 2012, \mnras, 419, 1413, \dodoi{10.1111/j.1365-2966.2011.19797.x}

\bibitem[{{Stupar} {et~al.}(2008){Stupar}, {Parker}, \& {Filipovi{\'c}}}]{Stupar2008}
{Stupar}, M., {Parker}, Q.~A., \& {Filipovi{\'c}}, M.~D. 2008, \mnras, 390, 1037, \dodoi{10.1111/j.1365-2966.2008.13761.x}

\bibitem[{{Stupar} {et~al.}(2007){Stupar}, {Parker}, {Filipovi{\'c}}, {Frew}, {Boji{\v{c}}i{\'c}}, \& {Aschenbach}}]{Stupar2007}
{Stupar}, M., {Parker}, Q.~A., {Filipovi{\'c}}, M.~D., {et~al.} 2007, \mnras, 381, 377, \dodoi{10.1111/j.1365-2966.2007.12296.x}

\bibitem[{{Stupar} {et~al.}(2018){Stupar}, {Parker}, \& {Frew}}]{Stupar2018}
{Stupar}, M., {Parker}, Q.~A., \& {Frew}, D.~J. 2018, \mnras, 479, 4432, \dodoi{10.1093/mnras/sty1684}

\bibitem[{{Sun} {et~al.}(2011){Sun}, {Reich}, {Wang}, {Han}, \& {Reich}}]{Sun2011}
{Sun}, X.~H., {Reich}, W., {Wang}, C., {Han}, J.~L., \& {Reich}, P. 2011, \aap, 535, A64, \dodoi{10.1051/0004-6361/201117679}

\bibitem[{{Tamura} \& {Weinberger}(1995)}]{TaWe1995}
{Tamura}, S., \& {Weinberger}, R. 1995, \aap, 298, 204

\bibitem[{{Thureau} {et~al.}(2009){Thureau}, {Monnier}, {Traub}, {Millan-Gabet}, {Pedretti}, {Berger}, {Garcia}, {Schloerb}, \& {Tannirkulam}}]{Thureau2009}
{Thureau}, N.~D., {Monnier}, J.~D., {Traub}, W.~A., {et~al.} 2009, \mnras, 398, 1309, \dodoi{10.1111/j.1365-2966.2009.14949.x}

\bibitem[{{van den Bergh}(1978)}]{vdb1978a}
{van den Bergh}, S. 1978, \apjs, 38, 119, \dodoi{10.1086/190549}

\bibitem[{{van den Bergh} {et~al.}(1973){van den Bergh}, {Marscher}, \& {Terzian}}]{vdb1973}
{van den Bergh}, S., {Marscher}, A.~P., \& {Terzian}, Y. 1973, \apjs, 26, 19, \dodoi{10.1086/190278}

\bibitem[{{van Dokkum}(2001)}]{vanDokkum2001}
{van Dokkum}, P.~G. 2001, \pasp, 113, 1420, \dodoi{10.1086/323894}

\bibitem[{{Walker} {et~al.}(2001){Walker}, {Zealey}, \& {Parker}}]{Walker2001}
{Walker}, A.~J., {Zealey}, W.~J., \& {Parker}, Q.~A. 2001, \pasa, 18, 259, \dodoi{10.1071/AS01063}

\bibitem[{{Wendker}(1968)}]{Wendker1968}
{Wendker}, H. 1968, \zap, 69, 392

\bibitem[{{Wendker}(1971)}]{Wendker1971}
{Wendker}, H.~J. 1971, \aap, 13, 65

\bibitem[{{Willis}(1973)}]{Willis1973}
{Willis}, A.~G. 1973, \aap, 26, 237

\bibitem[{{Zealey} {et~al.}(1979){Zealey}, {Elliott}, \& {Malin}}]{Zealey1979}
{Zealey}, W.~J., {Elliott}, K.~H., \& {Malin}, D.~F. 1979, \aaps, 38, 39

\end{thebibliography}

\end{document}